\documentclass[12pt,usenames,dvipsnames]{article}
\usepackage[utf8]{inputenc}
\usepackage[en-GB]{datetime2}

\title{SUSY Dimensional Reduction}
\author{Arcadia Fegebank \\22705104}
\date{\today}

\usepackage{import}
\usepackage{latexsym}
\usepackage{fancyhdr}
\usepackage{amsfonts}
\usepackage{amssymb}
\usepackage{amsmath}
\usepackage{amsthm}
\usepackage{mathtools}
\usepackage{slashed}

\usepackage{graphicx} 
\usepackage{indentfirst}
\usepackage{bbm}
\usepackage{enumerate}
\usepackage{appendix}
\usepackage{chngcntr}
\usepackage{etoolbox}
\usepackage{changepage}
\usepackage{slashed}

\usepackage{pdfpages}
\usepackage{verbatim}
\usepackage{mathrsfs}

\usepackage{dsfont}
\usepackage{cite}
\usepackage{xcolor}
\usepackage{enumitem}
\usepackage{ytableau}
\usepackage{tikz}
\usepackage[colorlinks=false,citecolor=blue, urlcolor=blue ]{hyperref}

\usepackage{chngcntr}

\counterwithin*{equation}{section}

\allowdisplaybreaks

\AtBeginEnvironment{subappendices}{%
\chapter*{Appendix}
\addcontentsline{toc}{chapter}{Appendices}
\counterwithin{figure}{section}
\counterwithin{table}{section}
}

\graphicspath{{images/}}

\parskip=\medskipamount

\def\a{\alpha}
\def\b{\beta}

\def\d{\delta}

\def\g{\gamma}
\def\G{\Gamma}

\def\j{\psi}

\def\l{\lambda}
\def\m{\mu}

\def\n{\nu}
\def\o{\omega}

\def\q{\theta}

\def\r{\rho}
\def\s{\sigma}

\def\x{\xi}
\def\z{\zeta}
\def\D{\Delta}
\def\F{\Phi}
\def\J{\Psi}
\def\L{\Lambda}

\def\P{\Pi}

\def\S{\Sigma}

\newcommand {\cA}{{\cal A}}
\newcommand {\cB}{{\cal B}}
\newcommand {\cC}{{\cal C}}
\newcommand {\cD}{{\cal D}}

\newcommand {\cH}{{\cal H}}

\newcommand {\cJ}{{\cal J}}
\newcommand {\cK}{{\cal K}}
\newcommand {\cL}{{\cal L}}
\newcommand {\cM}{{\cal M}}
\newcommand {\cN}{{\cal N}}
\newcommand {\cO}{{\cal O}}

\newcommand {\cV}{{\cal V}}
\newcommand {\cW}{{\cal W}}
\newcommand {\cX}{{\cal X}}

\newcommand{\ad}{{\dot{\alpha}}}                           
\newcommand{\bd}{{\dot{\beta}}}

\newcommand{\gd}{{\dot\g}}
                        
\newcommand{\ve}{\varepsilon}                          
\newcommand{\cDB}{{\bar\cD}}                        
\newcommand{\DB}{\bar{D}}

\newcommand{\ab}{{\a\b}}

\renewcommand{\aa}{{\a\ad}}

\newcommand{\pa}{\partial}

\def\rd{{\rm d}}
\def\ri{{\rm i}}
\def\re{{\rm e}}

\def\rc{{\rm c}}

\newcommand{\vf}{\varphi}

\newcommand{\be}{\begin{equation}}
\newcommand{\ee}{\end{equation}}
\newcommand{\bsubeq}{\begin{subequations}}
\newcommand{\esubeq}{\end{subequations}}
\newcommand{\ba}{\begin{align}}
\newcommand{\ea}{\end{align}}
\newcommand{\bea}{\begin{eqnarray}}
\newcommand{\eea}{\end{eqnarray}}
\newcommand{\non}{\nonumber}

\newcommand{\bm}[1]{\mbox{\boldmath$#1$}}

\def \un{\underline}

\begin{document}
\begin{titlepage}
\begin{flushright}
April, 2026 \\
\end{flushright}
\vspace{5mm}

\begin{center}
{\Large \bf 
 Gauge-invariant off-shell formulations for 3D massive higher-spin supermultiplets
}
\end{center}

\begin{center}

{\bf Evgeny I. Buchbinder, Arcadia J. Fegebank and Sergei M. Kuzenko} \\
\vspace{5mm}

\footnotesize{
{\it Department of Physics M013, The University of Western Australia\\
35 Stirling Highway, Perth W.A. 6009, Australia}}  
~\\
\vspace{2mm}
~\\
Email: \texttt{ 
evgeny.buchbinder@uwa.edu.au, arcadia.fegebank@research.uwa.edu.au, sergei.kuzenko@uwa.edu.au}\\
\vspace{2mm}

\end{center}
\begin{abstract}
\baselineskip=14pt
Making use of the known off-shell formulations for massless higher-spin ${\cal N}=1$ supermultiplets in four dimensions, gauge-invariant off-shell actions for massive higher-spin ${\cal N}=2$ supermultiplets in three dimensions (3D) are derived by Kaluza-Klein reduction in superspace. To illustrate the formalism, we also construct, for the first time, massive gauge-invariant 3D ${\cal N}=2$ supersymmetric counterparts of the linearised actions for the old and new minimal supergravity theories. 
Our off-shell ${\cal N}=2$ supermultiplets carry a non-zero central charge, and are formulated in 3D ${\cal N}=2$ central charge superspace. The models can be reduced to 3D ${\cal N}=1$ superspace, by integrating out two Grassmann variables, and then consistent reality conditions on the superfields can be imposed. As a result, only two supercharges remain unbroken.    
\end{abstract}
\vspace{5mm}

\vfill

\vfill
\end{titlepage}

\newpage
\renewcommand{\thefootnote}{\arabic{footnote}}
\setcounter{footnote}{0}
\tableofcontents{}
\vspace{1cm}
\bigskip\hrule

\allowdisplaybreaks

\section{Introduction}

In supersymmetric field theory, an interesting open problem is to develop a regular procedure to construct gauge-invariant off-shell formulations for massive higher-spin supermultiplets in three, four and five dimensions. In the framework of the 4D $\cN=1$ Poincar\'e supersymmetry, non-gauge off-shell formulations have been constructed for the massive gravitino multiplets (superspin-1) \cite{Ogievetsky:1976qb, Ferrara:1983gn, Buchbinder:2002tt, Buchbinder:2005je}, the massive graviton multiplet (superspin-3/2) \cite{Buchbinder:2002gh, Gregoire:2004ic, Buchbinder:2005je, Gates:2006cq}, and massive half-integer superspin ($s>3/2$) multiplets  \cite{Koutrolikos:2020tel}. 
 To the best of our knowledge, gauge-invariant off-shell formulations have been constructed only for the massive gravitino multiplet \cite{Altendorfer:1999mn, Gates:2005su} and the massive graviton multiplet coupled to two massive scalar multiplets \cite{Berkovits:1998ua}.
 
 In the non-supersymmetric case,  the first non-gauge actions for massive particles of arbitrary spin in four dimensions were derived in 1974 by Singh and Hagen \cite{SH1, SH2}. 
The 1983 work by Zinoviev \cite{Zinovev1983} was one of the earliest publications 
to initiate the program of constructing gauge-invariant re-formulations of the models for massive higher-spin particles (spin $s>2$).
Specifically, he constructed gauge-invariant actions for massive fields of spin 3/2, 2, 5/2 and 3 with a ``correct'' massless limit. 
 Since then, several  approaches have been developed to generate gauge-invariant models for massive higher-spin fields.  
 These can be split in the following three categories. 
 \begin{enumerate} 
 
 \item
 Kaluza-Klein reduction to $d$ dimensions of the massless actions in $(d+1)$ dimensions \cite{Fronsdal:1978, FF, Curtright:1979uz, deWit:1979sib, Vasiliev}. This approach, outlined  in \cite{Siegel:1985tw, Aragone:1987dtt},
  was fully developed by Pashnev for arbitrary integer spins \cite{Pashnev:1989} and more recently by Lindwasser for arbitrary spins \cite{Lindwasser:2023}.\footnote{Dimensional reduction was carried out in 1985 by Rindani and Sivakumar
 \cite{Rindani:1985pi} to re-derive the gauge-invariant actions constructed earlier by Zinoviev \cite{Zinovev1983}.}
 For a recent review of Pashnev's approach see \cite{Fegebank}.
 
 \item Massive Stueckelberg-type deformations of massless models \cite{Klishevich:1997pd, Klishevich:1998yt, Zinoviev:2001, Metsaev}. This approach, which is a generalisation of Zinoviev's work \cite{Zinovev1983}, 
 has found numerous applications, including the construction of massive higher-spin supermultiplets in four \cite{Zinoviev:2007js} and three \cite{Buchbinder:2015mta, Zinoviev:2023vvz} dimensions with on-shell supersymmetry.
 
 \item The BRST approach to massive higher-spin theories advocated in 
 \cite{Buchbinder:2005ua, Buchbinder:2006ge, Buchbinder:2007ix, Buchbinder:2008ss}. 
 The power of BRST techniques was recognised long ago in the framework of string field theory \cite{Siegel:1985tw, Siegel:1984ogw, Siegel:1984xd}. Since then, the BRST setting has been used for the construction of higher-spin actions, including the so-called triplet formulations for  massless 
\cite{Ouvry:1986dv, Francia:2002pt, ST, Barnich:2005bn, Sorokin:2008tf, Campoleoni:2012th, Agugliaro:2016ngl} and massive \cite{Hussain:1988uk, Pashnev:1997rm, Bekaert:2003uc} 
reducible higher spin fields. In this context it is also necessary to mention the important work by Metsaev  
\cite{Metsaev:2014vda, Metsaev:2015yyv, Metsaev:2015oza}  on  the BRST-FV approach to massive and massless higher-spin fields.
 
 \end{enumerate}

The different approaches described above often lead to equivalent gauge-invariant massive models. For example, it was demonstrated in \cite{Fegebank} that, in the integer-spin case, the Klishevich-Zinoviev theory \cite{Klishevich:1997pd}
as well as the models proposed in  \cite{Lindwasser:2023, Buchbinder:2008ss, Asano:2019smc} are equivalent  to 
the theory derived earlier by Pashnev \cite{Pashnev:1989} using the dimensional reduction procedure. 
Ref. \cite{Fegebank} also made use of the Klishevich-Zinoviev theory \cite{Klishevich:1997pd} to derive, for the first time,  
a unique generalisation of the Singh-Hagen model for a massive integer-spin field in $d>4 $ dimensions. 

Returning to our goal of developing a regular procedure to construct gauge-invariant off-shell formulations for massive higher-spin supermultiplets, it appears that only dimensional reduction provides a natural path to achieve it.\footnote{Metsaev's formulation in terms of light-cone gauge unconstrained superfields \cite{Metsaev:2019aig, Metsaev:2021bjh} is powerful for constructing all cubic interaction vertices. However, we are interested in developing an off-shell approach with manifest Poincar\'{e} supersymmetry.} 
All attempts to extend the formalism of  \cite{Klishevich:1997pd, Klishevich:1998yt, Zinoviev:2001, Metsaev} to superspace have so far failed.\footnote{Keeping in mind the equivalence of Pashnev's theory  \cite{Pashnev:1989} 
and the Klishevich-Zinoviev theory \cite{Klishevich:1997pd} mentioned above, there is a hope that supersymmetric dimensional reduction may provide hints as to how to extend Zinoviev's approach to superspace.}
Concerning the BRTS formalism, this setting is quite powerful  and, in principle, applicable to general gauge theories. 
So far it remains unclear, however, how to set it up in superspace. These are the reasons why in this paper we concentrate on carrying out Kaluza-Klein dimensional reduction in superspace. In this regard, it should be noticed that the $\cN=4$ super Yan-Mills theory \cite{Brink:1976bc, Gliozzi:1976qd} and the $\cN=8 $ supergravity \cite{Cremmer:1978ds, Cremmer:1979up}
were originally constructed by toroidal dimensional reduction from ten and eleven dimensions, respectively
(see \cite{Scherk:1978fh} for a pedagogical review).

An important comment is in order.
In three dimensions, massless higher-spin models (spin $s>1$) and their supersymmetric extensions do not have any propagating degrees of freedom. However, these models can be combined with linearised (super)conformal higher-spin actions to generate topologically massive higher-spin theories, following the philosophy of topologically massive gauge theories \cite{WS,JS,DJT1,DJT2,DeserKay}. In the non-supersymmetric case, such 
topologically massive models for higher-spin gauge fields in both Minkowski and AdS space were constructed in 
\cite{KP18}. In the $\cN$-extended supersymmetric case, topologically massive higher-spin models have the schematic form (with the mass parameter omitted) 
\bea
S^{( \cN|n )}_{\rm massive} [ H_{\alpha(n)} , \dots]  = S^{( \cN|n )}_{\rm massless} [ H_{\alpha(n)}, \dots ] +
S^{( \cN|n )} [ H_{\alpha(n)} ] ~.
\label{SuperTMHST}
\eea
Here $S^{( \cN|n )} [ H_{\alpha(n)} ] $ is the $\cN$-extended superconformal gauge-invariant action 
\eqref{NnCSG-action}. The massless action $S^{( \cN|n )}_{\rm massless} [ H_{\alpha(n)}, \dots ] $
depends not only on the superconformal prepotential $H_{\a(n)}$, but also on certain compensating supermultiplets. 
The latter actions were constructed in \cite{KT} for $\cN=1$ and in \cite{KO} in the $\cN=2$ case. 

An alternative realisation of the topologically massive $\cN$-extended supersymmetric higher-spin models is given in Appendix
\ref{AppendixA}, eq. \eqref{action-mass}. The models \eqref{SuperTMHST} and \eqref{action-mass} are higher-derivative theories, 
although the corresponding equations of motion are equivalent to standard massive equations with at most two spacetime derivatives. In the non-supersymmetric case, this fact was demonstrated in \cite{KP18}.
All massive theories proposed in the present paper are at most second-order in spacetime derivatives. 

The primary purpose of this paper is to carry out the dimensional reduction of the 4D $\cN=1$ massless half-integer superspin theories proposed in \cite{KPS}, see \cite{BK:1998} for a pedagogical review.\footnote{A similar analysis for the 4D $\cN=1$ massless integer superspin theories \cite{KS, Hutomo:2017nce} will be described elsewhere.} 
Ref. \cite{KPS} introduced two off-shell formulations for each massless half-integer superspin, the so-called transverse and the longitudinal formulations, which are related to each other by a superfield duality transformation involving a first-order parent action. 
In order to carry out the Kaluza-Klein reduction, it suffices to work with the parent model. 
We follow the procedure of \cite{GK} to recast this model in terms of oscillators, at which point we carry out a dimensional reduction.
The resulting model we derive describes a massive gauge theory of half-integer superspin in 3D $\cN=2$ central charge superspace, which we will denote by $\mathbb{M}^{3|4}_c$. We will carry out an analysis of the massless limit of such a theory and compare it to the known 3D $\cN=2$ massless higher-spin theory \cite{KO}. We will also consider a particular reformulation of this theory which is analogous to the transverse formulation of the massless theory. Finally, we will verify that the dimensionally reduced  equations of motion do in fact describe a massive half-integer superspin multiplet.

This paper is organised as follows. We outline the general procedure of dimensional reduction in section \ref{Dim Red Section}, considering three different cases. In section \ref{SUSY Reduction Section} we consider the treatment of reductions of the superspace $\mathbb{M}^{4|4}$ and derive a new model for massive linearised supergravity in 3D with $\cN=2$ supersymmetry via dimensional reduction as an illustrative example. 
Section \ref{Massless 4D Section} reviews the massless 4D $\cN=1$ supersymmetric higher-spin models.
Section \ref{Oscillator Section} reviews how the oscillator formalism of \cite{GK} is constructed for $\mathbb{M}^{4|4}$ as well as analysing how it is deformed under dimensional reduction. The purpose of this oscillator formalism is to make symmetrisation easier to manage in calculations. Section \ref{Model Section} is dedicated to the reduction of the model from \cite{KPS}. We include an analysis of the massless limit of the resulting model in section \ref{Massless Limit} and compare it with the known massless theory \cite{KO}. We briefly summarise our work in section \ref{Conclusion}. 

The main body of the paper is accompanied by four technical appendices. Appendix \ref{AppendixA}
provides a brief review of the  topologically massive $\cN$-extended supersymmetric higher-spin models.
Appendix \ref{Properties Section} outlines several properties of the $\mathbb{M}^{3|4}_c$ oscillators which are used in calculations throughout the paper. Appendix \ref{appendixC} provides the proof of the gauge invariance of the model derived in section 
\ref{New Minimal Reduction}.
Finally, appendix \ref{Gauge Invariance Section} contains a proof that the model produced in section \ref{Model Section} is indeed gauge invariant.


\section{Dimensional reduction} \label{Dim Red Section}

We start by outlining the general approach for Kaluza-Klein dimensional reduction of field theories on Minkowski space $\mathbb{M}^{d+1}$ parametrised by Cartesian coordinates $\bm{x}^\m$, with $\m=0, 1, \dots, d$.
The reduction is obtained  by compactifying one space dimension, 
\bea
    \mathbb{M}^{d+1}\rightarrow \mathbb{M}^{d}\times S^1~,
\eea
where $S^1$ is a circle of radius of ${m}^{-1}$, and $m$ is the mass parameter. The resulting space will be parametrised by
\bea
    \bm{x}^\m~\rightarrow~(x^a,y)~,\qquad a=0,1, \dots, ,d-1~,
    \label{22y}
\eea
where $x^a\equiv \bm{x}^a$ parametrise $\mathbb{M}^{d}$, while the variable  $y \equiv \bm{x}^{d}$ corresponds to $S^1$. 
In the context of supersymmetric field theories formulated in a superspace $\mathbb{M}^{d+1|\d}$, the dimensional reduction is
\bea
    \mathbb{M}^{d+1|\d}\rightarrow \mathbb{M}^{d|\d}\times S^1~,
\eea
where $\d$ is the number of fermionic dimensions.

It will be insightful for our purposes to outline how the dimensional reduction is carried out across three different cases of action functionals. We shall begin with the case that is the most straightforward to reduce, that of actions involving only unconstrained complex fields and their conjugates. Such actions are used to describe fields of half-integer spin, while actions of integer spin fall into the second case, where the actions involve only real fields. As we shall see, this second case of action can be complexified so that they take the form of the first case and can thus be dimensionally reduced more easily, before the reality conditions are restored. Finally, we consider the case of actions involving real fields coupled to complex fields, which are used in formulations of higher-spin supermultiplets. We shall see that this case can also be complexified to simplify the reduction procedure.


\subsection{Case 1: complex fields}

Consider a massless action $S[\bm{\vf},\bar{\bm{\vf}}]$ in $\mathbb{M}^{d+1}$ which is quadratic in a set of $p$ free-fields $\{\bm{\vf}^i(\bm{x})\}$ where $\bm{x}^\m$ is a point in $\mathbb{M}^{d+1}$. This action has the form
\bea\label{Complex Action}
    S[\bm{\vf},\bar{\bm{\vf}}]=\int_{\mathbb{M}^{d+1}}\bar{\bm{\vf}}^{i} E_i(\bm{\vf})~,
\eea
where
\bea
    E_i(\bm{\vf})=A_{ij}\bm{\vf}^j~,
\eea
for a Hermitian differential operator $A_{ij}$. The on-shell field equations are
\bea
    E_i(\bm{\vf})=A_{ij}\bm{\vf}^j=0~,\qquad \bar{E}_i(\bar{\bm{\vf}})=\bar{A}_{ij}\bar{\bm{\vf}}^j=0~.
\eea
The gauge transformations take the form
\bea
    \d\bm{\vf}^i(\bm{\xi})=R^{i}_{~\a}\bm{\xi}^{\a}~,\qquad \d\bar{\bm{\vf}}^i(\bar{\bm{\xi}})=\bar{R}^{i}_{~\a}\bar{\bm{\xi}}^{\a}~,
\eea
with $R_{ij}$ being a differential operator such that
\bea
    A_{ij}R^j_{~\a}\bm{\xi}^{\a}=0~,
\eea
given a set of $r$ gauge parameters $\{\bm{\xi}^{\a}(\bm{x})\}$.

General fields and gauge parameters on $\mathbb{M}^{d}\times S^1$ are represented  by Fourier series 
\bsubeq \label{seriesall}
\bea
    \bm{\vf}^i(x,y)&=&  \vf^i(x)+
    \sum_{k=1}^{\infty}\left(\re^{\ri m k y} \vf^i_k(x)+\re^{-\ri m k y } \vf^i_{-k}(x)\right)~,\label{vf series}\\
    \bm{\x}^\a(x,y)&=& \x^\a(x) +
    \sum_{k=1}^{\infty}\left(\re^{\ri m k y } \x^\a_k(x)+\re^{-\ri m k y } \x^\a_{-k}(x) \right)~,
\eea
\esubeq
where the $\vf$'s and $\xi$'s are the `reduced' fields and gauge parameters which are independent of $y$. 
Only the zero mode in \eqref{vf series}, $\vf^i(x)$, should be kept to reduce the massless theory on  $\mathbb{M}^{d+1}$ 
to a massless one on $\mathbb{M}^{d}$. For example, this is the approach which was used in \cite{KT-M11} to obtain all linearised off-shell actions for 3D $\cN=2$ supergravity\footnote{The off-shell formulations for general 3D $\cN=2$ supergravity-matter systems were constructed in \cite{KT-M11, KLT-M}.}  from the linearised actions \cite{BK:1998,GKP} for all off-shell formulations for 4D $\cN=1$ supergravity. 
The first excited Kaluza-Klein modes in \eqref{vf series}, corresponding to $k=\pm1$, should be used for obtaining massive theories in $d$ dimensions. In what follows we are only interested in these modes.

For unconstrained complex fields, we can always truncate the  series \eqref{seriesall} to 
\bsubeq\label{Truncated Fields}
\bea
    \bm{\vf}^i(x,y)&=&\re^{\ri m y } \vf^i(x)~,\non\\
    \bm{\x}^\a(x,y)&=&\re^{\ri m  y } \x^\a(x)~,
\eea
\esubeq
since the action is quadratic in $\bm{\vf}^i$ and the transformations linear in $\bm{\x}^\a$.
The integration measure is deformed as follows
\bea
    \int_{\mathbb{M}^{d+1}}\rightarrow \int_{\mathbb{M}^d}\oint \rd y~,\qquad \oint\rd y=\frac{m}{2\pi}\int^{\frac{\pi}{m}}_{-\frac{\pi}{m}}\rd y~.
\eea
The resulting action takes the form
\bea
    S[\vf,\bar{\vf}]=\int_{\mathbb{M}^{d}}\bar{\vf}^{i} \bigg(\oint \rd y~\re^{-\ri ym}A_{ij}\re^{\ri y m}\bigg)\vf^j~,
\eea
and is invariant under gauge transformations
\bea
    \d\vf^i(\xi)=\re^{-\ri y m}R^{i}_{~\a}\re^{\ri y m}\xi^{\a}~,\qquad \d\bar{\vf}^i(\bar{\xi})=\re^{\ri y m}\bar{R}^{i}_{~\a}\re^{-\ri y m}\bar{\xi}^{\a}~.
\eea


\subsection{Case 2: real fields}
Consider now the same situation as above but with the set of $p$ fields $\{\bm{\vf}^i(\bm{x})\}$ constrained to be real. The action is now
\bea\label{Real Action}
    S[\bm{\vf}]=\int_{\mathbb{M}^{d+1}}\bm{\vf}^{i} E_i(\bm{\vf})~,
\eea
with $E_i(\vf)$ as before. The gauge transformations, however, must also be constrained to be real, and so should be of the form\footnote{It is generally possible to replace the complex gauge parameter and its conjugate with a single real field. We do not do that here as to be consistent with the third case where doing so leads to complications.}
\bea\label{Real Variation}
    \d\bm{\vf}^i(\xi,\bar{\xi})=R^{i}_{~\a}\bm{\xi}^{\a}+\bar{R}^{i}_{~\a}\bar{\bm{\xi}}^{\a}~.
\eea
To dimensionally reduce the set of real fields $\{\vf^i(\bm{x})\}$, we can only truncate \eqref{vf series} up to two conjugate terms
\bea
    \bm{\vf}^i(x,y)=\re^{\ri m y } \vf^i(x)+\re^{-\ri m y } \bar{\vf}^i(x)~.
\eea

To simplify calculations, it is more convenient to `complexify' such fields prior to the reduction. This will allow us to instead make use of the simpler truncation in \eqref{Truncated Fields}. To accomplish this, we first replace the action \eqref{Real Action} with \eqref{Complex Action}. We then must relax the reality constraint on \eqref{Real Variation}. We do so by taking introducing a new gauge parameter $\bm{\j}_\a$ which replaces $\bar{\bm{\x}}_\a$. The complexified variation is
\bea\label{Complexified Variation}
    \d\bm{\vf}^i(\bm{\xi},\bm{\j})=R^{i}_{~\a}\bm{\xi}^{\a}+\bar{R}^{i}_{~\a}\bm{\j}^{\a}~,\qquad \d\bar{\vf}^i(\bar{\bm{\xi}},\bar{\bm{\j}})=R^{i}_{~\a}\bar{\bm{\j}}^{\a}+\bar{R}^{i}_{~\a}\bar{\bm{\x}}^{\a}~.
\eea
From here, we simply treat the procedure as we did for the unconstrained complex field action. 

\subsubsection{Maxwell action}

To illustrate the procedure described above, we will now consider the dimensional reduction of the Maxwell action in $\mathbb{M}^{d+1}$. We will label spacetime indices in $\mathbb{M}^{d+1}$ and $\mathbb{M}^{d}$ with Greek letters and Latin letters respectively. The action has the form
\bea
    S[\mathfrak{A}_\m]=-\frac{1}{4}\int\rd^{d+1} \bm{x}~\mathfrak{F}^{\m\n}\mathfrak{F}_{\m\n}~,\qquad \mathfrak{F}_{\m\n}=\pa_{\m}\mathfrak{A}_{\n}-\pa_{\n}\mathfrak{A}_{\m}~,
\eea
invariant under the gauge transformation
\bea
    \d \mathfrak{A}_\m=\pa_\m\bm{\x}~.
\eea
We can rewrite the action as
\bea
    S[\mathfrak{A}_\m]=\frac{1}{2}\int\rd^{d+1} \bm{x}~\mathfrak{A}^{\n}\pa^\m \mathfrak{F}_{\m\n}~,
\eea
which we complexify to get
\bea
    S[\mathfrak{A}_\m, \bar{\mathfrak{A}}_\m]=\frac{1}{2}\int\rd^{d+1} \bm{x}~\bar{\mathfrak{A}}^{\n}\pa^\m \mathfrak{F}_{\m\n}~,
\eea
which is invariant under
\bea
    \d \mathfrak{A}_\m=\pa_\m\bm{\x}~,\qquad \d \bar{\mathfrak{A}}_\m=\pa_\m\bar{\bm{\x}}~.
\eea
From here we reduce the field and gauge parameter\footnote{The factor of $\sqrt{-1}$ is included to ensure that the field $A$ can be made to be real when reality conditions are imposed.}:
\bea
    (\mathfrak{A}_\m)= \re^{\ri y m}(A_a,\ri A)~,\qquad \bm{\x}=\re^{\ri y m}\x~.
\eea
Inserting these into the action gives
\bea
    S[A_a, \bar{A}_a,A,\bar{A}]&=&\frac{1}{2}\int\rd^{d} x\bigg\{\bar{A}^{b}[\pa^a (\pa_{a}A_{b}-\pa_{b}A_{a})+\ri m (\ri m A_{b}-\ri\pa_{b}A)]\non\\
    &&-\ri\bar{A}[\pa^a (\ri\pa_{a}A-\ri m A_{a})+\ri m (-mA+ m A)]\bigg\}\non\\
    &=&\frac{1}{2}\int\rd^{d} x\bigg\{\bar{A}^{a}[(\square-m^2)A_{a}-\pa^b\pa_{a}A_{b}+m\pa_{a}A]\non\\
    &&+\bar{A}(\square A- m\pa^a A_{a})\bigg\}~,
\eea
which is invariant under
\bsubeq
\bea
    \d A_a=\pa_a\x~,\qquad \d \bar{A}_a=\pa_a\bar{\x}~,\\
    \d A=m\x~,\qquad \d \bar{A}=m\bar{\x}~.
\eea
\esubeq
We can then enforce the reality conditions
\bea
    A_a=\bar{A}_a~,\qquad A=\bar{A}~,
\eea
so that the action becomes
\bea
    S[A_a, A]&=&\frac{1}{2}\int\rd^{d} x\bigg\{A^{a}[(\square-m^2)A_{a}-\pa^b\pa_{a}A_{b}+2 m\pa_{a}A]-A\square A\bigg\}~,
\eea
with gauge transformations
\bea
    \d A_a=\pa_a\x~,\qquad \d A=m\x~.
\eea

We see in this example that the procedure allows us to deform a massless gauge-invariant theory in a way that endows it with mass while preserving the gauge invariance. The model described above is the well known Stuekelberg reformulation of the Proca action.


\subsubsection{Linearised gravity}

Let us now consider another important example, the reduction of the graviton action. The action in $d+1$-dimensions is:
\bea
    S[\mathfrak{h}_{\m\n}]=\frac{1}{2}\int \rd^{d+1} \bm{x}\big\{\mathfrak{h}^{\m\n}\square\mathfrak{h}_{\m\n}-2\mathfrak{h}^{\m\n}\pa_\m\pa^\l\mathfrak{h}_{\l\n}-\mathfrak{h}^\m_{~\m}\square\mathfrak{h}^\n_{~\n}+2\mathfrak{h}^{\m\n}\pa_\m\pa_\n\mathfrak{h}^\l_{~\l}\big\}~,
\eea
invariant under the gauge transformation
\bea
    \d \mathfrak{h}_{\m\n}=\pa_{(\m}\bm{\x}_{\n)}~.
\eea
The field $\mathfrak{h}_{\m\n}$ is symmetric and real and the gauge parameter $\bm{\xi}_\m$ is real. We start by complexifying according to our schema so that the action becomes\footnote{Here we have split the last term of the original action into two conjugate terms so that the new action is still real in the complexified theory.}
\begin{align}\label{Graviton Complexified Action}
    S[\mathfrak{h}_{\m\n},\mathfrak{\bar{h}}_{\m\n}]=\frac{1}{2}\int \rd^{d+1} \bm{x}\big\{\mathfrak{\bar{h}}^{\m\n}\square\mathfrak{h}_{\m\n}-2\mathfrak{\bar{h}}^{\m\n}\pa_\m\pa^\l\mathfrak{h}_{\l\n}-\mathfrak{\bar{h}}^\m_{~\m}\square\mathfrak{h}^\n_{~\n}+\mathfrak{\bar{h}}^{\m\n}\pa_\m\pa_\n\mathfrak{h}^\l_{~\l}+\mathfrak{\bar{h}}^\l_{~\l}\pa^\m\pa^\n\mathfrak{h}_{\m\n}\big\}~.
\end{align}
We then reduce the field and gauge parameter:
\bea
    (\mathfrak{h}_{\m\n})=\re^{\ri y m}(h_{ab},\ri h_a, h)~,\qquad \bm{\x}_\a=\re^{\ri y m}(\x_a,\ri\x)~.
\eea
Inserting these into \eqref{Graviton Complexified Action} gives us the reduced massive graviton action:
\bea
    &&S[h_{ab},\bar{h}_{ab},h_{a},\bar{h}_{a},h,\bar{h}]\non\\
    &&~~~~~~~=\frac{1}{2}\int \rd^{d}x\big\{\bar{h}^{ab}(\square-m^2) h_{ab}+\bar{h}^a\square h_a-2\bar{h}^{ab}\pa_a\pa^c h_{bc}-m\bar{h}^a\pa^b h_{ab}\non\\
    &&~~~~~~~~~~~~~~~~~~~~~+m\bar{h}^{ab}\pa_a h_b-\bar{h}^a\pa_a\pa^b h_b-\bar{h}^a_{~a}(\square-m^2)h^b_{~b}-\bar{h}\square h^a_{~a}\non\\
    &&~~~~~~~~~~~~~~~~~~~~~-\bar{h}^{a}_{~a}\square h+\bar{h}^{ab}\pa_a\pa_b h^{l}_{~l}+\bar{h}^{ab}\pa_a\pa_b h+\bar{h}^{c}_{~c}\pa^a\pa^b h_{ab}\non\\
    &&~~~~~~~~~~~~~~~~~~~~~+\bar{h}\pa^a\pa^b h_{ab}+m\bar{h}^a\pa_a h^b_{~b}-m\bar{h}^a_{~a}\pa^b h_b\big\}~,
\eea
which is invariant under the gauge transformations:
\bea\label{Massive Graviton Transformations}
    \d h_{ab}=\pa_{(a}\x_{b)}~,\qquad \d h_{a}=\frac{1}{2}(\pa_a\x+m\x_a)~,\qquad \d h=m\x~.
\eea

If we then impose the reality conditions:
\bea
    h_{ab}&=&\bar{h}_{ab}~,\qquad h_{a}=\bar{h}_{a}~, \qquad h=\bar{h}~,\non\\
    \x_{m}&=&\bar{\x}_{m}~, \qquad \x=\bar{\x}~,
\eea
the action becomes
\bea
    &&S[h_{ab},h_{a},h]
    =\frac{1}{2}\int \rd^{d}x\big\{h^{ab}(\square-m^2) h_{ab}+h^a\square h_a-2h^{ab}\pa_a\pa^c h_{bc}-2mh^a\pa^b h_{ab}\non\\
    &&~~~~~~~~~~~~~~~~~~~~~-h^a\pa_a\pa^b h_b-h^a_{~a}(\square-m^2)h^b_{~b}-2h\square h^a_{~a}\non\\
    &&~~~~~~~~~~~~~~~~~~~~~+2h^{ab}\pa_a\pa_b h^{c}_{~c}+2h^{ab}\pa_a\pa_b h+2mh^a\pa_a h^b_{~b}\big\}~,
\eea
invariant under the gauge transformations \eqref{Massive Graviton Transformations}. The above action coincides with the model for gauge-invariant massive gravitons \cite{Zinovev1983, Pashnev:1989}.


\subsection{Case 3: real fields coupled to complex fields} \label{Coupled Dim Red}

Suppose we now have an action in which $p$ real fields $\{\bm{\vf}^i(\bm{x})\}$ are coupled to $q$ complex fields $\{\bm{\o}^a(\bm{x})\}$ and their conjugates $\{\bar{\bm{\o}}^a(\bm{x})\}$. The action will be of the form
\bea\label{Coupled Action}
        S[\bm{\vf},\bm{\o}, \bar{\bm{\o}}]=\int_{\mathbb{M}^{d+1}}(\bm{\vf}^{i} E_i(\bm{\vf},\bm{\o},\bar{\bm{\o}})+\bar{\bm{\o}}^{a} F_a(\bm{\vf},\bm{\o},\bar{\bm{\o}})+\bm{\o}^{a} G_a(\bm{\vf},\bm{\o},\bar{\bm{\o}}))~,
\eea
where
\bsubeq \label{Coupled EoM}
\bea
        E_i(\bm{\vf},\bm{\o},\bar{\bm{\o}})&=&A_{ij}\bm{\vf}^j+B_{i a}\bm{\o}^a+\bar{B}_{i a}\bar{\bm{\o}}^a~,\\
        F_a(\bm{\vf},\bm{\o},\bar{\bm{\o}})&=&\bar{B}_{ia}\bm{\vf}^i+C_{ab}\bm{\o}^b+D_{ab}\bar{\bm{\o}}^b~,\\
        G_a(\bm{\vf},\bm{\o},\bar{\bm{\o}})&=&B_{ia}\bm{\vf}^i+D_{ab}\bm{\o}^b+C_{ab}\bar{\bm{\o}}^b~,
\eea
\esubeq
where $A_{ij}$, $C_{ab}$ and $D_{ab}$ are Hermitian.
The action is invariant under gauge transformations:
\bsubeq
\bea
        \d\bm{\vf}^i(\bm{\xi},\bar{\bm{\xi}})=R^{i}_{~\a}\bm{\xi}^{\a}+\bar{R}^{i}_{~\a}\bar{\bm{\x}}^{\a}~,~~~~~~~~\\
        \d\bm{\o}^a(\bm{\xi})=S^{a}_{~\a}\bm{\x}^{\a}~,\qquad \d\bar{\bm{\o}}^a(\bar{\bm{\xi}})=\bar{S}^{a}_{~\a}\bar{\bm{\x}}^{\a}~.
\eea
\esubeq

As before, it is desirable to complexify the fields $\bm{\vf}$. Doing so, however, requires the introduction of a new gauge parameter $\bm{\j}$. If we only make the replacement in $\d\bm{\vf}^i$ as we did in \eqref{Complexified Variation}, we see that it will break the gauge invariance in \eqref{Coupled EoM}. To preserve the gauge invariance, we must introduce another set of $q$ complex fields $\{\bm{\r}^a(\bm{x})\}$ and their conjugates $\{\bar{\bm{\r}}^a(\bm{x})\}$ with gauge transformations
\bea
    \d\bm{\r}^a(\j)=\bar{S}^{a}_{~\a}\bm{\j}^{\a}~,\qquad \d\bar{\bm{\r}}^a(\bar{\j})=S^{a}_{~\a}\bar{\bm{\j}}^{\a}~,
\eea
and take the following replacement in \eqref{Coupled EoM}
\bea
    \bar{\bm{\o}}^a\rightarrow \bm{\r}^a~,
\eea
while replacing the action with
\bea
    S[\bm{\vf},\bar{\bm{\vf}},\bm{\o}, \bar{\bm{\o}},\bm{\r},\bar{\bm{\r}}]=\int_{\mathbb{M}^{d+1}}(\bar{\bm{\vf}}^{i} E_i(\bm{\vf},\bm{\o},\bm{\r})+\bar{\o}^{a} F_a(\bm{\vf},\bm{\o},\bm{\r})+\bar{\r}^{a} G_a(\bm{\vf},\bm{\o},\bm{\r}))~.
\eea
We can then reduce this action as we did in the complex case.


\section{Reduction in superspace} \label{SUSY Reduction Section}

In order to reduce an action in the 4D $\cN=1$ superspace $\mathbb{M}^{4|4}$, 
parametrised by variables $\bm{z}^M = (\bm{x}^\m, \q^\a , \bar \q_\ad)$, 
we must first consider what happens to the algebra of covariant derivatives for this superspace  under dimensional reduction. 
We denote by $\mathfrak{D}_\a$ and $\bar{\mathfrak{D}}_\ad$
the 4D $\cN=1$ spinor covariant derivatives. They are known to obey the following graded commutation relations: 
\bea
    \{\mathfrak{D}_\a,\bar{\mathfrak{D}}_\ad\}=-2\ri\pa_\aa~,\qquad\{\mathfrak{D}_\a,\mathfrak{D}_\b\}=\{\bar{\mathfrak{D}}_\ad,\bar{\mathfrak{D}}_\bd\}=0~,\qquad 
    \partial_{\a\ad} = (\s^\m)_{\a\ad} \frac{\partial}{ \partial \bm{x}^\m}~.
\eea
For the dimensional reduction, we must first bring this algebra to 3D $\mathcal{N}=2$ superspace. 
As the coordinate for $S^1$ in \eqref{22y} we choose
$y \equiv \bm{x}^2$.  Thus the 3D gamma matrices are
\bsubeq\label{Reduced Pauli Matrices}
\bea\label{3D gamma}
    (\g_a):=(\mathds{1},\s_1,\s_3)~,\qquad a=0,1,3~,
\eea
which are real and symmetric, while $\s_2$ is replaced by
\bea\label{Sigma Replcae}
    (\s_2)_\aa\rightarrow \ri\ve_{\a\b}~.
\eea
\esubeq
The 4D spinor covariant derivatives turn into the operators
\bea \label{4D Covariant Deformation}
    \mathfrak{D}_\a \rightarrow \cD_\a:=D_\a-\bar{\theta}_\a\pa_{y}~,\qquad \bar{\mathfrak{D}}_\ad \rightarrow \cDB_\a:=\DB_\a-\theta_\a\pa_{y}~,
\eea
which satisfy the algebra
\bea \label{Dim Reduced Algebra Complex}
    \{\cD_\a,\cDB_\b\}= -2\ri\pa_{\ab}+2\ve_\ab\pa_{y}~,\qquad\{\cD_\a,\cD_\b\}=\{\cDB_\a,\cDB_\b\}=0~.
\eea
The spinor covariant derivatives $D_\a$ and $\bar D_\a$  satisfy the grade commutation relations
\bea
    \{D_\a,\DB_\b\}= -2\ri\pa_{\ab}~,\qquad\{D_\a,D_\b\}=\{\DB_\a,\DB_\b\}=0~,
\eea
and correspond to the 3D $\cN=2$ Minkowski superspace, ${\mathbb M}^{3|4}$.
The derivative $\pa_{y}$ appearing in \eqref{Dim Reduced Algebra Complex}
is a central charge. We see that the reduced theory will have $\cN=2$ supersymmetry with a central charge. All massive $\cN=2$ supersymmetric models discussed in this paper will realise $\cN=2$ supersymmetry with central charge.
The corresponding central charge superspace will be denoted ${\mathbb M}^{3|4}_c$.


\subsection{Vector supermultiplet} \label{Vector Supermultiplet}

Let us begin by applying our reduction procedure to the simple theory of the massless vector supermultiplet. We start with the action in $\mathbb{M}^{4|4}$:
\bea \label{Massless vector multiplet action 4D}
    S[\mathfrak{V}]=\frac{1}{8}\int\rd^{4|4}\bm{z}~\mathfrak{V} ~\mathfrak{D}^\a\mathfrak{\DB}^2 \mathfrak{D}_\a \mathfrak{V}~,
\eea
invariant under the transformation
\bea
    \d \mathfrak{V}=\bm{\L}+\bar{\bm{\L}}~,\qquad \mathfrak{\DB}_\ad\bm{\L}=0~.
\eea
We then complexify the action to get
\bea
    S[\mathfrak{V},\bar{\mathfrak{V}}]=\frac{1}{8}\int\rd^{4|4}\bm{z}~\bar{\mathfrak{V}} ~\mathfrak{D}^\a\mathfrak{\DB}^2 \mathfrak{D}_\a \mathfrak{V}~,
\eea
while the gauge transformation becomes
\bea \label{Vector Mult Gauge Transformation}
    \d \mathfrak{V}=\bm{\L}+\bm{\r}~,\qquad \mathfrak{\DB}_\ad\bm{\L}=\mathfrak{D}_\a\bm{\r}=0~.
\eea

We dimensionally reduce the field and gauge parameters:
\bea
    \mathfrak{V}(\bm{z}):=\re^{\ri m y} V(z)~,\qquad \bm{\L}(\bm{z}):=\re^{\ri m y}\L(z)~,\qquad \bar{\bm{\r}}(\bm{z}):=\re^{\ri m y}\r(z)~,
\eea
so that the action becomes
\bea
    S[V, \bar{V}]=\frac{1}{8}\oint\rd y\int\rd^{3|4}z~\re^{-\ri ym} \bar{V} \cD^\a\cDB^2\cD_\a\re^{\ri ym}  V~,
\eea
and the transformation \eqref{Vector Mult Gauge Transformation} becomes
\bea
    \d V=\L+\r~,\qquad \re^{-\ri ym}\cDB_\a\re^{\ri ym}\bm{\L}=\re^{-\ri ym}\cD_\a\re^{\ri ym}\r=0~.
\eea
We can then integrate out $y$ and replace everywhere $\pa_y\rightarrow\ri m$ to get the action
\bea\label{Massive Vector Action}
    S[V, \bar{V}]=\frac{1}{8}\int\rd^{3|4}z~ \bar{V} \cD^\a\cDB^2\cD_\a  V~,
\eea
invariant under the gauge transformation
\bea
    \d V=\L+\r~,\qquad \cDB_\a\bm{\L}=\cD_\a\r=0~.
\eea
The above model describes a massive vector supermultiplet in $\mathbb{M}^{3|4}_c$.


\subsection{Linearised old minimal supergravity}

With this established, let us apply the Kaluza-Klein reduction procedure to the linearised action for old minimal supergravity\footnote{The old minimal formulation for 4D $\cN =1$ supergravity was constructed by several groups  \cite{Siegel77-77,WZ,old1,old2}.} 
(for the technical details, see \cite{BK:1998}). In this way we will obtain a new gauge model -- massive 3D $\cN=2$ linearised supergravity. 

The linearised action for old minimal supergravity action in $\mathbb{M}^{4|4}$ is given by
\bea
    S^{(\rm{I})}&=&\int\rd ^{4|4}\bm{z}\bigg\{-\frac{1}{16}\mathfrak{H}^{\a\ad}\mathfrak{D}^\b\bar{\mathfrak{D}}^2 \mathfrak{D}_\b \mathfrak{H}_{\a\ad}+\frac{1}{48}([\mathfrak{D}_\a,\bar{\mathfrak{D}}_\ad]\mathfrak{H}^{\a\ad})^2-\frac{1}{4}(\pa_{\a\ad}\mathfrak{H}^{\a\ad})^2\non\\
    &&~~~~~~~~~~~-3\bar{\bm{\s}}\bm{\s}-\ri(\bm{\s}-\bar{\bm{\s}})\pa_{\a\ad}\mathfrak{H}^{\a\ad}\bigg\}~,
\eea
where $\mathfrak{H}_{\a\ad}$ is real and $\bm{\s}$ is chiral:
\bea
    \mathfrak{\DB}_\ad\bm{\s}=0~.
\eea
The theory possesses the gauge invariance described by
\bea
    \d \mathfrak{H}_{\a\ad}=\bar{\mathfrak{D}}_{\ad}\mathfrak{L}_\a-\mathfrak{D}_\a\bar{\mathfrak{L}}_\ad~,\qquad \d\bm{\s}=-\frac{1}{12}\bar{\mathfrak{D}}^2\mathfrak{D}^\a \mathfrak{L}_\a~.
\eea
The action can be rearranged to get
\bea
    S^{(\rm{I})}=\frac{1}{2}\int\rd ^{4|4}\bm{z}\{\mathfrak{H}^{\a\ad} \mathfrak{E}_{\a\ad}+\bar{\bm{\s}}\mathfrak{E}_1+\bm{\s} \mathfrak{E}_2\}~,
\eea
where
\bsubeq
\bea
    \mathfrak{E}_{\a\ad}&=&-\frac{1}{8}\mathfrak{D}^\b\bar{\mathfrak{D}}^2\mathfrak{D}_\b \mathfrak{H}_{\a\ad}+\frac{1}{24}[\mathfrak{D}_\a,\bar{\mathfrak{D}}_\ad][\mathfrak{D}_\b,\bar{\mathfrak{D}}_\bd]\mathfrak{H}^{\b\bd}\non\\
    &&+\frac{1}{2}\pa_{\a\ad}\pa_{\b\bd}\mathfrak{H}^{\b\bd}+\ri\pa_{\a\ad}(\bm{\s}-\bar{\bm{\s}})~,\\
    \mathfrak{E}_1&=&\ri\pa_{\a\ad}\mathfrak{H}^{\a\ad}-3\bm{\s}~,\qquad \mathfrak{E}_2=\bar{(\mathfrak{E}_1)}=-\ri\pa_{\a\ad}\mathfrak{H}^{\a\ad}-3\bar{\bm{\s}}~.
\eea
\esubeq
The equations of motion are then simply
\bea
    \mathfrak{D}^2\mathfrak{E}_1=\bar{\mathfrak{D}}^2 \mathfrak{E}_2=0~,\qquad \mathfrak{E}_{\a\ad}=0~.
\eea
The massless dimensional reduction of this theory was carried out in \cite{KT-M11}.

According to our approach, we introduce the complexified action
\bea\label{Compl. SUGRA Type I}
    S^{(\rm{I})}=\frac{1}{2}\int\rd ^{4|4}\bm{z}\{\bar{\mathfrak{H}}^{\a\ad} \mathfrak{E}_{\a\ad}+\bar{\bm{\s}}\mathfrak{E}_1+\bar{\bm{\j}} \mathfrak{E}_2\}~,
\eea
where
\bsubeq
\bea
    \mathfrak{E}_{\a\ad}&=&-\frac{1}{8}\mathfrak{D}^\b\bar{\mathfrak{D}}^2\mathfrak{D}_\b \mathfrak{H}_{\a\ad}+\frac{1}{24}[\mathfrak{D}_\a,\bar{\mathfrak{D}}_\ad][\mathfrak{D}_\b,\bar{\mathfrak{D}}_\bd]\mathfrak{H}^{\b\bd}\non\\
    &&+\frac{1}{2}\pa_{\a\ad}\pa_{\b\bd}\mathfrak{H}^{\b\bd}+\ri\pa_{\a\ad}(\bm{\s}-\bm{\j})~,\\
    \mathfrak{E}_1&=&\ri\pa_{\a\ad}\mathfrak{H}^{\a\ad}-3\bm{\s}~,\qquad \mathfrak{E}_2=-\ri\pa_{\a\ad}\mathfrak{H}^{\a\ad}-3\bm{\j}~.
\eea
\esubeq
The action \eqref{Compl. SUGRA Type I} is invariant under gauge transformations:
\bea
    \d \mathfrak{H}_{\a\ad}=\bar{\mathfrak{D}}_{\ad}\mathfrak{L}_\a-\mathfrak{D}_\a \mathfrak{M}_\ad~,\qquad \d\bm{\s}=-\frac{1}{12}\bar{\mathfrak{D}}^2\mathfrak{D}^\a \mathfrak{L}_\a~,\qquad \d\bm{\j} = -\frac{1}{12}\mathfrak{D}^2 \bar{\mathfrak{D}}_\ad \mathfrak{M}^\ad~,
\eea
where $\bm{\s}$ is chiral and $\bm{\j}$ is antichiral. To return to the original action,we take the reality conditions
\bea
    \bar{\mathfrak{H}}_{\a\ad}=\mathfrak{H}_{\a\ad}~,\qquad\bm{\j}=\bar{\bm{\s}}~,\qquad \mathfrak{M}_\a=\bar{\mathfrak{L}}_\a~.
\eea

When taking the dimensional reduction, the dotted indices are replaced by undotted indices, hence, we have
\bea
    \mathfrak{H}_{\a\ad}\rightarrow\mathfrak{H}_{\ab}~.
\eea
We can then decompose this object into irreducible components as follows:
\bea
    \mathfrak{H}_{\ab}=\mathfrak{H}_{(\ab)}+(\sigma_2)_{\ab}\mathfrak{H}=\mathfrak{H}_{(\ab)}+\ri\ve_{\ab}\mathfrak{H}~.
\eea
With this in mind, we can now dimensionally reduce the action by taking
\bsubeq
\bea
    &&\mathfrak{H}_\ab(\bm{z})=\re^{\ri ym}H_\ab(z)+\ri\ve_\ab \re^{\ri ym}H(z)~,~~~~~\\
    &&\bm{\s}(\bm{z})=\re^{\ri ym}\s(z)~,\qquad \bm{\j}(\bm{z})=\re^{\ri ym}\j(z)~,\\
    &&\mathfrak{L}_\a(\bm{z})=\re^{\ri ym}L_\a(z)~,\qquad \mathfrak{M}_\a(\bm{z})=\re^{\ri ym}M_\a(z)~.
\eea
\esubeq
The action becomes
\bea
    S^{(\rm{I})}=\frac{1}{2}\int\rd ^{3|4}z\oint\rd y\{\bar{H}^{\a\b} E_{\a\b}+\bar{H}E_0+\bar{\s}E_1+\bar{\j}E_2\}~,
\eea
where
\bsubeq
\bea
    E_{\a\b}&=&-\frac{1}{8}\re^{-\ri ym}\cD^\g\cDB^2\cD_\g\re^{\ri ym} H_{\ab}+\frac{1}{24}\re^{-\ri ym}[\cD_{(\a},\cDB_{\b)}][\cD_\g,\cDB_\r]\re^{\ri ym}H^{\g\r}\non\\
    &&+\frac{\ri}{24}\re^{-\ri ym}[\cD_{(\a},\cDB_{\b)}][\cD^\r,\cDB_\r]\re^{\ri ym}H+\frac{1}{2}\pa_{\ab}\pa_{\g\r}H^{\g\r}\non\\
    &&-\ri\re^{-\ri ym}\pa_{y}\pa_{\ab}\re^{\ri ym}H-\ri\pa_{\ab}(\s-\j)~,~~~~~\\
    E_0&=&\frac{1}{4}\re^{-\ri ym}\cD^\g\cDB^2\cD_\g\re^{\ri ym} H+\frac{\ri}{24}\re^{-\ri ym}[\cD^\a,\cDB_{\a}][\cD_\g,\cDB_\r]\re^{\ri ym}H^{\g\r}\non\\
    &&-\frac{1}{24}\re^{-\ri ym}[\cD^\a,\cDB_{\a}][\cD^\r,\cDB_\r]\re^{\ri ym}H+\ri\re^{-\ri ym}\pa_{y}\pa_{\g\r}\re^{\ri ym}H^{\g\r}\non\\
    &&+2\re^{-\ri ym}(\pa_{y})^2\re^{\ri ym}H-2\re^{-\ri ym}\pa_{y}\re^{\ri ym}(\s-\j)~,~~~~~\\
    E_1&=&\ri\pa_{\a\b} H^{\a\b}+2\re^{-\ri ym}\pa_{y} \re^{\ri ym}H-3\s~,\\
    E_2&=&-\ri\pa_{\a\b}H^{\a\b}-2\re^{-\ri ym}\pa_{y}\re^{\ri ym} H-3\j~,
\eea
\esubeq
where the action is invariant under
\bsubeq
\bea
    \d H_{\a\b}&=&\re^{-\ri ym}\cDB_{(\a}\re^{\ri ym}L_{\b)}-\re^{-\ri ym}\cD_{(\a}\re^{\ri ym}M_{\b)}~, \non\\
    \d H&=&\frac{\ri}{2}(\re^{-\ri ym}\cDB^{\a}\re^{\ri ym}L_{\a}-\re^{-\ri ym}\cD_{\a}\re^{\ri ym}M^{\a})~, \non\\
    \d\s&=&-\frac{1}{12}\re^{-\ri ym}\cDB^2\cD^\a\re^{\ri ym} L_\a~,\qquad \d\j = -\frac{1}{12}\re^{-\ri ym}\cD^2 \cDB_\a\re^{\ri ym} M^\a~,
\eea
\esubeq
with constraints
\bea
    \re^{-\ri ym}\cDB_\a\re^{\ri ym}\s=0~,\qquad \re^{-\ri ym}\cD_\a\re^{\ri ym}\j=0~.
\eea

We can eliminate the exponentials by replacing everywhere $\pa_{y}$ with $\ri m$ to get
\bea\label{Old Linearised SUGRA Massive Action}
    S^{(\rm{I})}=\frac{1}{2}\int\rd ^{3|4}z\{\bar{H}^{\a\b} E_{\a\b}+\bar{H}E_0+\bar{\s}E_1+\bar{\j}E_2\}~,
\eea
where
\bsubeq
\bea
    E_{\a\b}&=&-\frac{1}{8}\cD^\g\cDB^2\cD_\g H_{\ab}+\frac{1}{24}[\cD_{(\a},\cDB_{\b)}][\cD_\g,\cDB_\r]H^{\g\r}+\frac{\ri}{24}[\cD_{(\a},\cDB_{\b)}][\cD^\r,\cDB_\r]H\non\\
    &&+\frac{1}{2}\pa_{\ab}\pa_{\g\r}H^{\g\r}+m\pa_{\ab}H-\ri\pa_{\ab}(\s-\j)~,~~~~~\\
    E_0&=&\frac{1}{4}\cD^\g\cDB^2\cD_\g H+\frac{\ri}{24}[\cD^\a,\cDB_{\a}][\cD_\g,\cDB_\r]H^{\g\r}-\frac{1}{24}[\cD^\a,\cDB_{\a}][\cD^\r,\cDB_\r]H\non\\
    &&-m\pa_{\g\r}H^{\g\r}-2m^2H-2\ri m(\s-\j)~,~~~~~\\
    E_1&=&\ri\pa_{\a\b}H^{\a\b}+2\ri mH-3\s~,\qquad E_2=-\ri\pa_{\a\b}H^{\a\b}-2\ri mH-3\j~.
\eea
\esubeq
The action is now invariant under
\bsubeq
\bea
    \d H_{\a\b}&=&\cDB_{(\a}L_{\b)}-\cD_{(\a}M_{\b)}~, \\
    \d H&=&\frac{\ri}{2}(\cDB^{\a}L_{\a}-\cD_{\a}M^{\a})~, \\
    \d\s&=&-\frac{1}{12}\cDB^2\cD^\a L_\a~,\qquad \d\j = -\frac{1}{12}\cD^2 \cDB_\a M^\a~,
\eea
\esubeq
with constraints
\bea
    \cDB_\a\s=0~,\qquad \cD_\a\j=0~.
\eea

We see that the resulting theory is realised in terms of the complex fields $\{H_\ab,\s,\j\}$ with central charge $\ri m$ and their conjugates with central charge $-\ri m$. At this stage, unlike the non-supersymmetric theory discussed earlier, one cannot impose reality conditions since theories of $3D$ $\cN=2$ supersymmetry with central charge are realised in terms complex fields. 

We see then that the procedure we have described is a powerful tool which allows for the derivation of new massive gauge-invariant theories such as the above model describing massive linearised supergravity in $\mathbb{M}^{3|4}_c$.


\subsection{Linearised new minimal supergravity} \label{New Minimal Reduction}

We now will demonstrate the same process for a dual theory, the linearised new minimal supergravity.\footnote{The new minimal formulation for 4D $\cN =1$ supergravity was constructed in\cite{SohniusW1,SohniusW3}.}  
The action in this formulation is given by (for the technical details, see \cite{BK:1998})
\bea\label{Real New Minimal Linearised SUGRA}
    S^{(\rm{II})}[\mathfrak{H}_{\aa},\mathfrak{F}]&=&\int \rd^{4|4}\bm{z}~\bigg\{-\frac{1}{16}\mathfrak{H}^\aa \mathfrak{D}^\g\mathfrak{\DB}^2 \mathfrak{D}_\g \mathfrak{H}_\aa-\frac{1}{4}(\pa_\aa \mathfrak{H}^\aa)^2\non\\
    &&~~~~~~~~~~~+\frac{1}{16}([\mathfrak{D}_\a,\mathfrak{\DB}_\ad]\mathfrak{H}^\aa)^2+\frac{1}{2}\mathfrak{F}[\mathfrak{D}_\a,\mathfrak{\DB}_\ad]\mathfrak{H}^\aa+\frac{3}{2}\mathfrak{F}^2\bigg\}~.
\eea
It is invariant under gauge transformations of the form
\bsubeq\label{Type II Transformation}
\bea
    \d \mathfrak{H}_\aa=\mathfrak{\DB}_\ad \mathfrak{L}_\a-\mathfrak{D}_\a \bar{\mathfrak{L}}_\ad~,\\
    \d\mathfrak{F}=\frac{1}{4}(\mathfrak{D}^\a\mathfrak{\DB}^2 \mathfrak{L}_\a+\mathfrak{\DB}_\ad \mathfrak{D}^2\bar{\mathfrak{L}}^\ad)~.
\eea
\esubeq
Both superfields $\mathfrak{H_{\a\ad}}$ and $\mathfrak{F}$ are real. Additionally, the field $\mathfrak{F}$ is linear, i.e.
\bea
    \mathfrak{D}^2\mathfrak{F}=\mathfrak{\DB}^2\mathfrak{F}=0~.
\eea
We rewrite the action \eqref{Real New Minimal Linearised SUGRA} as
\bea
    S^{(II)}=\frac{1}{2}\int \rd^{4|4}\bold{z}\{\mathfrak{H}^{\a\ad}\mathfrak{E}_{\a\ad}+\mathfrak{F}\mathfrak{E}\}~,
\eea
where
\bsubeq
\bea
    \mathfrak{E}_{\a\ad}&=&-\frac{1}{8}\mathfrak{D}^\g\mathfrak{\DB}^2\mathfrak{D}_\g\mathfrak{H}_{\a\ad}+\frac{1}{2}\pa_{\a\ad}\pa^{\g\gd}\mathfrak{H}_{\g\gd}+\frac{1}{8}[\mathfrak{D}_\a,\mathfrak{\DB}_\ad][\mathfrak{D}^\g,\mathfrak{\DB}^\gd]\mathfrak{H}_{\g\gd}+\frac{1}{2}[\mathfrak{D}_\a,\mathfrak{\DB}_\ad]\mathfrak{F}~,~~~~~~\\
    \mathfrak{E}&=&\frac{1}{2}[\mathfrak{D}^\a,\mathfrak{\DB}^\ad]\mathfrak{H}_{\a\ad}+3\mathfrak{F}~.
\eea
\esubeq
The massless dimensional reduction of this theory was carried out in \cite{KT-M11}.

We see that the above theory is of Case 2, hence, we need only relax the reality conditions by replacing the action with
\bea
    S^{(II)}=\frac{1}{2}\int \rd^{4|4}\bm{z}\{\bar{\mathfrak{H}}^{\a\ad}\mathfrak{E}_{\a\ad}+\bar{\mathfrak{F}}\mathfrak{E}\}~,
\eea
while the gauge transformations \eqref{Type II Transformation} become
\bsubeq
\bea
    \d \mathfrak{H}_\aa=\mathfrak{\DB}_\ad \mathfrak{L}_\a-\mathfrak{D}_\a \mathfrak{M}_\ad~,\\
    \d\mathfrak{F}=\frac{1}{4}(\mathfrak{D}^\a\mathfrak{\DB}^2 \mathfrak{L}_\a+\mathfrak{\DB}_\ad \mathfrak{D}^2\mathfrak{M}^\ad)~.
\eea
\esubeq
Both $\mathfrak{F}$ and its conjugate are linear superfields.

For the dimensional reduction we take the decompositions:
\bsubeq
\bea
    \mathfrak{H}_\ab(\bm{z})&=&\re^{\ri ym}H_\ab(z)+\ri\ve_\ab \re^{\ri ym}H(z)~,\\
    \mathfrak{F}(\bm{z})&=&\re^{\ri ym}F(z)~,\qquad \mathfrak{L}_\a(\bm{z})=\re^{\ri ym}L_\a(z)~,\qquad \mathfrak{M}_\a(\bm{z})=\re^{\ri ym}M_\a(z)~.
\eea
\esubeq
The action then becomes
\bea
    S^{(II)}=\frac{1}{2}\int \rd^{3|4}z\oint\rd y\{\bar{H}^{\ab}E_{\ab}+\bar{H}E_{0}+\bar{F}E_{1}\}~,
\eea
where
\bsubeq
\bea
    E_{\ab}&=&-\frac{1}{8}\re^{-\ri y m}\cD^\g\cDB^2 \cD_\g \re^{\ri x_2m}H_\ab+\frac{1}{2}\re^{-\ri y m}\pa_\ab(\pa^{\r\g}\re^{\ri x_2m}H_{\r\g}-2\pa_{x_2}\re^{\ri x_2m}H)\non\\
    &&+\frac{1}{8}\re^{-\ri y m}[\cD_\a,\cDB_\b]([\cD^\g,\cDB^\r]\re^{\ri x_2m}H_{\g\r}+\ri\re ^{-\ri y m}[\cD^\g,\cDB_\g]\re^{\ri x_2m}H)\non\\
    &&+\frac{1}{2}\re^{-\ri y m}[\cD_\a,\cDB_\b]\re^{\ri x_2m}F~,\\
    E_0&=& \frac{1}{4} \re^{-\ri y m}\cD^\g\cDB^2 \cD_\g \re^{\ri x_2m}H-\re^{-\ri y m}\pa_{x_2}( \re^{\ri x_2m}\pa^{\r\g}H_{\r\g}-2\pa_{x_2}\re^{\ri x_2m}H)\non\\
    &&+\frac{1}{8}\ri\re^{-\ri y m}[\cD^\a,\cDB_\a]([\cD^\g,\cDB^\r]\re^{\ri x_2m}H_{\g\r}+\ri[\cD^\g,\cDB_\g]\re^{\ri x_2m}H)\non\\
    &&+\frac{1}{2}\ri\re^{-\ri ym}[\cD^\a,\cDB_\a]\re^{\ri x_2m}F~,\\
    E_1&=&\frac{1}{2}\re^{-\ri y m}[\cD_\a,\cDB_\b]\re^{\ri x_2m}H^\ab+\frac{\ri}{2}\re ^{-\ri y m}[\cD^\a,\cDB_\a]\re^{\ri x_2m}H+3F~.
\eea
\esubeq
The above action is invariant under transformations:
\bsubeq
\bea
    \d H_{\a\b}&=&\re^{-\ri ym}\cDB_{(\a}\re^{\ri ym}L_{\b)}-\re^{-\ri ym}\cD_{(\a}\re^{\ri ym}M_{\b)}~, \\
    \d H&=&\frac{\ri}{2}(\re^{-\ri ym}\cDB^{\a}\re^{\ri ym}L_{\a}-\re^{-\ri ym}\cD_{\a}\re^{\ri ym}M^{\a})~, \\
    \d F&=&\frac{1}{4}\re^{-\ri y m}(\cD^\a\cDB^2 \re^{\ri y m}L_\a+\cDB_\b \cD^2\re^{\ri y m}M^\b)~,
\eea
\esubeq
and the field $F$ is constrained:
\bea
    \cD^2\re^{\ri y m}F=\cDB^2\re^{\ri y m}F=0~.
\eea

Integrating out the $y$ coordinate and replacing $\pa_{y}\rightarrow \ri m$ gives the action
\bea\label{New Linearised SUGRA Massive Action}
    S^{(II)}=\frac{1}{2}\int \rd^{3|4}z\{\bar{H}^{\ab}E_{\ab}+\bar{H}E_{0}+\bar{F}E_{1}\}~,
\eea
where
\bsubeq
\bea
    E_{\ab}&=&-\frac{1}{8}\cD^\g\cDB^2 \cD_\g H_\ab+\frac{1}{2}\pa_\ab(\pa^{\r\g}H_{\r\g}-2\ri mH)\non\\
    &&+\frac{1}{8}[\cD_\a,\cDB_\b]([\cD^\g,\cDB^\r]H_{\g\r}+\ri[\cD^\g,\cDB_\g]H)\non\\
    &&+\frac{1}{2}[\cD_\a,\cDB_\b]F~,\\
    E_0&=& \frac{1}{4} \cD^\g\cDB^2 \cD_\g H-\ri m( \pa^{\r\g}H_{\r\g}-2\ri mH)\non\\
    &&+\frac{1}{8}\ri[\cD^\a,\cDB_\a]([\cD^\g,\cDB^\r]H_{\g\r}+\ri[\cD^\g,\cDB_\g]H)\non\\
    &&+\frac{\ri}{2}[\cD^\a,\cDB_\a]F~,\\
    E_1&=&\frac{1}{2}[\cD_\a,\cDB_\b]H^\ab+\frac{\ri}{2}[\cD^\a,\cDB_\a]H+3F~,
\eea
\esubeq
invariant under gauge transformations:
\bsubeq\label{New Linearised Massive SUGRA Transformations}
\bea
    \d H_{\a\b}&=&\cDB_{(\a}L_{\b)}-\cD_{(\a}M_{\b)}~, \\
    \d H&=&\frac{\ri}{2}(\cDB^{\a}L_{\a}-\cD_{\a}M^{\a})~, \\
    \d F&=&\frac{1}{4}(\cD^\a\cDB^2 L_\a+\cDB_\a \cD^2M^\a)~,
\eea
\esubeq
while the field $F$ obeys the constraints:
\bea
    \cD^2F=\cDB^2F=0~.
\eea
We thus generate a new model which describes 3D $\cN=2$ new minimal linearised massive supergravity.


\subsection{Parent model}
We note that the two above formulations for linearised supergravity are connected by a parent model from which both can be derived. We shall now see what we get if we apply the dimensional reduction procedure to this parent model.

The parent model is described by the action \cite{BK:1998}
\bea
    S[\mathfrak{H}_{\aa},\mathfrak{U}, \bm{\s}]&=&\int \rd^{4|4}\bm{z}~\bigg\{-\frac{1}{16}\mathfrak{H}^\aa \mathfrak{D}^\g\mathfrak{\DB}^2 \mathfrak{D}_\g \mathfrak{H}_\aa-\frac{1}{4}(\pa_\aa \mathfrak{H}^\aa)^2+\frac{1}{16}([\mathfrak{D}_\a,\mathfrak{\DB}_\ad]\mathfrak{H}^\aa)^2\non\\
    &&~~~~~~~~~~~+\frac{1}{2}\mathfrak{U}[\mathfrak{D}_\a,\mathfrak{\DB}_\ad]\mathfrak{H}^\aa-3(\bm{\s}+\bar{\bm{\s}})\mathfrak{U}+\frac{3}{2}\mathfrak{U}^2\bigg\}~,
\eea
which is invariant under the following transformations:
\bsubeq
\bea
    \d \mathfrak{H}_\aa=\mathfrak{\DB}_\ad \mathfrak{L}_\a-\mathfrak{D}_\a \bar{\mathfrak{L}}_\ad~,\\
    \d\mathfrak{U}=\frac{1}{4}(\mathfrak{D}^\a\mathfrak{\DB}^2 \mathfrak{L}_\a+\mathfrak{\DB}_\ad \mathfrak{D}^2\bar{\mathfrak{L}}^\ad)~,\\
    \d\bm{\s}=-\frac{1}{12}\bar{\mathfrak{D}}^2\mathfrak{D}^\a \mathfrak{L}_\a~.
\eea
\esubeq
The fields $\mathfrak{H}_{\a\ad}$ and $\mathfrak{U}$ are real while $\bm{\s}$ is chiral:
\bea
    \mathfrak{\DB}_\ad\bm{\s}=0~.
\eea
This model is complexified as in Case 3 from Section \ref{Coupled Dim Red} which leads us to the action:
\bea
    S&=&\frac{1}{2}\int \rd^{4|4}\bm{z}~\{\bar{\mathfrak{H}}^\aa \mathfrak{E}_\aa+\bar{\mathfrak{U}}\mathfrak{E}_1+\bar{\bm{\s}}\mathfrak{E}_{2}+\bar{\bm{\j}}\mathfrak{E}_{3}\}~,
\eea
where
\bsubeq
\bea
    \mathfrak{E}_{\a\ad}&=&-\frac{1}{8}\mathfrak{D}^\g\mathfrak{\DB}^2\mathfrak{D}_\g\mathfrak{H}_{\a\ad}+\frac{1}{2}\pa_{\a\ad}\pa^{\g\gd}\mathfrak{H}_{\g\gd}+\frac{1}{8}[\mathfrak{D}_\a,\mathfrak{\DB}_\ad][\mathfrak{D}^\g,\mathfrak{\DB}^\gd]\mathfrak{H}_{\g\gd}+\frac{1}{2}[\mathfrak{D}_\a,\mathfrak{\DB}_\ad]\mathfrak{U}~,~~~~~~\\
    \mathfrak{E}_1&=&\frac{1}{2}[\mathfrak{D}^\a,\mathfrak{\DB}^\ad]\mathfrak{H}_{\a\ad}-3(\s+\j)+3\mathfrak{U}~,\qquad \mathfrak{E}_2=\mathfrak{E}_3=-3\mathfrak{U}~,
\eea
\esubeq
constrained:
\bea
    \bar{\mathfrak{D}}_\ad\bm{\s}=0~,\qquad \mathfrak{D}_\a\bm{\j}=0~.
\eea
The action is now invariant under transformations:
\bsubeq
\bea
    \d \mathfrak{H}_\aa=\mathfrak{\DB}_\ad \mathfrak{L}_\a-\mathfrak{D}_\a \bar{\mathfrak{M}}_\ad~,\\
    \d\mathfrak{U}=\frac{1}{4}(\mathfrak{D}^\a\mathfrak{\DB}^2 \mathfrak{L}_\a+\mathfrak{\DB}_\ad \mathfrak{D}^2\bar{\mathfrak{M}}^\ad)~,\\
    \d\bm{\s}=-\frac{1}{12}\bar{\mathfrak{D}}^2\mathfrak{D}^\a \mathfrak{L}_\a~,\qquad \d\bm{\j}=-\frac{1}{12}\mathfrak{D}^2\bar{\mathfrak{D}}_\ad \mathfrak{M}^\ad~.
\eea
\esubeq

We now take the reduction by replacing the fields with
\bsubeq
\bea
    \mathfrak{H}_\ab(\bm{z})&=&\re^{\ri ym}H_\ab(z)+\ri\ve_\ab \re^{\ri ym}H(z)~,\qquad \mathfrak{U}(\bold{z})=\re^{\ri ym}F(z)~,\\
    \bm{\s}(\bm{z})&=&\re^{\ri ym}\s(z)~,\qquad \bm{\j}(\bm{z})=\re^{\ri ym}\j(z)~,\\
    \mathfrak{L}_\a(\bm{z})&=&\re^{\ri ym}L_\a(z)~,\qquad \mathfrak{M}_\a(\bm{z})=\re^{\ri ym}M_\a(z)~,
\eea
\esubeq
so that the action becomes
\bea
    S^{(II)}=\frac{1}{2}\int \rd^{3|4}z\oint\rd y\{\bar{H}^{\ab}E_{\ab}+\bar{H}E_{0}+\bar{U}E_{1}+\bar{\s}E_2+\bar{\j}E_3\}~,
\eea
where
\bsubeq
\bea
    E_{\ab}&=&-\frac{1}{8}\re^{-\ri y m}\cD^\g\cDB^2 \cD_\g \re^{\ri x_2m}H_\ab+\frac{1}{2}\re^{-\ri y m}\pa_\ab(\pa^{\r\g}\re^{\ri x_2m}H_{\r\g}-2\pa_{x_2}\re^{\ri x_2m}H)\non\\
    &&+\frac{1}{8}\re^{-\ri y m}[\cD_\a,\cDB_\b]([\cD^\g,\cDB^\r]\re^{\ri x_2m}H_{\g\r}+\ri\re ^{-\ri y m}[\cD^\g,\cDB_\g]\re^{\ri x_2m}H)\non\\
    &&+\frac{1}{2}\re^{-\ri y m}[\cD_\a,\cDB_\b]\re^{\ri x_2m}U~,\\
    E_0&=& \frac{1}{4} \re^{-\ri y m}\cD^\g\cDB^2 \cD_\g \re^{\ri x_2m}H-\re^{-\ri y m}\pa_{x_2}( \re^{\ri x_2m}\pa^{\r\g}H_{\r\g}-2\pa_{x_2}\re^{\ri x_2m}H)\non\\
    &&+\frac{1}{8}\ri\re^{-\ri y m}[\cD^\a,\cDB_\a]([\cD^\g,\cDB^\r]\re^{\ri x_2m}H_{\g\r}+\ri[\cD^\g,\cDB_\g]\re^{\ri x_2m}H)\non\\
    &&+\frac{1}{2}\ri\re^{-\ri ym}[\cD^\a,\cDB_\a]\re^{\ri x_2m}U~,\\
    E_1&=&\frac{1}{2}\re^{-\ri y m}[\cD_\a,\cDB_\b]\re^{\ri x_2m}H^\ab+\frac{\ri}{2}\re ^{-\ri y m}[\cD^\a,\cDB_\a]\re^{\ri x_2m}H+3U-3(\s+\j)~,\\
    E_2&=&E_3=-3U~.
\eea
\esubeq
The above action is invariant under transformations:
\bsubeq
\bea
    \d H_{\a\b}&=&\re^{-\ri ym}\cDB_{(\a}\re^{\ri ym}L_{\b)}-\re^{-\ri ym}\cD_{(\a}\re^{\ri ym}M_{\b)}~, \\
    \d H&=&\frac{\ri}{2}(\re^{-\ri ym}\cDB^{\a}\re^{\ri ym}L_{\a}-\re^{-\ri ym}\cD_{\a}\re^{\ri ym}M^{\a})~, \\
    \d U&=&\frac{1}{4}\re^{-\ri y m}(\cD^\a\cDB^2 \re^{\ri y m}L_\a+\cDB_\b \cD^2\re^{\ri y m}M^\b)~,\\
    \d\s&=&-\frac{1}{12}\re^{-\ri ym}\cDB^2\cD^\a \re^{\ri ym}L_\a~,\qquad \d\j = -\frac{1}{12}\re^{-\ri ym}\cD^2 \cDB_\a\re^{\ri ym} M^\a~,
\eea
\esubeq
with constraints
\bea
    \cDB_\a\re^{\ri ym}\s=0~,\qquad \cD_\a\re^{\ri ym}\j=0~.
\eea
We can then integrate out the $y$ coordinate and replace $\pa_{y}\rightarrow \ri m$ to get the action
\bea
    S^{(II)}=\frac{1}{2}\int \rd^{3|4}z\{\bar{H}^{\ab}E_{\ab}+\bar{H}E_{0}+\bar{U}E_{1}+\bar{\s}E_2+\bar{\j}E_3\}~,
\eea
where
\bsubeq
\bea
    E_{\ab}&=&-\frac{1}{8}\cD^\g\cDB^2 \cD_\g H_\ab+\frac{1}{2}\pa_\ab(\pa^{\r\g}H_{\r\g}-2\ri mH)\non\\
    &&+\frac{1}{8}[\cD_\a,\cDB_\b]([\cD^\g,\cDB^\r]H_{\g\r}+\ri[\cD^\g,\cDB_\g]H)\non\\
    &&+\frac{1}{2}[\cD_\a,\cDB_\b]U~,\\
    E_0&=& \frac{1}{4} \cD^\g\cDB^2 \cD_\g H-\ri m(\pa^{\r\g}H_{\r\g}-2\ri mH)\non\\
    &&+\frac{1}{8}[\cD^\a,\cDB_\a]([\cD^\g,\cDB^\r]H_{\g\r}+\ri[\cD^\g,\cDB_\g]H)\non\\
    &&+\frac{1}{2}\ri[\cD^\a,\cDB_\a]U~,\\
    E_1&=&\frac{1}{2}[\cD_\a,\cDB_\b]H^\ab+\frac{\ri}{2}[\cD^\a,\cDB_\a]H+3U-3(\s+\j)~,\\
    E_2&=&E_3=-3U~,
\eea
\esubeq
which is invariant under transformations:
\bsubeq
\bea
    \d H_{\a\b}&=&\cDB_{(\a}L_{\b)}-\cD_{(\a}M_{\b)}~, \\
    \d H&=&\frac{\ri}{2}(\cDB^{\a}L_{\a}-\cD_{\a}M^{\a})~, \\
    \d U&=&\frac{1}{4}(\cD^\a\cDB^2 L_\a+\cDB_\b \cD^2M^\b)~,\\
    \d\s&=&-\frac{1}{12}\cDB^2\cD^\a L_\a~,\qquad \d\j = -\frac{1}{12}\cD^2 \cDB_\a M^\a~,
\eea
\esubeq
with constraints
\bea
    \cDB_\a\s=0~,\qquad \cD_\a\j=0~.
\eea

It is worth pointing out that by integrating out $U$ from the above model, we recover the action \eqref{Old Linearised SUGRA Massive Action}. If we instead represent  $\s$ in terms of an unconstrained prepotential $\S$ in the form 
\bea
    \s=\cDB^2\S~,
\eea
and also rename $U\rightarrow F$, we then can see that $F$ becomes linear upon integrating out $\S$, which gives us the action \eqref{New Linearised SUGRA Massive Action}. Hence, we see that by reducing the parent model of a theory we can construct a more general massive action which contains the derived dual massive models.


\section{Massless higher half-integer superspin in 4D} \label{Massless 4D Section}

We shall now review the theory for massless higher half-integer supermultiplets in $\mathbb{M}^{4|4}$ \cite{KPS}. We will begin with describing the so-called transverse and longitudinal formulations of these models before moving to the parent model which connects each formulation.

\subsection{Parent model in 4D}
For the massless theory of higher half-integer superspins \cite{KPS} of $Y=(s+1/2)$ in $\mathbb{M}^{4|4}$ there exist two different formulations which are connected by an parent action. These formulations are known as the transverse and longitudinal formulations, and they involve the supermultiplets
\bea
    \{\mathfrak{H}_{\a(s)\ad(s)},\bm{\G}_{\a(s-1)\ad(s-1)},\bar{\bm{\G}}_{\a(s-1)\ad(s-1)}\}
\eea
and
\bea
    \{\mathfrak{H}_{\a(s)\ad(s)},\mathfrak{G}_{\a(s-1)\ad(s-1)},\bar{\mathfrak{G}}_{\a(s-1)\ad(s-1)}\}
\eea
respectively. The field $\mathfrak{H}_{\a(s)\ad(s)}$ is a real field, while the remaining fields obey the constraints
\bea
    \mathfrak{\DB}^{\ad_1}\bm{\G}_{\a(s-1)\ad(s-1)}=0~,\qquad \mathfrak{\DB}_{(\ad_1}\mathfrak{G}_{\a(s-1)\ad_2\dots\ad_s)}=0~,
\eea
and consequently
\bea
    \mathfrak{\DB}^2\bm{\G}_{\a(s-1)\ad(s-1)}=0~,\qquad \mathfrak{\DB}^2\mathfrak{G}_{\a(s-1)\ad(s-1)}=0~.
\eea
The fields $\bm{\G}_{\a(s-1)\ad(s-1)}$ and $\mathfrak{G}_{\a(s-1)\ad(s-1)}$ are respectively called transversely linear and longitudinally linear. 

The gauge transformations of the fields $H_{\a(s)\ad(s)}$, $\bm{\G}_{\a(s-1)\ad(s-1)}$ and $\mathfrak{G}_{\a(s-1)\ad(s-1)}$ are defined:
\bsubeq
\bea
    \d \mathfrak{H}_{\a(s)\ad(s)}&=&\mathfrak{\DB}_{(\ad_1}\mathfrak{L}_{\a(s)\ad_2\dots\ad_s)}-\mathfrak{D}_{(\a_1}\bar{\mathfrak{L}}_{\a_2\dots\a_s)\ad(s)}~, \\
    \d \mathfrak{G}_{\a(s-1)\ad(s-1)}&=&-\frac{1}{4}\mathfrak{\DB}^2\mathfrak{D}^{\a_s}\mathfrak{L}_{\a(s)\ad(s-1)}+\ri(s-1)\pa^{\b\bd}\mathfrak{\DB}_{(\ad_1}\mathfrak{L}_{\b\a(s-1)\ad_2\dots\ad_{s-1})\bd}~,\\
    \d \bm{\G}_{\a(s-1)\bd(s-1)}&=&-\frac{1}{4}\mathfrak{\DB}^{\bd_s}\mathfrak{D}^2\bar{\mathfrak{L}}_{\a(s-1)\bd(s)}~,
\eea
\esubeq
where $\mathfrak{L}_{\a(s)\ad(s-1)}$ are unconstrained fields. The actions for the transverse and linear formulations are respectively:
\bsubeq
\bea
    S^{\bot}&=&\bigg(-\frac{1}{2}\bigg)^s\int\rd^{4|4}\bm{z}\bigg\{\frac{1}{8}\mathfrak{H}^{\a(s)\ad(s)}\mathfrak{D}^\g\mathfrak{\DB}^2\mathfrak{D}_\g \mathfrak{H}_{\a(s)\ad(s)}\non\\
    &&+\bigg(\mathfrak{H}^{\b\a(s-1)\bd\ad(s-1)}\mathfrak{D}_\a \mathfrak{\DB}_\ad \bm{\G}_{\a(s-1)\ad(s-1)}+\bar{\bm{\G}}^{\a(s-1)\ad(s-1)}\bm{\G}_{\a(s-1)\ad(s-1)}\non \\
    &&+\frac{s+1}{s}\bm{\G}^{\a(s-1)\ad(s-1)}\bm{\G}_{\a(s-1)\ad(s-1)}+\rc.\rc.\Big)\bigg\}~,\label{Transverse Action}
\eea
and
\bea
    S^{\|}&=&\bigg(-\frac{1}{2}\bigg)^s\int\rd^{4|4}\bm{z}\bigg\{\frac{1}{8}\mathfrak{H}^{\a(s)\ad(s)}\mathfrak{D}^\g\mathfrak{\DB}^2\mathfrak{D}_\g \mathfrak{H}_{\a(s)\ad(s)}\non\\
    &&-\frac{s}{8(2s+1)}[\mathfrak{D}_\b,\mathfrak{\DB}_\bd]\mathfrak{H}^{\b\a(s-1)\bd\ad(s-1)}[\mathfrak{D}^\g,\mathfrak{\DB}^\gd]\mathfrak{H}_{\g\a(s-1)\gd\ad(s-1)}\non\\
    &&+\frac{s}{2}\pa_{\b\bd}\mathfrak{H}^{\b\a(s-1)\bd\ad(s-1)}\pa^{\g\gd}\mathfrak{H}_{\g\a(s-1)\gd\ad(s-1)}\non\\
    &&+\bigg(\frac{2\ri s}{2s+1}\pa_{\b\bd}\mathfrak{H}^{\b\a(s-1)\bd\ad(s-1)} \mathfrak{G}_{\a(s-1)\ad(s-1)}+\frac{1}{2s+1}\bar{\mathfrak{G}}^{\a(s-1)\ad(s-1)}\mathfrak{G}_{\a(s-1)\ad(s-1)}\non \\
    &&-\frac{s+1}{s(2s+1)}\mathfrak{G}^{\a(s-1)\ad(s-1)}\mathfrak{G}_{\a(s-1)\ad(s-1)}+\rc.\rc.\Big)\bigg\}~.\label{Longitudinal Action}
\eea
\esubeq
As mentioned earlier, these two models are dual to each other. They turn out to be connected by the parent action
\bea\label{Action 4D Massless Real}
    S&=&\bigg(-\frac{1}{2}\bigg)^s\int\rd^{4|4}\bm{z}\bigg\{\frac{1}{8}\mathfrak{H}^{\a(s)\ad(s)}\mathfrak{D}^\g\mathfrak{\DB}^2\mathfrak{D}_\g \mathfrak{H}_{\a(s)\ad(s)}\non\\
    &&+\bigg(\mathfrak{H}^{\b\a(s-1)\bd\ad(s-1)}\mathfrak{D}_\b \mathfrak{\DB}_\bd \mathfrak{V}_{\a(s-1)\ad(s-1)}-\frac{2}{s}\mathfrak{G}^{\a(s-1)\ad(s-1)}\mathfrak{V}_{\a(s-1)\ad(s-1)}\non \\
    &&+\bar{\mathfrak{V}}^{\a(s-1)\ad(s-1)}\mathfrak{V}_{\a(s-1)\ad(s-1)}+\frac{s+1}{s}\mathfrak{V}^{\a(s-1)\ad(s-1)}\mathfrak{V}_{\a(s-1)\ad(s-1)}+\rc.\rc.\Big)\bigg\}~,
\eea
where
\be
    \mathfrak{H}_{\a(s)\ad(s)}=\bar{\mathfrak{H}}_{\a(s)\ad(s)}~,\qquad \mathfrak{\DB}_{(\ad_1}\mathfrak{G}_{\a(s-1)\ad_{2}\dots\ad_{s})}=0~,
\ee
and $\mathfrak{V}_{\a(s-1)\ad(s-1)}$ is unconstrained. This model exhibits the gauge transformations:
\bsubeq
\label{Higher Superspin Gauge Transformations}
\bea
    \d \mathfrak{H}_{\a(s)\ad(s)}&=&\mathfrak{\DB}_{(\ad_1}\mathfrak{L}_{\a(s)\ad_2\dots\ad_s)}-\mathfrak{D}_{(\a_1}\bar{\mathfrak{L}}_{\a_2\dots\a_s)\ad(s)}~, \\
    \d \mathfrak{G}_{\a(s-1)\ad(s-1)}&=&-\frac{1}{4}\mathfrak{\DB}^2\mathfrak{D}^{\a_s}\mathfrak{L}_{\a(s)\ad(s-1)}+\ri(s-1)\pa^{\b\bd}\mathfrak{\DB}_{(\ad_1}\mathfrak{L}_{\b\a(s-1)\ad_2\dots\ad_{s-1})\bd}~,\\
    \d \mathfrak{V}_{\a(s-1)\bd(s-1)}&=&-\frac{1}{4}\mathfrak{\DB}^{\bd_s}\mathfrak{D}^2\bar{\mathfrak{L}}_{\a(s-1)\bd(s)}~.
\eea
\esubeq
To recover the action \eqref{Transverse Action}, we simply integrate out $\mathfrak{G}_{\a(s-1)\ad(s-1)}$ and relabel $\mathfrak{V}_{\a(s-1)\ad(s-1)}$ by $\bm{\G}_{\a(s-1)\ad(s-1)}$, and to recover the action \eqref{Longitudinal Action}, we instead integrate out $\mathfrak{V}_{\a(s-1)\ad(s-1)}$.

We see from the above review that there are three possible models to dimensionally reduced in order to construct massive higher half-integer superspin theories in $\mathbb{M}^{3|4}_c$. Given that both the transverse and longitudinal models can be readily derived from the parent model, it is sensible to focus our approach on the latter. In doing so, we aim to create a more general theory from which contains all possible reduced models. With this in mind, we now move to the complexification of the action.


\subsection{Complexification of action}
As discussed in section \ref{Dim Red Section}, it is useful to dimensionally reduce models including real fields. To achieve this, it is useful to rewrite \eqref{Action 4D Massless Real} as
\bea
    S&=&\bigg(-\frac{1}{2}\bigg)^s\int\rd^{4|4}\bm{z}\bigg\{\mathfrak{H}^{\a(s)\ad(s)} \mathfrak{E}^{(0)}_{\a(s)\ad(s)}+\bar{\mathfrak{G}}^{\a(s-1)\ad(s-1)}\mathfrak{E}^{(1)}_{\a(s-1)\ad_{(s-1)}} \non \\
    &&~~~~~~~~~~~~~~~~~~~~~~+\mathfrak{G}^{\a(s-1)\ad(s-1)}\mathfrak{E}^{{(2)}}_{\a(s-1)\ad_{(s-1)}}+\bar{\mathfrak{V}}^{\a(s-1)\ad(s-1)}\mathfrak{E}^{(3)}_{\a(s-1)\ad_{(s-1)}}\non\\
    &&~~~~~~~~~~~~~~~~~~~~~~+\mathfrak{V}^{\a(s-1)\ad(s-1)}\mathfrak{E}^{(4)}_{\a(s-1)\ad_{(s-1)}}\bigg\}~,
\eea
where
\bea
    \mathfrak{E}^{(0)}_{\a(s)\bd(s)}&:=& \frac{1}{8}\mathfrak{D}^\g\mathfrak{\DB}^2\mathfrak{D}_\g \mathfrak{H}_{\a(s)\bd(s)}+\frac{1}{2}((\mathfrak{D} \mathfrak{\DB} \mathfrak{V})_{\a(s)\bd(s)}-(\mathfrak{\DB} \mathfrak{D}\bar{\mathfrak{V}})_{\a(s)\bd(s)})~,\non\\
    \mathfrak{E}^{(1)}_{\a(s-1)\bd(s-1)}&:=& -\frac{1}{s}\mathfrak{V}_{\a(s-1)\bd(s-1)}~,\qquad \mathfrak{E}^{(2)}_{\a(s-1)\bd(s-1)}:= -\frac{1}{s}\bar{\mathfrak{V}}_{\a(s-1)\bd(s-1)}~,\non\\
    \mathfrak{E}^{(3)}_{\a(s-1)\bd(s-1)}&:=& \frac{1}{2}\mathfrak{D}^\g \mathfrak{\DB}^\gd \mathfrak{H}_{\g\a(s-1)\gd\ad(s-1)} +\frac{s+1}{s}\bar{\mathfrak{V}}_{\a(s-1)\bd(s-1)}+\mathfrak{V}_{\a(s-1)\bd(s-1)}-\frac{1}{s}\bar{\mathfrak{G}}_{\a(s-1)\bd(s-1)}~,\non\\
    \mathfrak{E}^{(4)}_{\a(s-1)\bd(s-1)}&:=& -\frac{1}{2} \mathfrak{\DB}^\gd \mathfrak{D}^\g \mathfrak{H}_{\g\a(s-1)\gd\bd(s-1)} +\frac{s+1}{s}\mathfrak{V}_{\a(s-1)\bd(s-1)}+\bar{\mathfrak{V}}_{\a(s-1)\bd(s-1)}-\frac{1}{s}\mathfrak{G}_{\a(s-1)\bd(s-1)}~.\non\\
\eea

We note that this theory is of the same general form as outlined in subsection \ref{Coupled Dim Red}. To complexify this, we relax the reality condition on $\mathfrak{H}_{\a(s)\ad(s)}$, and make the following replacements in $\mathfrak{E}^{(0)}_{\a(s)\ad(s)}$, $\mathfrak{E}^{(1)}_{\a(s-1)\ad(s-1)}$, $\mathfrak{E}^{(2)}_{\a(s-1)\ad(s-1)}$, $\mathfrak{E}^{(3)}_{\a(s-1)\ad(s-1)}$ and $\mathfrak{E}^{(4)}_{\a(s-1)\ad(s-1)}$:
\bea
    \bar{\mathfrak{G}}_{\a(s-1)\ad(s-1)} \rightarrow \mathfrak{J}_{\a(s-1)\ad(s-1)}~,\\
    \bar{\mathfrak{V}}_{\a(s-1)\ad(s-1)} \rightarrow \mathfrak{W}_{\a(s-1)\ad(s-1)}~,
\eea
as well as substituting into the action
\bea
    \mathfrak{G}^{\a(s-1)\ad(s-1)}\mathfrak{E}^{(2)}_{\a(s-1)\ad_{(s-1)}}\rightarrow \bar{\mathfrak{J}}^{\a(s-1)\ad(s-1)}\mathfrak{E}^{(2)}_{\a(s-1)\ad_{(s-1)}}~,\\
    \mathfrak{V}^{\a(s-1)\ad(s-1)}\mathfrak{E}^{(4)}_{\a(s-1)\ad_{(s-1)}}\rightarrow \bar{\mathfrak{W}}^{\a(s-1)\ad(s-1)}\mathfrak{E}^{(4)}_{\a(s-1)\ad_{(s-1)}}~,
\eea
where $\mathfrak{J}_{\a(s-1)\ad(s-1)}$ is longitudinally linear in $\theta_\a$, i.e.
\bea
    \mathfrak{D}_{(\a_1}\mathfrak{J}_{\a_2\dots\a_{s})\ad(s-1)}=0~.
\eea
Therefore the new action is given by
\bea \label{4D Action Complex Field}
    \bm{S}&=&\bigg(-\frac{1}{2}\bigg)^s\int\rd^{4|4}\bm{z}\bigg\{\mathfrak{H}^{\a(s)\ad(s)} \mathfrak{E}^{(0)}_{\a(s)\ad(s)}+\bar{\mathfrak{G}}^{\a(s-1)\ad(s-1)}\mathfrak{E}^{(1)}_{\a(s-1)\ad_{(s-1)}} \non \\
    &&~~~~~~~~~~~~~~~~~~~~~~+\bar{\mathfrak{J}}^{\a(s-1)\ad(s-1)}\mathfrak{E}^{{(2)}}_{\a(s-1)\ad_{(s-1)}}+\bar{\mathfrak{V}}^{\a(s-1)\ad(s-1)}\mathfrak{E}^{(3)}_{\a(s-1)\ad_{(s-1)}}\non\\
    &&~~~~~~~~~~~~~~~~~~~~~~+\bar{\mathfrak{W}}^{\a(s-1)\ad(s-1)}\mathfrak{E}^{(4)}_{\a(s-1)\ad_{(s-1)}}\bigg\}~,
\eea
where
\bea
    \mathfrak{E}^{(0)}_{\a(s)\bd(s)}&:=& \frac{1}{8}\mathfrak{D}^\g\mathfrak{\DB}^2\mathfrak{D}_\g \mathfrak{H}_{\a(s)\bd(s)}+\bigg(\frac{1}{2}(\mathfrak{D} \mathfrak{\DB} \mathfrak{V})_{\a(s)\bd(s)}-\frac{1}{2}(\mathfrak{\DB} \mathfrak{D}\mathfrak{W})_{\a(s)\bd(s)}\bigg)~,\non\\
    \mathfrak{E}^{(1)}_{\a(s-1)\bd(s-1)}&:=& -\frac{1}{s}\mathfrak{V}_{\a(s-1)\bd(s-1)}~,\qquad \mathfrak{E}^{(2)}_{\a(s-1)\bd(s-1)}:= -\frac{1}{s}\mathfrak{W}_{\a(s-1)\bd(s-1)}~,\non\\
    \mathfrak{E}^{(3)}_{\a(s-1)\bd(s-1)}&:=& \frac{1}{2}\mathfrak{D}^\g \mathfrak{\DB}^\gd \mathfrak{H}_{\g\a(s-1)\gd\ad(s-1)} +\frac{s+1}{s}\mathfrak{W}_{\a(s-1)\bd(s-1)}+\mathfrak{V}_{\a(s-1)\bd(s-1)}-\frac{1}{s}\mathfrak{J}_{\a(s-1)\bd(s-1)}~,\non\\
    \mathfrak{E}^{(4)}_{\a(s-1)\bd(s-1)}&:=& -\frac{1}{2} \mathfrak{\DB}^\gd \mathfrak{D}^\g \mathfrak{H}_{\g\a(s-1)\gd\bd(s-1)} +\frac{s+1}{s}\mathfrak{V}_{\a(s-1)\bd(s-1)}+\mathfrak{W}_{\a(s-1)\bd(s-1)}-\frac{1}{s}\mathfrak{G}_{\a(s-1)\bd(s-1)}~.\non\\
\eea
The action $\bm{S}$ is invariant under the following gauge transformations
\bsubeq
\bea
    \d \mathfrak{H}_{\a(s)\bd(s)}&=&\mathfrak{\DB}_{(\bd_1}\mathfrak{L}_{\a(s)\bd_2\dots\bd_s)}-\mathfrak{D}_{(\a_1}\mathfrak{M}_{\a_2\dots\a_s)\bd(s)}~, \\
    \d \mathfrak{G}_{\a(s-1)\ad(s-1)}&=&-\frac{1}{4}\mathfrak{\DB}^2\mathfrak{D}^{\a_s}\mathfrak{L}_{\a(s)\ad(s-1)}+\ri(s-1)\pa^{\b\bd}\mathfrak{\DB}_{(\ad_1}\mathfrak{L}_{\b\a(s-1)\ad_2\dots\ad_{s-1})\bd},\\
    \d \mathfrak{J}_{\a(s-1)\ad(s-1)}&=&\frac{1}{4}\mathfrak{D}^2\mathfrak{\DB}^{\ad_s}\mathfrak{M}_{\a(s-1)\ad(s)}+\ri(s-1)\pa^{\b\bd}\mathfrak{D}_{(\a_1}\mathfrak{M}_{\a_2\dots\a_{s-1})\b\ad(s-1)\bd}~,\\
    \d \mathfrak{V}_{\a(s-1)\bd(s-1)}&=&-\frac{1}{4}\mathfrak{\DB}^{\bd_s}\mathfrak{D}^2\mathfrak{M}_{\a(s-1)\bd(s)}~,\\
    \d \mathfrak{W}_{\a(s-1)\bd(s-1)}&=&\frac{1}{4}\mathfrak{D}^{\bd_s}\mathfrak{\DB}^2\mathfrak{L}_{\a(s)\bd(s-1)}~.
\eea
\esubeq
To return to the original action, we simply impose the reality conditions
\bea
    \mathfrak{H}_{\a(s)\bd(s)}=\bar{\mathfrak{H}}_{\a(s)\bd(s)}~,\qquad \mathfrak{J}_{\a(s-1)\bd(s-1)}=\bar{\mathfrak{G}}_{\a(s-1)\bd(s-1)}~,\non\\
    \mathfrak{W}_{\a(s-1)\bd(s-1)}=\bar{\mathfrak{V}}_{\a(s-1)\bd(s-1)}~,\qquad \mathfrak{M}_{\a(s-1)\bd(s)}=\bar{\mathfrak{L}}_{\a(s-1)\bd(s)}~.
\eea

To dimensionally reduce the above theory will involve symmetrisation across many indices. It is useful to introduce a formalism which simplifies the process of symmetrisation in order to reduce the tediousness of calculations while carrying out the reduction procedure.


\section{Oscilator formalism for superfields} \label{Oscillator Section}
When considering a dimensional reduction of higher-(super)spin models, it is often useful to have a schema which makes the process of symmetrisation easier. This is particularly the case for the reduction $\mathbb{M}^{4|4}\rightarrow \mathbb{M}^{3|4}_c\times S^1$, since the inherited index structure possesses two separate sets of indices which are separately symmetrised. In the work of Pashnev \cite{Pashnev:1989}, an oscillator formalism was introduced where fields were replaced by bras and kets in a Fock space. In this way, the oscillators encoded the symmetrisation of fields and ensured that all operators which depended on such oscillators did not break the symmetry of a field when acting on a bra or ket-state. A similar formalism was introduced in \cite{GK} for 4D $\cN=1$ superspace where the Fock space exhibits two independent sets of creation and annihilation operators. One oscillator set encodes the symmetry of undotted indices, and the other encodes that of dotted indices. We will now review this latter formalism and consider its deformation under dimensional reduction.


\subsection{$\mathbb{M}^{4|4}$}
To construct an oscillator formalism for superfields in $\mathbb{M}^{4|4}$, we must opt to use spinor valued ladder operators as opposed to tensor valued ladder operators. To do so, we establish a pair of bosonic annihilation and creation operators $a_\a$ and $b_\a$ which transform in the left spinor representation $(1/2,0)$ with commutators
\bsubeq\label{Ladder Commutators}
\bea
   [a_\a,a_\b]=[b_\a,b_\b]=0~,\qquad [a_\a,b_\b] = [b_\a,a_\b] = \ve_{\a\b}~, \qquad\a,\b=1,2~. 
\eea
We also introduce a pair of bosonic annihilation and creation operators $ a'_\ad$ and $ b'_\ad$, transforming in the right spinor representation $(0,1/2)$ with commutators
\bea
   [ a'_\ad, a'_\bd]=[ b'_\ad, b'_\bd]=0~,\qquad  [ a'_\ad, b'_\bd] = [ b'_\ad, a'_\bd] = \ve_{\ad\bd}~, \qquad\ad,\bd=1,2~. 
\eea
\esubeq
The vacuum ket-state $|0\rangle$ and vacuum bra-state $\langle0|$ of the Fock space are defined such that
\bea
    a_\a|0\rangle= a'_\ad|0\rangle=\langle0|b_\a=\langle0| b'_\ad=0~,\qquad \langle0|0\rangle=1~.
\eea
We then construct operators $\mathfrak{B}=b^\a \mathfrak{D}_\a$, $ \bar{\mathfrak{A}}'= a'^\ad \mathfrak{\DB}_\ad$, $\mathfrak{A}=a^\a \mathfrak{D}_\a$ and $ \bar{\mathfrak{B}}'= b'^\ad \mathfrak{\DB}_\ad$ with the properties
\bea
    \mathfrak{B}^2=\mathfrak{A}^2= \bar{\mathfrak{A}}'^2= \bar{\mathfrak{B}}'^2=0~,\qquad \{\mathfrak{B},\mathfrak{A}\}=\mathfrak{D}^2~, \qquad \qquad \{ \bar{\mathfrak{A}}', \bar{\mathfrak{B}}'\}=-\mathfrak{\DB}^2~.
\eea
Consider then a superfield $\bm{\J}_{\a(j)\ad(k)}$. We associate with it a ket-state in the Fock space
\bea
    |\bm{\J}_{(j,k)}\rangle:=\bm{\J}_{\a(j)\ad(k)}b^{\a_1}\dots b^{\a_{j}} b'^{\ad_1}\dots b'^{\ad_{k}}|0\rangle~,
\eea
while its conjugate $\bar{\bm{\J}}_{\a(k)\ad(j)}$ is associated with a bra-state
\bea
    \langle\bm{\J}_{(j,k)}|:=\langle0|a^{\a_1}\dots a^{\a_{k}} a'^{\ad_1}\dots a'^{\ad_{j}}\bar{\bm{\J}}_{\a(k)\ad(j)}~.
\eea
The action of the operators on a ket state correspond to the following actions on superfields:
\bsubeq
\bea
    \mathfrak{B}|\bm{\J}_{(j,k)}\rangle\rightarrow \mathfrak{D}_{(\a_1}\bm{\J}_{\a_2\dots\a_{j+1})\ad(k)}~,\qquad \mathfrak{A}|\bm{\J}_{(j,k)}\rangle\rightarrow j\mathfrak{D}^{\a_j}\bm{\J}_{\a(j)\ad(k)}~,\\
     \bar{\mathfrak{B}}'|\bm{\J}_{(j,k)}\rangle\rightarrow \mathfrak{\DB}_{(\ad_1}\bm{\J}_{\a(j)\ad_2\dots\ad_{k+1})}~,\qquad  \bar{\mathfrak{A}}'|\bm{\J}_{(j,k)}\rangle\rightarrow k\mathfrak{\DB}^{\ad_k}\bm{\J}_{\a(j)\ad(k)}~,
\eea
while the action on bra states corresponds to
\bea
    \langle\bm{\J}_{(j,k)}|\mathfrak{B}&\rightarrow& -(-1)^{\ve(\bm{\J}_{(j,k)})}k\mathfrak{D}^{\a_k}\bar{\bm{\J}}_{\a(k)\ad(j)}~,\\
    \langle\bm{\J}_{(j,k)}|\mathfrak{A}&\rightarrow& -(-1)^{\ve(\bm{\J}_{(j,k)})}\mathfrak{D}_{(\a_1}\bar{\bm{\J}}_{\a_2\dots\a_{k+1)}\ad(j)}~,\\
    \langle\bm{\J}_{(j,k)}| \bar{\mathfrak{B}}'&\rightarrow& -(-1)^{\ve(\bm{\J}_{(j,k)})}j\mathfrak{\DB}^{\ad_j}\bar{\bm{\J}}_{\a(k)\ad(j)}~,\\
    \langle\bm{\J}_{(j,k)}| \bar{\mathfrak{A}}'&\rightarrow& -(-1)^{\ve(\bm{\J}_{(j,k)})}\mathfrak{\DB}_{(\ad_1}\bar{\bm{\J}}_{\a(k)\ad_1\dots\ad_{k+1})}~,
\eea
\esubeq
up to some surface terms. When we are working in coordinates present in the action measure, we can simply ignore these surface terms.
We introduce number operators
\bea
    N = b_\a a^\a~,\qquad N|\bm{\J}_{(j,k)}\rangle = j|\bm{\J}_{(j,k)}\rangle~,\qquad \langle\bm{\J}_{(j,k)}|N =\langle\bm{\J}_{(j,k)}|k~,\\
     N' =  b'_\ad  a'^{\ad} ~,\qquad  N'|\bm{\J}_{(j,k)}\rangle=k|\bm{\J}_{(j,k)}\rangle~,\qquad \langle\J_{(j,k)}| N' =\langle\bm{\J}_{(j,k)}|j~,
\eea
from which we find
\bea
    \mathfrak{B}\mathfrak{A} = -\frac{1}{2}N \mathfrak{D}^2~,\qquad  \bar{\mathfrak{B}}'~ \bar{\mathfrak{A}}' = \frac{1}{2} N' \mathfrak{\DB}^2~.
\eea
We then determine the remaining operator algebra:
\bea
    \{\mathfrak{B}, \bar{\mathfrak{B}}'\} = -2\ri b^\a b'^\ad\pa_{\a\ad}~,\qquad \{\mathfrak{B}, \bar{\mathfrak{A}}'\} = -2\ri b^\a a'^\ad\pa_{\a\ad}~,\\
    \{\mathfrak{A}, \bar{\mathfrak{B}}'\} = -2\ri a^\a b'^\ad\pa_{\a\ad}~,\qquad \{\mathfrak{A}, \bar{\mathfrak{A}}'\} = -2\ri a^\a a'^\ad\pa_{\a\ad}~.
\eea

Finally, we note that
\bea
    \mathfrak{D}^\a\mathfrak{\DB}^2 \mathfrak{D}_\a|\bm{\J}_{(j,k)}\rangle= \frac{1}{j+1}(\mathfrak{B}\{ \bar{\mathfrak{A}}', \bar{\mathfrak{B}}'\}\mathfrak{A}-\mathfrak{A}\{ \bar{\mathfrak{A}}', \bar{\mathfrak{B}}'\}\mathfrak{B})|\bm{\J}_{(j,k)}\rangle~,
\eea
for which it is convenient to introduce the operator
\bea
    \mathfrak{C}:=\mathfrak{D}^\a\mathfrak{\DB}^2 \mathfrak{D}_\a~.
\eea

Finally, we shall discuss how to take an inner product in the field formalism to the oscillator formalism. Consider two superfields $\F_{\a(j)\ad(k)}$ and $\J_{\a(k)\ad(j)}$ and the inner product $\langle\J_{(k,j)}|\F_{(j,k)}\rangle$. Our goal is to see how $\langle\J_{(k,j)}|\F_{(j,k)}\rangle$ is related to the inner product $\bar{\J}^{\a(j)\ad(k)}\F_{\a(j)\ad(k)}$. We expand the bra and ket states to get
\bea
    \langle\J_{(k,j)}|\F_{(j,k)}\rangle&=&\langle0|a^{\a_1}\dots a^{\a_{j}}a'^{\ad_{1}}\dots a'^{\ad_{k}}\bar{\J}_{\a(j)\ad(k)}\non\\
    &&~~~~\times\F_{\b(j)\bd(k)}b^{\b_{1}}\dots b^{\b_{j}}b'^{\bd_{1}}\dots b'^{\bd_{k}}|0\rangle~,
\eea
and then make use of the reduced form of the commutators \eqref{Ladder Commutators} to get
\bea
    \langle\J_{(k,j)}|\F_{(j,k)}\rangle&=&j!k!\bar{\J}_{\a(j)\ad(k)}\F_{\b(j)\bd(k)}\ve^{\b_1\a_1}\dots\ve^{\b_{j}\a_{j}}\ve^{\bd_1\ad_1}\dots\ve^{\bd_k\ad_k}\langle0|0\rangle~,\non\\
    &=&j!k!\bar{\J}^{\a(j)\ad(k)}\F_{\a(j)\ad(k)}
\eea
hence, to replace an inner product in the field formalism with one in the oscillator formalism, we take
\bea \label{Field to Oscillator Inner Product}
    \bar{\J}^{\a(j)\ad(k)}\F_{\a(j)\ad(k)}=\frac{1}{j!k!}\langle\J_{(k,j)}|\F_{(j,k)}\rangle~.
\eea

\subsection{From $\mathbb{M}^{4|4}$ to $\mathbb{M}^{3|4}_c$}\label{3D General} \label{3D Oscillator Formalism}
We now shall consider how the operators and states deform under dimensional reduction. The deformation of covariant derivatives will be as in \eqref{4D Covariant Deformation}, while the second set of creation and annihilation operators will be
\bea
    a'_\ad\rightarrow a'_\a~,\qquad b'_\ad\rightarrow b'_\a~.
\eea
The operators defined above will then be deformed:
\bsubeq \label{Reduced Operators}
\bea
    \mathfrak{A}\rightarrow\cA=a^\a\cD_\a~,\qquad \mathfrak{B}\rightarrow\cB=b^\a\cD_\a~,\non\\
    \bar{\mathfrak{A}}'\rightarrow\bar{\cA}'=a'^\a\cDB_\a~,\qquad \bar{\mathfrak{B}}'\rightarrow\bar{\cB}'=b'^\a\cDB_\a~,\qquad \mathfrak{C}\rightarrow\cC=\cD^\a\cDB^2\cD_\a~.
\eea
In addition, the following differential operators arise
\bea
    \cA'&=&a'^\a\cD_\a~,\qquad \cB'=b'^\a\cD_\a~,\qquad \bar{\cA}=a^\a\cDB_\a~,\qquad \bar{\cB}=b^\a\cDB_\a~,\non\\
    P_1 &:=& - b^\a a^\b p_\ab~,\qquad P_2 := -   b'^\b a'^\a p_\ab~,\qquad p_\ab=-\ri\pa_\ab~,\non\\
    P_3&:=&  -  a'^\a a^\b p_\ab~,\qquad P_4 := -   b'^\b b^\a p_\ab~,\qquad P_5:=  -  a'^\a  a'^\b p_\ab~,\qquad P_6 := -   b'^\a  b'^\b p_\ab~,\non\\
    P_7&:=&  - a^\a  b'^\b p_\ab~,\qquad P_8 := -   a'^\b b^\a p_\ab~,\qquad P_{9}:=  - a^\a a^\b p_\ab~,\qquad P_{10} := -  b^\a b^\b p_\ab~,\non\\
    \cK&:=& \cD^\a\cDB_\a~,\qquad \bar{\cK}:= \cDB_\a\cD^\a~.\qquad~~
\eea
\esubeq
The reduction of a ket state is then taken as follows
\bea
    |\bm{\J}_{(j,k)}\rangle=\re^{\ri ym}|\J_{(j,k)}\rangle=\re^{\ri ym}\J_{\a(j)\b(k)}b^{\a_1}\dots b^{\a_j}b'^{\b_1}\dotsb'^{\b_k}|0\rangle~.
\eea
It is important to note that the field $\J_{\a(j)\b(k)}$ is not an irreducible field. Rather it is a field within the spin representation $\cD^{(j/2)}\otimes\cD^{(k/2)}$. We can then make use of the decomposition
\bea
    \cD^{(j/2)}\otimes\cD^{(k/2)}=\bigoplus_{l=0}^{\min\{j,k\}}\cD^{(|j-k|/2+l)}~,
\eea
to see that the field can be decomposed into a sum of $\min\{j,k\}$ irreducible fields. It is not difficult to see that these irreducible parts of $\J_{\a(j)\b(k)}$ will be given by symmetrised traces of the field, i.e. by $\J^{\g(l)}_{~~~~\a(j+k-2l)\g(l)}$. 
We can take the reduction to be
\begin{align}
    |\bm{\J}_{(j,k)}\rangle=\sum_{l=0}^{min\{j,k\}}\re^{\ri ym}(\s^2)_{\a_{1}\b_{1}}\dots(\s^2)_{\a_l\b_l}\F_{(\a_{l+1}\dots\a_{j}\b_{l+1}\dots \b_{k})}b^{\a_1}\dots b^{\a_j}b'^{\b_1}\dotsb'^{\b_k}|0\rangle~,
\end{align}
where $\F_{\a(j+k-2l)}$ is given by
\bea
    \F_{\a(j+k-2l)}=(-\ri)^l\frac{\binom{j}{l}\binom{k}{l}}{\binom{j+k-l+1}{l}}\J^{\g(l)}_{~~~~\a(j+k-2l)\g(l)}~.
\eea
We can then then insert $(\s^2)_\ab=\ri\ve _{\ab}$ to get
\bea
    |\bm{\J}_{(j,k)}\rangle=\sum_{l=0}^{min\{j,k\}}\re^{\ri ym}(\ri b^{\g}b'_\g)^l\F_{(\a_{l+1}\dots\a_{j}\b_{l+1}\dots \b_{k})}b^{\a_{l+1}}\dots b^{\a_j}b'^{\b_{l+1}}\dotsb'^{\b_k}|0\rangle~.
\eea
We can then note the irreducible ket states
\bea
    |\F_{(j-l,k-l)}\rangle=\F_{(\a_{l+1}\dots\a_{j}\b_{l+1}\dots \b_{k})}b^{\a_{l+1}}\dots b^{\a_j}b'^{\b_{l+1}}\dotsb'^{\b_k}|0\rangle~,
\eea
and introduce the operators
\bea
    V:=a^\a a'_\a~,\qquad W:=b_\a b'^\a~,
\eea
to get
\bea
    |\bm{\J}_{(j,k)}\rangle=\sum_{l=0}^{min\{j,k\}}\re^{\ri ym}(-\ri W)^l|\F_{(j-l,k-l)}\rangle~.\label{Irred Field Reduction}
\eea

It is worth noting that while the field $|\F_{(j-l,k-l)}\rangle$ are symmetric, they do not remain so under the operators we have introduced. The reason for this is that the two sets of oscillators we have inherited from $\mathbb{M}^{4|4}$ correspond to two different symmetrisations across one type of index. It is thus useful to devise a method of removing one of these oscillator sets from our calculations.

To achieve this, we introduce the following operators which arise in 3D
\bea \label{Swap Operators}
    T:=a^\a b'_\a~,\qquad U:= a'^\a b_\a~,
\eea
whose commutator is
\bea
    [U,T]=N- N'~,
\eea
and which act on bra and ket states such that
\bsubeq
\bea
    T|\F_{(j,k)}\rangle=j|\F_{(j-1,k+1)}\rangle \rightarrow j\d^{\a_j}_{(\b_1|}\F_{\a(j)|\b_2\dots\b_{k+1})}~,\\
    \langle\F_{(j,k)}|T=\langle\F_{(j-1,k+1)}|j \rightarrow j\d^{\b_j}_{(\a_1}\bar{\F}_{\a_2\dots\a_{k+1})\b(j)}~,\\
    U|\F_{(j,k)}\rangle=k|\F_{(j+1,k-1)}\rangle \rightarrow k\d^{\b_k}_{(\a_1}\F_{\a_2\dots\a_{j+1})\b(k)}~,\\
    \langle\F_{(j,k)}|U=\langle\F_{(j+1,k-1)}|k \rightarrow k\d^{\a_k}_{(\b_1|}\bar{\F}_{\a(k)|\b_2\dots\b_{j+1})}~.
\eea
\esubeq
We see then that the operators \eqref{Swap Operators} allow us to move a state's dependence on one set of oscillators to another. It follows that given a totally symmetric field $|\F_{(j,k)}\rangle$, we have
\bea\label{3D Symmetrisation}
    |\F_{(j,k)}\rangle = \frac{k!}{(j+k)!}U^j|\F_{(0,j+k)}\rangle~.
\eea
With the above, we can treat $|\F_{(0,j+k)}\rangle$ as the fundamental objects in $\mathbb{R}^{3|4}$, since it only depends on one set of creation and annihilation operators, $b'_\a$ and $a'_\a$. Inserting the above into the decomposition \eqref{Irred Field Reduction} gives us
\bsubeq
\bea\label{Ket Full Decomp}
    |\bm{\J}_{(j,k)}\rangle=\sum_{l=0}^{min\{j,k\}}\re^{\ri ym}\frac{(k-l)!}{(j+k-2l)!}(-\ri W)^lU^{j-l}|\F_{(0,j+k-2l)}\rangle
\eea
as a means to decompose ket-states into their irreducible components. A similar analysis for the bra-states leads to the decomposition
\bea\label{Bra Full Decomp}
    \langle\bm{\J}_{(j,k)}|=\sum_{l=0}^{min\{j,k\}}\re^{-\ri ym}\frac{(j-l)!}{(j+k-2l)!}\langle\F_{(j+k-2l,0)}|T^{k-l}(\ri V)^l~.
\eea
\esubeq

We shall now see that any inner product of bra and ket states can be written entirely in terms of one set of oscillators. To see how this is done, we first consider the action of the various operators on $|\F_{(0,k)}\rangle$:
\bea\label{Operator Action 3D}
    V|\F_{(0,k)}\rangle&=&P_3|\F_{(0,k)}\rangle=P_1|\F_{(0,k)}\rangle=P_7|\F_{(0,k)}\rangle=P_9|\F_{(0,k)}\rangle=\cA|\F_{(0,k)}\rangle=0~,\non\\
    \cB'|\F_{(0,k)}\rangle&\rightarrow& \cD_{(\a_1}\F_{\a_2\dots\a_{k+1})}~,\qquad \bar{\cB}'|\F_{(0,k)}\rangle\rightarrow \cDB_{(\a_1}\F_{\a_2\dots\a_{k+1})}~,\non\\
    \cA'|\F_{(0,k)}\rangle&\rightarrow& k\cD^{\a_{k}}\F_{\a(k)}~,\qquad \bar{\cA}'|\F_{(0,k)}\rangle\rightarrow k\cDB^{\a_{k}}\F_{\a(k)}~,\non\\
    \cB|\F_{(0,k)}\rangle&=&\frac{1}{k+1}(U\cB'- W\cA')|\F_{(0,k)}\rangle~,\qquad \bar{\cB}|\F_{(0,k)}\rangle=\frac{1}{k+1}(U\bar{\cB}'- W\bar{\cA}')|\F_{(0,k)}\rangle~,\non\\
    P_2|\F_{(0,k)}\rangle&\rightarrow& \ri k\pa_{(\a_1}^\g\F_{\a_2\dots\a_{k})\g}~,\non\\
    P_5|\F_{(0,k)}\rangle&\rightarrow& \ri k(k-1)\pa^{\a_1\a_2}\F_{\a(k)}~,\qquad P_6|\F_{(0,k)}\rangle\rightarrow \ri\pa_{(\g\r}\F_{\a_1\dots\a_{k})}~,\non\\
    P_4|\F_{(0,k)}\rangle&=&\frac{1}{k+2}(U P_6- WP_2)|\F_{(0,k)}\rangle~,\qquad P_8|\F_{(0,k)}\rangle=\frac{1}{k}(U P_2- WP_5)|\F_{(0,k)}\rangle~,\non\\
    \cK|\F_{(0,k)}\rangle &=&\frac{1}{k+1}(\cA'\bar{\cB}'-\cB'\bar{\cA}')|\F_{(0,k)}\rangle~,\qquad \bar{\cK}|\F_{(0,k)}\rangle =\frac{1}{k+1}(\bar{\cB}'\cA'-\bar{\cA}'\cB')|\F_{(0,k)}\rangle~,\non\\
    P_{10}|\F_{(0,k)}\rangle&=&\frac{1}{k+1}\bigg[\frac{1}{k+2}U^2P_6+\frac{1}{k} W^2P_5-\frac{2(k+1)}{k(k+2)}U WP_2\bigg]|\F_{(0,k)}\rangle~.
\eea
We see from the above, that any operator acting on $|\F_{(0,k)}\rangle$ can be decomposed in terms of the operators $U, W,\cA ', \cB',\bar{\cA}',\bar{\cB}',P_2,P_5$ and $P_6$. With this established, let us consider the inner product of two symmetric fields $\F_{\a(2s-2k)}$ and $\J_{\a(2s-2k)}$
\bea
    \langle\J_{(s-k,s-k)}|O_0|\F_{(s-k,s-k)}\rangle~,
\eea
where $O_0$ is some operator which is a combination of those defined in \eqref{Reduced Operators}\footnote{We consider for simplicity only inner products of fields with the same rank, however it can be shown that the below procedure is valid for fields of different rank.}. We will show how this inner product which depends on both sets of oscillators can be rewritten as an inner product depending only on the oscillators $a'_\a$ and $b'_\a$. We can substitute \eqref{3D Symmetrisation} to get
\bea
    \frac{(s-k)!}{(2s-2k)!}\langle\J_{(s-k,s-k)}|O_0U^{s-k}|\F_{(0,2s-2k)}\rangle~.
\eea
We can then use the algebra to pull $U^{s-k}$ through $O_0$ to get
\bea
    \frac{(s-k)!}{(2s-2k)!}\langle\J_{(s-k,s-k)}|([O_0,U^{s-k}]+U^{s-k} O_0)|\F_{(0,2s-2k)}\rangle~.
\eea
We can see from the algebra that the commutator will have the form
\bea
    [O_0,U^{s-k}]=\sum_{i=1}^{s-k}U^{s-k-i}O_i~,
\eea
where $O_i$ are some combinations of operators excluding $U$. We can then write the object as
\bea
    \sum_{i=1}^{s-k}\frac{(s-k)!}{(2s-2k)!}\langle\J_{(s-k,s-k)}|U^{s-k-i}O_i|\F_{(0,2s-2k)}\rangle~,
\eea
which we can rewrite as
\bea
    \sum_{i=1}^{s-k}\frac{((s-k)!)^2}{(2s-2k)!i!}\langle\J_{(2s-2k-i,i)}|O_i|\F_{(0,2s-2k)}\rangle~.
\eea
From here, we note that we can use \eqref{Operator Action 3D} to see that the non-vanishing contributions from $O_i$ will be those that depend only on $U, W,\cA', \cB', \bar{\cA}',\bar{\cB}',P_2,P_5$ and $P_6$. We can use the algebra to bring all $U$ and $ W$ operators to the left, i.e. we can take
\bea
    O_i(U, W,\cA', \cB',\bar{\cA}',\bar{\cB}',P_2,P_5,P_6)=\sum_{l=0}^{i} U^{i-l} W^{l} O'_{il}(\cA', \cB',\bar{\cA}',\bar{\cB}',P_2,P_5,P_6)~,
\eea
since $U$ and $ W$ commute. We can then substitute the above to get
\bea
    \sum_{i=1}^{s-k}\sum_{l=0}^i\frac{((s-k)!)^2}{(2s-2k)!i!}\langle\J_{(2s-2k-i,i)}|U^{i-l} W^lO'_{il}|\F_{(0,2s-2k)}\rangle\non\\
    = \sum_{i=1}^{s-k}\sum_{l=0}^i\frac{((s-k)!)^2}{(2s-2k)!l!}\langle\J_{(2s-2k-l,l)}| W^lO'_{il}|\F_{(0,2s-2k)}\rangle\non\\
    = \sum_{i=1}^{s-k}\frac{((s-k)!)^2}{(2s-2k)!}\langle\J_{(2s-2k,0)}|O'_{i0}|\F_{(0,2s-2k)}\rangle\non\\
    = \frac{((s-k)!)^2}{(2s-2k)!}\langle\J_{(2s-2k,0)}|\bigg(\sum_{i=1}^{s-k}O'_{i0}\bigg)|\F_{(0,2s-2k)}\rangle~,
\eea
which now only involves one set of creation and annihilation operators, $b'_\a$ and $a'_\a$. At this stage, it makes sense to redenote
\bea\label{Redenote States}
    |\F_{(0,2s-2k)}\rangle=|\F_{(2s-2k)}\rangle~, \qquad \langle\J_{(2s-2k,0)}|=\langle\J_{(2s-2k)}|~.
\eea


\section{Massive higher half-integer superspin in 3D} \label{Model Section}
With the oscillator formalism introduced, we will now move to the main result of this paper, that is the dimensional reduction of higher half-integer superspin superfields from $\mathbb{M}^{4|4}\rightarrow\mathbb{M}^{3|4}_c$. To achieve this, we shall recast the model described in section \ref{Massless 4D Section} in terms of the oscillator formalism. We shall then carry out the dimensional reduction according to section \ref{SUSY Reduction Section} to derive a new model for massive gauge-invariant half-integer spin supermultiplets in $\mathbb{M}^{3|4}_c$. We then integrate out the fields descending from the 4D longitudinally-linear superfields to generate a reduced analogue for the transverse formulation of the 4D model. Finally, we prove that the on-shell equations lead to the expected equations of motion for massive half-integer superspin fields under dimensional reduction.


\subsection{Massless model in oscillator formalism}
Following the substitution procedure of \eqref{Field to Oscillator Inner Product}, the action \eqref{4D Action Complex Field} becomes
\bea\label{4D massless action oscillator}
    \bm{S}&=&\bigg(-\frac{1}{2}\bigg)^s\bigg(\frac{1}{s!}\bigg)^2\int\rd^{4|4}\bm{z}\bigg\{\langle \mathfrak{H}_{(s,s)}| \mathfrak{E}^{(0)}_{(s,s)}\rangle+s^2\langle \mathfrak{G}_{(s-1,s-1)}| \mathfrak{E}^{(1)}_{(s-1,s-1)}\rangle\non\\
    &&~~~~~~~~~~+s^2\langle \mathfrak{J}_{(s-1,s-1)}| \mathfrak{E}^{(2)}_{(s-1,s-1)}\rangle+s^2\langle \mathfrak{V}_{(s-1,s-1)}| \mathfrak{E}^{(3)}_{(s-1,s-1)}\rangle+s^2\langle \mathfrak{W}_{(s-1,s-1)}| \mathfrak{E}^{(4)}_{(s-1,s-1)}\rangle\bigg\}~,\non\\
\eea
where
\bea\label{Higher Spin EoM Complex Basis}
    |\mathfrak{E}_{(s,s)}\rangle&=& \frac{1}{8}\mathfrak{C}|\mathfrak{H}_{(s,s)}\rangle+\frac{1}{2}(\mathfrak{B} \bar{\mathfrak{B}}'|\mathfrak{V}_{(s-1,s-1)}\rangle- \bar{\mathfrak{B}}'\mathfrak{B}|\mathfrak{W}_{(s-1,s-1)}\rangle)~,\non\\
    |\mathfrak{E}^{(1)}_{(s-1,s-1)}\rangle&=&-\frac{1}{s}|\mathfrak{W}_{(s-1,s-1)}\rangle~,\qquad |\mathfrak{E}^{(2)}_{(s-1,s-1)}\rangle=-\frac{1}{s}|\mathfrak{V}_{(s-1,s-1)}\rangle~,\non\\
    |\mathfrak{E}^{(3)}_{(s-1,s-1)}\rangle&=& \frac{1}{2s^2}\mathfrak{A} \bar{\mathfrak{A}}'| \mathfrak{H}_{(s,s)}\rangle +\frac{s+1}{s}|\mathfrak{W}_{(s-1,s-1)}\rangle+|\mathfrak{V}_{(s-1,s-1)}\rangle-\frac{1}{s}|\mathfrak{J}_{(s-1,s-1)}\rangle~,\non\\
    |\mathfrak{E}^{(4)}_{(s-1,s-1)}\rangle&=& -\frac{1}{2s^2}  \bar{\mathfrak{A}}'\mathfrak{A} |\mathfrak{H}_{(s,s)}\rangle +\frac{s+1}{s}|\mathfrak{V}_{(s-1,s-1)}\rangle+|\mathfrak{W}_{(s-1,s-1)}\rangle-\frac{1}{s}|\mathfrak{G}_{(s-1,s-1)}\rangle~.~~~~~~
\eea
The action in the oscillator formalism is invariant under the gauge transformations
\bsubeq
\label{Complex Higher Superspin Gauge Transformations}
\bea
    \d |\mathfrak{H}_{(s,s)}\rangle&=& \bar{\mathfrak{B}}'|\mathfrak{L}_{(s,s-1)}\rangle-\mathfrak{B}|\mathfrak{M}_{(s-1,s)}\rangle~, \\
    \d |\mathfrak{G}_{(s-1,s-1)}\rangle&=&-\frac{1}{4s}\mathfrak{\DB}^2\mathfrak{A}|\mathfrak{L}_{(s,s-1)}\rangle-\frac{1}{2s} \bar{\mathfrak{B}}'\{\mathfrak{A}, \bar{\mathfrak{A}}'\}|\mathfrak{L}_{(s,s-1)}\rangle,\\
    \d |\mathfrak{J}_{(s-1,s-1)}\rangle&=&\frac{1}{4s}\mathfrak{D}^2 \bar{\mathfrak{A}}' |\mathfrak{M}_{(s-1,s)}\rangle-\frac{1}{2s}\mathfrak{B} \{\mathfrak{A}, \bar{\mathfrak{A}}'\}|\mathfrak{M}_{(s-1,s)}\rangle~,\\
    \d |\mathfrak{V}_{(s-1,s-1)}\rangle&=&-\frac{1}{4s} \bar{\mathfrak{A}}'\mathfrak{D}^2|\mathfrak{M}_{(s-1,s)}\rangle~,\\
    \d |\mathfrak{W}_{(s-1,s-1)}\rangle&=&\frac{1}{4s}\mathfrak{A}\mathfrak{\DB}^2|\mathfrak{L}_{(s,s-1)}\rangle~.
\eea
\esubeq
We can rewrite these in terms of two longitudinally linear gauge parameters:
\bea
    |\mathfrak{l}_{(s,s)}\rangle:= \bar{\mathfrak{B}}'|\mathfrak{L}_{(s,s-1)}\rangle~,\qquad |\mathfrak{m}_{(s,s)}\rangle:=-\mathfrak{B}|\mathfrak{M}_{(s-1,s)}\rangle~,
\eea
so that the gauge transformations are
\bsubeq
\label{Complex Higher Superspin Gauge Transformations Long Linear}
\bea
    \d |\mathfrak{H}_{(s,s)}\rangle&=&|\mathfrak{l}_{(s,s)}\rangle+|\mathfrak{m}_{(s,s)}\rangle~, \\
    \d |\mathfrak{G}_{(s-1,s-1)}\rangle&=&\frac{1}{2s(s+1)}\mathfrak{A} \bar{\mathfrak{A}}'|\mathfrak{l}_{(s,s)}\rangle-\frac{1}{2s}\{\mathfrak{A}, \bar{\mathfrak{A}}'\}|\mathfrak{l}_{(s,s)}\rangle,\\
    \d |\mathfrak{J}_{(s-1,s-1)}\rangle&=&-\frac{1}{2s(s+1)} \bar{\mathfrak{A}}'\mathfrak{A} |\mathfrak{m}_{(s,s)}\rangle+\frac{1}{2s} \{\mathfrak{A}, \bar{\mathfrak{A}}'\}|\mathfrak{m}_{(s,s)}\rangle~,\\
    \d |\mathfrak{V}_{(s-1,s-1)}\rangle&=&\frac{1}{2s(s+1)} \bar{\mathfrak{A}}'\mathfrak{A}|\mathfrak{m}_{(s,s)}\rangle~,\\
    \d |\mathfrak{W}_{(s-1,s-1)}\rangle&=&-\frac{1}{2s(s+1)}\mathfrak{A} \bar{\mathfrak{A}}'|\mathfrak{l}_{(s,s)}\rangle~.
\eea
\esubeq

In oscillator form, the constraints on $|\mathfrak{G}_{(s-1,s-1)}\rangle$ and $|\mathfrak{J}_{(s-1,s-1)}\rangle$ are
\bea \label{Constraints 4D Oscillator}
     \bar{\mathfrak{B}}'|\mathfrak{G}_{(s-1,s-1)}\rangle=0~,\qquad \mathfrak{B}|\mathfrak{J}_{(s-1,s-1)}\rangle=0~.
\eea


\subsection{Dimensional reduction}
We now move to the dimensional reduction of the model above. We start by reducing the fields and gauge parameters ket-states according to \eqref{Ket Full Decomp} and absorbing the coefficients into the fields to get:
\bea\label{Fields Decomposed Fully No Coeff}
    |\mathfrak{H}_{(s,s)}\rangle&=&\re^{\ri y m}\sum_{k=0}^s(-\ri  W)^{k}U^{s-k}| H_{(2s-2k)}\rangle~,\non\\
    |\mathfrak{G}_{(s-1,s-1)}\rangle&=&\re^{\ri y m}\sum_{k=1}^{s}(-\ri  W)^{k-1}U^{s-k}| G_{(2s-2k)}\rangle~,\non\\
    |\mathfrak{J}_{(s-1,s-1)}\rangle&=&\re^{\ri y m}\sum_{k=1}^{s}(-\ri  W)^{k-1}U^{s-k}| J_{(2s-2k)}\rangle~,\non\\
    |\mathfrak{V}_{(s-1,s-1)}\rangle&=&\re^{\ri y m}\sum_{k=1}^{s}(-\ri  W)^{k-1}U^{s-k}| V_{(2s-2k)}\rangle~,\non\\
    |\mathfrak{W}_{(s-1,s-1)}\rangle&=&\re^{\ri y m}\sum_{k=1}^{s}(-\ri  W)^{k-1}U^{s-k}| W_{(2s-2k)}\rangle~,\non\\
    |\mathfrak{l}_{(s,s)}\rangle&=&\re^{\ri y m}\sum_{k=0}^{s}(-\ri  W)^{k}U^{s-k}| l_{(2s-2k)}\rangle~,\non\\
    |\mathfrak{m}_{(s,s)}\rangle&=&\re^{\ri y m}\sum_{k=0}^{s}(-\ri  W)^{k}U^{s-k}| m_{(2s-2k)}\rangle~,
\eea
and doing the same for their bra-states according to \eqref{Bra Full Decomp} to get
\bea
    \langle\mathfrak{H}_{(s,s)}|&=&\re^{-\ri y m}\sum_{k=0}^s\langle H_{(2s-2k)}|T^{s-k}(\ri  V)^{k}~,\non\\
    \langle\mathfrak{G}_{(s-1,s-1)}|&=&\re^{-\ri y m}\sum_{k=1}^{s}\langle G_{(2s-2k)}|T^{s-k}(\ri  V)^{k}~,\non\\
    \langle\mathfrak{J}_{(s-1,s-1)}|&=&\re^{-\ri y m}\sum_{k=1}^{s}\langle J_{(2s-2k)}|T^{s-k}(\ri  V)^{k}~,\non\\
    \langle\mathfrak{V}_{(s-1,s-1)}|&=&\re^{-\ri y m}\sum_{k=1}^{s}\langle V_{(2s-2k)}|T^{s-k}(\ri  V)^{k}~,\non\\
    \langle\mathfrak{W}_{(s-1,s-1)}|&=&\re^{-\ri y m}\sum_{k=1}^{s}\langle W_{(2s-2k)}|T^{s-k}(\ri  V)^{k}~,\non\\
    \langle\mathfrak{l}_{(s,s)}|&=&\re^{-\ri y m}\sum_{k=0}^{s}\langle l_{(2s-2k)}|T^{s-k}(\ri  V)^{k}~,\non\\
    \langle\mathfrak{m}_{(s,s)}|&=&\re^{-\ri y m}\sum_{k=0}^{s}\langle m_{(2s-2k)}|T^{s-k}(\ri  V)^{k}~,
\eea
where we have made use of the redenotation \eqref{Redenote States}. Upon these substitutions, we see that the massive action for higher half-integer spin gauge supermultiplets is given by
\bea\label{Action Recap}
    \bm{S}&=&\sum_{k=0}^s\bigg(-\frac{1}{2}\bigg)^s\bigg(\frac{1}{s!}\bigg)^2\int\rd^{3|4}z\bigg\{\langle  H_{(2s-2k)}|  E^{(0)}_{(2s-2k)}\rangle+s^2\langle  G_{(2s-2k)}|  E^{(1)}_{(2s-2k)}\rangle\non\\
    &&~~~~~~~~~~~~~~~~~~~+s^2\langle  J_{(2s-2k)}|  E^{(2)}_{(2s-2k)}\rangle+s^2\langle  V_{(2s-2k)}|  E^{(3)}_{(2s-2k)}\rangle\non\\
    &&~~~~~~~~~~~~~~~~~~~+s^2\langle  W_{(2s-2k)}|  E^{(4)}_{(2s-2k)}\rangle\bigg\}~,
\eea
where
\bea\label{EoM Higher Spin}
    | E^{(0)}_{(2s-2k)}\rangle&:=& \oint\rd y~\re^{-\ri ym}~T^{s-k}(\ri V)^k|\mathfrak{E}^{(0)}_{(s,s)}\rangle~,\non\\
    | E^{(1)}_{(2s-2k)}\rangle&:=& \oint\rd y~\re^{-\ri ym}~T^{s-k}(\ri V)^{k-1}|\mathfrak{E}^{(1)}_{(s-1,s-1)}\rangle~,\non\\
    | E^{(2)}_{(2s-2k)}\rangle&:=& \oint\rd y~\re^{-\ri ym}~T^{s-k}(\ri V)^{k-1}|\mathfrak{E}^{(2)}_{(s-1,s-1)}\rangle~,\non\\
    | E^{(3)}_{(2s-2k)}\rangle&:=& \oint\rd y~\re^{-\ri ym}~T^{s-k}(\ri V)^{k-1}|\mathfrak{E}^{(3)}_{(s-1,s-1)}\rangle~,\non\\
    | E^{(4)}_{(2s-2k)}\rangle&:=& \oint\rd y~\re^{-\ri ym}~T^{s-k}(\ri V)^{k-1}|\mathfrak{E}^{(4)}_{(s-1,s-1)}\rangle~,
\eea
where it is assumed that
\bea
    \langle G_{(s,s)}|=0~,\qquad \langle J_{(s,s)}|=0~,\qquad \langle V_{(s,s)}|=0~,\qquad \langle W_{(s,s)}|=0~.
\eea
We now move to finding the expressions \eqref{EoM Higher Spin}. By rewriting \eqref{Higher Spin EoM Complex Basis} in terms of our 3D operators, inserting in \eqref{Fields Decomposed Fully No Coeff} and making use of the algebra to rearrange the operators, we arrive at
\bea\label{Higher Spin EoM Real Basis}
    |\mathfrak{E}^{(0)}_{(s,s)}\rangle&=& \re^{\ri y m}\sum_{k=0}^s(-\ri)^k\bigg\{\frac{1}{8} W^kU^{s-k}\cC | H_{(2s-2k)}\rangle-\frac{\ri W^kU^{s-k}}{4}(\cK| V_{(2s-2k)}\rangle-\bar{\cK}| W_{(2s-2k)}\rangle)\non\\
    &&+\frac{\ri W^{k-1}U^{s-k+1}}{4(2s-2k+1)}(\cB'\bar{\cB}'| V_{(2s-2k)}\rangle-\bar{\cB}'\cB'| W_{(2s-2k)}\rangle)\non\\
    &&-\frac{\ri W^{k+1}U^{s-k-1}}{4(2s-2k+1)}(\cA'\bar{\cA}'| V_{(2s-2k)}\rangle-\bar{\cA}'\cA'| W_{(2s-2k)}\rangle)\bigg\}~,\non\\
    |\mathfrak{E}^{(1)}_{(s-1,s-1)}\rangle&=& -\frac{1}{s}\re^{\ri y m}\sum_{k=1}^s(-\ri  W)^{k-1}U^{s-k}| W_{(2s-2k)}\rangle~,\non\\
    |\mathfrak{E}^{(2)}_{(s-1,s-1)}\rangle&=& -\frac{1}{s}\re^{\ri y m}\sum_{k=1}^s(-\ri  W)^{k-1}U^{s-k}| V_{(2s-2k)}\rangle~,\non\\
    |\mathfrak{E}^{(3)}_{(s-1,s-1)}\rangle&=& \re^{\ri y m}\sum_{k=0}^s(-\ri)^{k-1}\non\\
    &&\times\bigg\{\frac{1}{4s^2(2s-2k+1)}\bigg[ \ri k(2s-2k+1)(2s-k+1) W^{k-1}U^{s-k}\cK\non\\
    &&+\ri k(k-1) W^{k-2}U^{s-k+1}\cB'\bar{\cB}'-\ri(2s-k)(2s-k+1) W^kU^{s-k-1}\cA'\bar{\cA}'\bigg] | H_{(2s-2k)}\rangle \non\\
    &&+\frac{s+1}{s} W^{k-1}U^{s-k}| W_{(2s-2k)}\rangle+ W^{k-1}U^{s-k}| V_{(2s-2k)}\rangle-\frac{1}{s} W^{k-1}U^{s-k}|J_{(2s-2k)}\rangle\bigg\}~,\non\\
    |\mathfrak{E}^{(4)}_{(s-1,s-1)}\rangle&=& \re^{\ri y m}\sum_{k=0}^s(-\ri)^{k-1}\non\\
    &&\times\bigg\{\frac{1}{4s^2(2s-2k+1)}\bigg[-\ri k(2s-2k+1)(2s-k+1) WU^{s-k}\bar{\cK}\non\\
    &&-\ri k(k-1) W^{k-2}U^{s-k+1}\bar{\cB}'\cB'+\ri(2s-k)(2s-k+1) W^kU^{s-k-1}\bar{\cA}'\cA'\bigg] | H_{(2s-2k)}\rangle \non\\
    &&+\frac{s+1}{s} W^{k-1}U^{s-k}| V_{(2s-2k)}\rangle+ W^{k-1}U^{s-k}| W_{(2s-2k)}\rangle-\frac{1}{s} W^{k-1}U^{s-k}|G_{(2s-2k)}\rangle\bigg\}~,\non\\
\eea
where we note $| V_{(2s)}\rangle=0$, $| W_{(2s)}\rangle=0$, $| G_{(2s)}\rangle=0$ and $| J_{(2s)}\rangle=0$. Inserting these into \eqref{EoM Higher Spin} leaves us with
\bsubeq \label{EoM Dim Reduced HSUSY Recap}
\bea
    | E^{(0)}_{(2s-2k)}\rangle &=& \frac{k!(2s-k+1)!}{(2s-2k+1)}\bigg\{\frac{1}{8}\cC| H_{(2s-2k)}\rangle-\frac{\ri}{4}(\cK| V_{(2s-2k)}\rangle-\bar{\cK}| W_{(2s-2k)}\rangle)\non\\
    &&+\frac{1}{4(2s-2k-1)}(\cB'\bar{\cB}'| V_{(2s-2k-2)}\rangle-\bar{\cB}'\cB'| W_{(2s-2k-2)}\rangle)\non\\
    &&+\frac{1}{4(2s-2k+3)}(\cA'\bar{\cA}'| V_{(2s-2k+2)}\rangle-\bar{\cA}'\cA'| W_{(2s-2k+2)}\rangle)\bigg\}+\cO(W)~,~~~~~~\\
    | E^{(1)}_{(2s-2k)}\rangle &=& -\frac{1}{s}\frac{(k-1)!(2s-k)!}{(2s-2k+1)}| W_{(2s-2k)}\rangle+\cO(W)~,\\
    | E^{(2)}_{(2s-2k)}\rangle &=& -\frac{1}{s}\frac{(k-1)!(2s-k)!}{(2s-2k+1)}| V_{(2s-2k)}\rangle+\cO(W)~,\\
    | E^{(3)}_{(2s-2k)}\rangle &=& \frac{(k-1)!(2s-k)!}{(2s-2k+1)}\non\\
    &&\times \bigg\{\frac{k(2s-k+1)}{4s^2}\ri \cK| H_{(2s-2k)}\rangle+\frac{k(k+1)}{4s^2(2s-2k-1)}\cB'\bar{\cB}'| H_{(2s-2k-2)}\rangle\non\\
    &&+\frac{(2s-k+1)(2s-k+2)}{4s^2(2s-2k+3)}\cA'\bar{\cA}'| H_{(2s-2k+2)}\rangle \non\\
    &&+\frac{s+1}{s}| W_{(2s-2k)}\rangle+| V_{(2s-2k)}\rangle-\frac{1}{s}| J_{(2s-2k)}\rangle\bigg\}+\cO(W)~, \\
    | E^{(4)}_{(2s-2k)}\rangle&=& \frac{(k-1)!(2s-k)!}{(2s-2k+1)}\non\\
    &&\times \bigg\{-\frac{k(2s-k+1)}{4s^2}\ri \bar{\cK}| H_{(2s-2k)}\rangle-\frac{k(k+1)}{4s^2(2s-2k-1)}\bar{\cB}'\cB'| H_{(2s-2k-2)}\rangle\non\\
    &&-\frac{(2s-k+1)(2s-k+2)}{4s^2(2s-2k+3)}\bar{\cA}'\cA'| H_{(2s-2k+2)}\rangle \non\\
    &&+\frac{s+1}{s}| V_{(2s-2k)}\rangle+| W_{(2s-2k)}\rangle-\frac{1}{s}| G_{(2s-2k)}\rangle\bigg\}+\cO(W)~,
\eea
\esubeq
where we note that kets $|\F_{(0,k)}\rangle$ with $k\leq0$ vanish. We see that the $\cO( W)$ terms in \eqref{EoM Dim Reduced HSUSY Recap} are annihilated in the action and can therefore be ignored. We now move to finding gauge invariance. We have the reduced form of \eqref{Complex Higher Superspin Gauge Transformations Long Linear}
\bea
    \d|\mathfrak{H}_{(s,s)}\rangle &=& |\mathfrak{l}_{(s,s)}\rangle+|\mathfrak{m}_{(s,s)}\rangle\non\\
    &=&\sum_{k=0}^s(-\ri W)^{k}U^{s-k}\bigg[| l_{(2s-2k)}\rangle +| m_{(2s-2k)}\rangle\bigg]~,\non\\
    \d|\mathfrak{V}_{(s-1,s-1)}\rangle &=&\frac{1}{2s(s+1)}\bar{\cA}'\cA|\mathfrak{m}_{(s,s)}\rangle\non\\
    &=&\sum_{k=0}^s\frac{1}{2s(s+1)}\bar{\cA}'A(-\ri W)^{k}U^{s-k}| m_{(2s-2k)}\rangle\non\\
    &=&\sum_{k=0}^s\frac{1}{4s(s+1)}\bigg[\frac{(2s-k)(2s-k+1)}{2s-2k+1}(-\ri W)^{k}U^{s-k-1}\bar{\cA}'\cA'\non\\
    &&+\frac{k(k-1)}{(2s-2k+1)}(-\ri W)^{k-2}U^{s-k+1}\bar{\cB}'\cB'+\ri k(2s-k+1)(-\ri W)^{k-1}U^{s-k}\bar{\cK}\bigg]| m_{(2s-2k)}\rangle~,\non\\
    \d|\mathfrak{W}_{(s-1,s-1)}\rangle &=&-\frac{1}{2s(s+1)}\cA\bar{\cA}'|\mathfrak{l}_{(s,s)}\rangle\non\\
    &=&-\sum_{k=0}^s\frac{1}{2s(s+1)}\cA\bar{\cA}'(-\ri W)^{k}U^{s-k}| l_{(2s-2k)}\rangle\non\\
    &=&-\sum_{k=0}^s\frac{1}{4s(s+1)}\bigg[\frac{(2s-k)(2s-k+1)}{2s-2k+1}(-\ri W)^{k}U^{s-k-1}\cA'\bar{\cA}'\non\\
    &&+\frac{k(k-1)}{(2s-2k+1)}(-\ri W)^{k-2}U^{s-k+1}\cB'\bar{\cB}'+\ri k(2s-k+1)(-\ri W)^{k-1}U^{s-k}\cK\bigg]| l_{(2s-2k)}\rangle~,\non\\
    \d|\mathfrak{G}_{(s-1,s-1)}\rangle &=&\frac{1}{2s(s+1)}\cA\bar{\cA}'|\mathfrak{l}_{(s,s)}\rangle-\frac{1}{2s}\{\cA,\bar{\cA}'\}|\mathfrak{l}_{(s,s)}\rangle\non\\
    &=&\sum_{k=0}^s\frac{1}{2s}\bigg(\frac{1}{(s+1)}\cA\bar{\cA}'-\{\cA,\bar{\cA}'\}\bigg)(-\ri W)^{k}U^{s-k}| l_{(2s-2k)}\rangle\non\\
    &=&\sum_{k=1}^s\frac{1}{2s}\bigg[\frac{(2s-k)(2s-k+1)}{2(2s-2k+1)(s+1)}(-\ri W)^{k}U^{s-k-1}\cA'\bar{\cA}'\non\\
    &&+\frac{k(k-1)}{2(s+1)(2s-2k+1)}(-\ri W)^{k-2}U^{s-k+1}\cB'\bar{\cB}'+\ri \frac{k(2s-k+1)}{2(s+1)}(-\ri W)^{k-1}U^{s-k}\cK\non\\
    &&+\frac{(2s-k)(2s-k+1)}{s(2s-2k+1)}(-\ri W)^{k}U^{s-k-1}P_5+\frac{k(k-1)}{s(2s-2k+1)}(-\ri W)^{k-2}U^{s-k+1}P_6\non\\
    &&+ m\frac{k(2s-k+1)}{s}(-\ri W)^{k-1}U^{s-k}\bigg]| l_{(2s-2k)}\rangle~,\non\\
    \d|\mathfrak{J}_{(s-1,s-1)}\rangle &=&-\frac{1}{2s(s+1)}\bar{\cA}'\cA |\mathfrak{m}_{(s,s)}\rangle+\frac{1}{2s}\{\cA,\bar{\cA}'\}|\mathfrak{m}_{(s,s)}\rangle\non\\
    &=&\sum_{k=0}^s\frac{1}{2s}\bigg(\{\cA,\bar{\cA}'\}-\frac{1}{(s+1)}\bar{\cA}'\cA\bigg)(-\ri W)^{k-1}U^{s-k}| m_{(2s-2k)}\rangle\non\\
    &=&\sum_{k=0}^s\frac{1}{2s}\bigg[-\frac{(2s-k)(2s-k+1)}{2(2s-2k+1)(s+1)}(-\ri W)^{k}U^{s-k-1}\bar{\cA}'\cA'\non\\
    &&-\frac{k(k-1)}{2(s+1)(2s-2k+1)}(-\ri W)^{k-2}U^{s-k+1}\bar{\cB}'\cB'-\ri \frac{k(2s-k+1)}{2(s+1)}(-\ri W)^{k-1}U^{s-k}\bar{\cK}\non\\
    &&-\frac{(2s-k)(2s-k+1)}{s(2s-2k+1)}(-\ri W)^{k}U^{s-k-1}P_5-\frac{k(k-1)}{s(2s-2k+1)}(-\ri W)^{k-2}U^{s-k+1}P_6\non\\
    &&- m\frac{2k(2s-k+1)}{s}(-\ri W)^{k-1}U^{s-k}\bigg]| m_{(2s-2k)}\rangle~.
\eea
Making use of the decompositions \eqref{Fields Decomposed Fully No Coeff}, we derive the gauge transformations:
\bsubeq \label{Gauge Variation Recap}
\bea
    \d| H_{(2s-2k)}\rangle &=& (-1)^k\frac{2s-2k+1}{k!(2s-k+1)!}\oint \rd y~\re^{-\ri y m}(\ri V)^kT^{s-k}\d|H_{(s,s)}\rangle-\cO( W)\non\\
    &=& (-1)^k\frac{(2s-2k+1)}{k!(2s-k+1)!}(\ri V)^kT^{s-k}\sum_{k'=0}^s(-\ri W)^{k'}U^{s-k'}\non\\
    &&\times \bigg[| l_{(2s-2k)}\rangle +| m_{(2s-2k)}\rangle\bigg]-\cO( W)\non\\
    &=& | l_{(2s-2k)}\rangle +| m_{(2s-2k)}\rangle~,\\
    \d| V_{(2s-2k)}\rangle &=& (-1)^{k-1}\frac{(2s-2k+1)}{(k-1)!(2s-k)!}\oint \rd y~\re^{-\ri y m}(\ri V)^{k-1}T^{s-k}\d|V_{(s-1,s-1)}\rangle-\cO( W)\non\\
    &=& \frac{(2s-k+1)(2s-k+2)}{4s(s+1)(2s-2k+3)}\bar{\cA}'\cA'| m_{(2s-2k+2)}\rangle\non\\
    &&+\frac{k(k+1)}{4s(s+1)(2s-2k-1)}\bar{\cB}'\cB'| m_{(2s-2k-2)}\rangle\non\\
    &&+\ri \frac{k(2s-k+1)}{4s(s+1)}\bar{\cK}| m_{(2s-2k)}\rangle~,\\
    \d| W_{(2s-2k)}\rangle &=& (-1)^{k-1}\frac{(2s-2k+1)}{(k-1)!(2s-k)!}\oint \rd y~\re^{-\ri y m}(\ri V)^{k-1}T^{s-k}\d|W_{(s-1,s-1)}\rangle-\cO( W)\non\\
    &=& -\frac{(2s-k+1)(2s-k+2)}{4s(s+1)(2s-2k+3)}\cA'\bar{\cA}'| l_{(2s-2k+2)}\rangle\non\\
    &&-\frac{k(k+1)}{4s(s+1)(2s-2k-1)}\cB'\bar{\cB}'| l_{(2s-2k-2)}\rangle\non\\
    &&-\ri \frac{k(2s-k+1)}{4s(s+1)}\cK| l_{(2s-2k)}\rangle~,\\
    \d| G_{(2s-2k)}\rangle &=& (-1)^{k-1}\frac{(2s-2k+1)}{(k-1)!(2s-k)!}\oint \rd y~\re^{-\ri y m}(\ri V)^{k-1}T^{s-k}\d|G_{(s-1,s-1)}\rangle-\cO( W)\non\\
    &=& \frac{(2s-k+1)(2s-k+2)}{4s(s+1)(2s-2k+3)}\cA'\bar{\cA}'| l_{(2s-2k+2)}\rangle\non\\
    &&+\frac{k(k+1)}{4s(s+1)(2s-2k-1)}\cB'\bar{\cB}'| l_{(2s-2k-2)}\rangle\non\\
    &&+\ri \frac{k(2s-k+1)}{4s(s+1)}\cK| l_{(2s-2k)}\rangle+\frac{(2s-k+1)(2s-k+2)}{2s(2s-2k+3)}P_5| l_{(2s-2k+2)}\rangle\non\\
    &&+\frac{k(k+1)}{2s(2s-2k-1)}P_6| l_{(2s-2k-2)}\rangle+m \frac{k(2s-k+1)}{s}| l_{(2s-2k)}\rangle~,\\
    \d| J_{(2s-2k)}\rangle &=& (-1)^{k-1}\frac{(2s-2k+1)}{(k-1)!(2s-k)!}\oint \rd y~\re^{-\ri y m}(\ri V)^{k-1}T^{s-k}\d|J_{(s-1,s-1)}\rangle-\cO( W)\non\\
    &=& -\frac{(2s-k+1)(2s-k+2)}{4s(s+1)(2s-2k+3)}\bar{\cA}'\cA'| m_{(2s-2k+2)}\rangle\non\\
    &&-\frac{k(k+1)}{4s(s+1)(2s-2k-1)}\bar{\cB}'\cB'| m_{(2s-2k-2)}\rangle\non\\
    &&-\ri \frac{k(2s-k+1)}{4s(s+1)}\bar{\cK}| m_{(2s-2k)}\rangle-\frac{(2s-k+1)(2s-k+2)}{2s(2s-2k+3)}P_5| m_{(2s-2k+2)}\rangle\non\\
    &&-\frac{k(k+1)}{2s(2s-2k-1)}P_6| m_{(2s-2k-2)}\rangle-m \frac{k(2s-k+1)}{s}| m_{(2s-2k)}\rangle~.
\eea
\esubeq
Finally, we note that the dimensional reduction of the constraints \eqref{Constraints 4D Oscillator} leads to
\bea
    \bar{\cB}'|\mathfrak{G}_{(s-1,s-1)}\rangle=0~,\qquad \cB|\mathfrak{J}_{(s-1,s-1)}\rangle=0~.
\eea
Inserting \eqref{Fields Decomposed Fully No Coeff} into these constraints imply that
\bsubeq \label{Constraints I want Recap}
\bea\label{Constraint I want 1 Recap}
    \bar{\cB}'| G_{(2s-2)}\rangle =0~,\non\\
    \bar{\cA}'| G_{(2s-2k)}\rangle=\ri~\frac{2s-2k+1}{2s-2k-1}\bar{\cB}'| G_{(2s-2k-2)}\rangle~,\qquad 1\leq k \leq s-1~,
\eea
and
\bea\label{Constraint I want 2 Recap}
    \cB'| J_{(2s-2)}\rangle =0~,\non\\
    \cA'| J_{(2s-2k)}\rangle=-\ri~\frac{2s-2k+1}{2s-2k-1}\cB'| J_{(2s-2k-2)}\rangle~,\qquad 1\leq k \leq s-1~.
\eea
\esubeq
If we hit \eqref{Constraint I want 1 Recap} with $\bar{\cB}'$ and \eqref{Constraint I want 2 Recap} with $\cB'$, we simply get
\bsubeq \label{Linear Constraints}
\bea
    \cDB^2| G_{(2s-2k)}\rangle=0~,\qquad \cD^2| J_{(2s-2k)}\rangle=0~,\qquad 1\leq k \leq s~.
\eea
\esubeq
The constraints allow us to write $| G_{(2s-2k)}\rangle$ and $| J_{(2s-2k)}\rangle$ in terms of prepotentials:
\bsubeq\label{Field Prepotentials}
\bea
    | G_{(2s-2)}\rangle&=&\bar{\cB}'| g_{(2s-3)}\rangle~,\qquad | J_{(2s-2)}\rangle=\cB'| j_{(2s-3)}\rangle~,\non\\
    | G_{(0)}\rangle&=&\ri\bar{\cA}'| g_{(1)}\rangle~,\qquad | J_{(0)}\rangle=-\ri\cA'| j_{(1)}\rangle~,\non\\
    | G_{(2s-2k)}\rangle&=&\bar{\cB}'| g_{(2s-2k-1)}\rangle+\ri\bar{\cA}'| g_{(2s-2k+1)}\rangle~,\qquad 2\leq k\leq s-1~,\non\\
    | J_{(2s-2k)}\rangle&=&\cB'| j_{(2s-2k-1)}\rangle-\ri\cA'| j_{(2s-2k+1)}\rangle~,\qquad 2\leq k\leq s-1~.
\eea
\esubeq
We also have for the gauge parameters:
\bea
    \bar{\cB}'|\mathfrak{l}_{(s,s)}\rangle=0~,\qquad \cB|\mathfrak{m}_{(s,s)}\rangle=0~,
\eea
hence,
\bsubeq 
\bea
    \bar{\cB}'| l_{(2s)}\rangle =0~,\non\\
    \bar{\cA}'| l_{(2s-2k)}\rangle=\ri~\frac{2s-2k+1}{2s-2k-1}\bar{\cB}'| l_{(2s-2k-2)}\rangle~,\qquad 0\leq k \leq s-1~,
\eea
and
\bea
    \cB'| m_{(2s)}\rangle =0~,\non\\
    \cA'| m_{(2s-2k)}\rangle=-\ri~\frac{2s-2k+1}{2s-2k-1}\cB'| m_{(2s-2k-2)}\rangle~,\qquad 0\leq k \leq s-1~.
\eea
\esubeq

It is shown in Appendix \ref{Gauge Invariance Section} that the action \eqref{Action Recap} is indeed invariant under the transformations \eqref{Gauge Variation Recap}.
\subsection{Transverse formulation}
For the 4D massless action \eqref{4D massless action oscillator}, it is possible to integrate out the fields $|\mathfrak{G}_{(s-1,s-1)}\rangle$ and $|\mathfrak{J}_{(s-1,s-1)}\rangle$ to obtain a model containing only $|\mathfrak{H}_{(s,s)}\rangle$ and transverse fields $|\mathfrak{V}_{(s-1,s-1)}\rangle$ and $|\mathfrak{W}_{(s-1,s-1)}\rangle$. We shall make use of a similar process to construct a `transverse' model for \eqref{Action Recap}. To do so, we must first decompose $\langle  G_{(2s-2k)}|$ and $\langle J_{(2s-2k)}|$. The action becomes
\bea
    \bm{S}&=&\sum_{k=0}^s\bigg(-\frac{1}{2}\bigg)^s\bigg(\frac{1}{s!}\bigg)^2\int\rd^{3|4}z\bigg\{\langle  H_{(2s-2k)}|  E^{(0)}_{(2s-2k)}\rangle\non\\
    &&~~~~~~~~~~~~~~~~~~~-s^2\langle  g_{(2s-2k-1)}| (\cA'|  E^{(1)}_{(2s-2k)}\rangle-\ri\cB'|  E^{(1)}_{(2s-2k-2)}\rangle)\non\\
    &&~~~~~~~~~~~~~~~~~~~-s^2\langle  j_{(2s-2k-1)}| (\bar{\cA}'|  E^{(2)}_{(2s-2k)}\rangle+\ri\bar{\cB}'|  E^{(2)}_{(2s-2k-2)}\rangle)\non\\
    &&~~~~~~~~~~~~~~~~~~~+s^2\langle  V_{(2s-2k)}|  E^{(3)}_{(2s-2k)}\rangle+s^2\langle  W_{(2s-2k)}|  E^{(4)}_{(2s-2k)}\rangle\bigg\}~.
\eea
Integrating out the prepotentials $\langle g_{(2s-2k-1)}|$ and $\langle j_{(2s-2k-1)}|$ gives the constraints
\bsubeq
\bea
    \frac{2s-k}{2s-2k+1}\bar{\cA}'| V_{(2s-2k)}\rangle=-\ri\frac{k}{2s-2k-1}\bar{\cB}'| V_{(2s-2k-2)}\rangle\label{Constraints 1}~,\\
    \frac{2s-k}{2s-2k+1}\cA'| W_{(2s-2k)}\rangle=\ri\frac{k}{2s-2k-1}\cB'| W_{(2s-2k-2)}\rangle~, \label{Constraints 2}
\eea
which imply
\bea\label{Linear Constraint Transverse}
    \cDB^2| V_{(2s-2k)}\rangle=0~,\qquad\cD^2| W_{(2s-2k)}\rangle=0~.
\eea
\esubeq
We see then that the transverse action is
\bea
    \bm{S}^\perp&=&\sum_{k=0}^s\bigg(-\frac{1}{2}\bigg)^s\bigg(\frac{1}{s!}\bigg)^2\int\rd^{3|4}z\bigg\{\langle  H_{(2s-2k)}|  E^{(0)}_{(2s-2k)}\rangle+s^2\langle  V_{(2s-2k)}|  E^{(1)}_{(2s-2k)}\rangle\non\\
    &&~~~~~~~~~~~~~~~~~~~+s^2\langle  W_{(2s-2k)}|  E^{(2)}_{(2s-2k)}\rangle\bigg\}~,
\eea
where
\bsubeq
\bea
    | E^{(0)}_{(2s-2k)}\rangle &=& \frac{k!(2s-k+1)!}{(2s-2k+1)}\bigg\{\frac{1}{8}\cC| H_{(2s-2k)}\rangle-\frac{\ri}{4}(\cK| V_{(2s-2k)}\rangle-\bar{\cK}| W_{(2s-2k)}\rangle)\non\\
    &&+\frac{1}{4(2s-2k-1)}(\cB'\bar{\cB}'| V_{(2s-2k-2)}\rangle-\bar{\cB}'\cB'| W_{(2s-2k-2)}\rangle)\non\\
    &&+\frac{1}{4(2s-2k+3)}(\cA'\bar{\cA}'| V_{(2s-2k+2)}\rangle-\bar{\cA}'\cA'| W_{(2s-2k+2)}\rangle)\bigg\}~,\\
    | E^{(1)}_{(2s-2k)}\rangle &=& \frac{(k-1)!(2s-k)!}{(2s-2k+1)}\non\\
    &&\times \bigg\{\frac{k(2s-k+1)}{4s^2}\ri \cK| H_{(2s-2k)}\rangle+\frac{k(k+1)}{4s^2(2s-2k-1)}\cB'\bar{\cB}'| H_{(2s-2k-2)}\rangle\non\\
    &&+\frac{(2s-k+1)(2s-k+2)}{4s^2(2s-2k+3)}\cA'\bar{\cA}'| H_{(2s-2k+2)}\rangle \non\\
    &&+\frac{s+1}{s}| W_{(2s-2k)}\rangle+| V_{(2s-2k)}\rangle\bigg\}~, \\
    | E^{(2)}_{(2s-2k)}\rangle&=& \frac{(k-1)!(2s-k)!}{(2s-2k+1)}\non\\
    &&\times \bigg\{-\frac{k(2s-k+1)}{4s^2}\ri \bar{\cK}| H_{(2s-2k)}\rangle-\frac{k(k+1)}{4s^2(2s-2k-1)}\bar{\cB}'\cB'| H_{(2s-2k-2)}\rangle\non\\
    &&-\frac{(2s-k+1)(2s-k+2)}{4s^2(2s-2k+3)}\bar{\cA}'\cA'| H_{(2s-2k+2)}\rangle \non\\
    &&+\frac{s+1}{s}| V_{(2s-2k)}\rangle+| W_{(2s-2k)}\rangle\bigg\}~. 
\eea
\esubeq
By constructing general solutions to \eqref{Linear Constraint Transverse}, we can decompose $| V_{(2s-2k)}\rangle$ and $| W_{(2s-2k)}\rangle$:
\bsubeq
\bea
    | V_{(2s-2k)}\rangle=a_k\bar{\cA}'| v_{(2s-2k+1)}\rangle+b_k\bar{\cB}'| v_{(2s-2k-1)}\rangle~,\\
    | W_{(2s-2k)}\rangle=c_k\cA'| w_{(2s-2k+1)}\rangle+d_k\cB'| w_{(2s-2k-1)}\rangle~,
\eea
\esubeq
where $a_k,$ $b_k$, $c_k$ and $d_k$ are some constants while $| v_{(2s-2k+1)}\rangle$ and $| w_{(2s-2k+1)}\rangle$ are unconstrained prepotentials. Inserting these into \eqref{Constraints 1} and \eqref{Constraints 2} gives
\bea
    \frac{b_k}{a_{k+1}}=\frac{k}{2s-k}\ri~,\qquad \frac{d_k}{c_{k+1}}=-\frac{k}{2s-k}\ri~,
\eea
hence, we can take
\bsubeq
\bea
    | V_{(2s-2k)}\rangle=(2s-k+1)\bar{\cA}'| v_{(2s-2k+1)}\rangle+\ri k\bar{\cB}'| v_{(2s-2k-1)}\rangle~,\\
    | W_{(2s-2k)}\rangle=(2s-k+1)\cA'| w_{(2s-2k+1)}\rangle-\ri k\cB'| w_{(2s-2k-1)}\rangle~.
\eea
\esubeq

We see above how a transverse model can be derived from our massive parent model. If one instead integrated out the $|V_{(2s-2k)}\rangle$ and $|W_{(2s-2k)}\rangle$, they would find a massive analogue to the massless longitudinal action. The resulting action will be more complicated than the above action. Moreover, it may be possible to redefine fields and instead integrate out a different set of fields to construct simpler transverse and longitudinal actions which have forms more similar to their massless counterparts. Such an analysis is outside the scope of this paper and will appear in a future work.


\subsection{Equations of motion}
To see how the produced action will describe an irreducible superspin $s+\frac{1}{2}$ superfield, we return to the 4D case and consider how the dimensional reduction occurs on-shell. For our purposes, it will actually be easier to return to the field formalism. We note that the theory \eqref{4D Action Complex Field} exhibits two gauge invariant superfield strengths \cite{KPS}
which are defined by
\bsubeq
\bea
    \mathfrak{W}_{\a(2s+1)}:=\mathfrak{\DB}^2\pa_{(\a_1}^{\bd_1}\dots\pa_{\a_s}^{\bd_s}\mathfrak{D}_{\a_{s+1}}\mathfrak{H}_{\a_{s+2}\dots\a_{2s+1})\bd(s)}~,\\
    \bar{\mathfrak{W}}_{\ad(2s+1)}:=\mathfrak{D}^2\pa_{(\ad_1}^{\b_1}\dots\pa_{\ad_s}^{\b_s}\mathfrak{\DB}_{\ad_{s+1}}\bar{\mathfrak{H}}_{\b(s)\ad_{s+2}\dots\ad_{2s+1})}~.
\eea
\esubeq
These superfield strengths are chiral and antichiral respectively,
\bea
    \mathfrak{\DB}_\bd \mathfrak{W}_{\a(2s+1)}=0~,\qquad \mathfrak{D}_\b \bar{\mathfrak{W}}_{\ad(2s+1)}=0~.
\eea
On the mass shell, they satisfy the following equations
\bea
    \mathfrak{D}^\b \mathfrak{W}_{\b\a(2s)}=0~,\qquad \mathfrak{\DB}^\bd \bar{\mathfrak{W}}_{\bd\ad(2s)}=0~.
\eea
We can then combine these with the (anti)chirality conditions to get
\bea
    \mathfrak{\DB}_{\bd}\mathfrak{D}^\b \mathfrak{W}_{\b\a(2s)}=0~,\qquad \mathfrak{D}_{\b}\mathfrak{\DB}^\bd \bar{\mathfrak{W}}_{\bd\ad(2s)}=0~,
\eea
which then lead to
\bea\label{EoM Irred}
    \ri\pa_{\bd}^\b \mathfrak{W}_{\b\a(2s)}=0~,\qquad \ri\pa_{\b}^\bd \bar{\mathfrak{W}}_{\bd\ad(2s)}=0~,
\eea
as the on-shell equations. We now consider what happens if we dimensionally reduce these superfield strengths. Since they are already symmetric, their decomposition is simply
\bsubeq
\bea
    \mathfrak{W}_{\a(2s+1)}=\re^{\ri ym} \cW_{\a(2s+1)}~,\\
    \bar{\mathfrak{W}}_{\ad(2s+1)}=\re^{-\ri ym}\bar{\cW}_{\a(2s+1)}~.
\eea
\esubeq
With these, \eqref{EoM Irred} becomes
\bsubeq
\bea
    (\ri\pa_{\g}^\b-\pa_{y}\d_\g^\b) \re^{\ri ym} \cW_{\b\a(2s)}=0~,\\
    (\ri\pa_{\g}^\b+\pa_{y}\d_\g^\b) \re^{-\ri ym}\bar{\cW}_{\b\a(2s)}=0~,
\eea
\esubeq
which then leads to
\bsubeq
\bea
    (\pa_{\g}^\b-m\d_\g^\b)  \cW_{\b\a(2s)}=0~,\\
    (\pa_{\g}^\b- m\d_\g^\b) \bar{\cW}_{\b\a(2s)}=0~.
\eea
\esubeq
Here the equation on $\cW_{a(2s+1)}$ is equivalent to 
\bsubeq
\bea
\pa^{\b\g} \cW_{\b\g\a_1 \cdots \a_{2s-1}} &=&0~,  \\
\pa^\b{}_{(\a_1} \cW_{\a_2 \dots \a_{2s+1})\b} &=& m  \cW_{\a_1 \dots \a{2s+1}}~.
\eea 
\end{subequations}
Hence the theory exhibits the correct on-shell equations for a massive theory of half-integer superspin.


\section{Massless higher half-integer superspin in 3D} \label{Massless Limit}
The off-shell formulation for massless higher half-integer spin supermultiplets in 3D $\cN=2$ superspace was found several years ago in \cite{KO}. To ensure the theory we have derived is consistent with this model, we dedicate this section to analysing the massless limit of the dimensionally reduced action \eqref{Action Recap}. Taking the massless limit of \eqref{Action Recap} we obtain
\bea\label{Action Massless}
    \bm{S}&=&\sum_{k=0}^s\bigg(-\frac{1}{2}\bigg)^s\bigg(\frac{1}{s!}\bigg)^2\int\rd^{3|4}z\bigg\{\langle  H_{(2s-2k)}|  E^{(0)}_{(2s-2k)}\rangle+s^2\langle  G_{(2s-2k)}|  E^{(1)}_{(2s-2k)}\rangle\non\\
    &&~~~~~~~~~~~~~~~~~~~+s^2\langle  J_{(2s-2k)}|  E^{(2)}_{(2s-2k)}\rangle+s^2\langle  V_{(2s-2k)}|  E^{(3)}_{(2s-2k)}\rangle\non\\
    &&~~~~~~~~~~~~~~~~~~~+s^2\langle  W_{(2s-2k)}|  E^{(4)}_{(2s-2k)}\rangle\bigg\}~,
\eea
where
\bsubeq \label{EoM Dim Reduced Massless}
\bea
    | E^{(0)}_{(2s-2k)}\rangle &=& \frac{k!(2s-k+1)!}{(2s-2k+1)}\bigg\{\frac{1}{8}C| H_{(2s-2k)}\rangle-\frac{\ri}{4}(K| V_{(2s-2k)}\rangle-\bar{K}| W_{(2s-2k)}\rangle)\non\\
    &&+\frac{1}{4(2s-2k-1)}( B' \bar{B}'| V_{(2s-2k-2)}\rangle- \bar{B}' B'| W_{(2s-2k-2)}\rangle)\non\\
    &&+\frac{1}{4(2s-2k+3)}( A' \bar{A}'| V_{(2s-2k+2)}\rangle- \bar{A}' A'| W_{(2s-2k+2)}\rangle)\bigg\}~,\\
    | E^{(1)}_{(2s-2k)}\rangle &=& -\frac{1}{s}\frac{(k-1)!(2s-k)!}{(2s-2k+1)}| V_{(2s-2k)}\rangle~,\\
    | E^{(2)}_{(2s-2k)}\rangle &=& -\frac{1}{s}\frac{(k-1)!(2s-k)!}{(2s-2k+1)}| W_{(2s-2k)}\rangle~,\\
    | E^{(3)}_{(2s-2k)}\rangle &=& \frac{(k-1)!(2s-k)!}{(2s-2k+1)}\non\\
    &&\times \bigg\{\frac{k(2s-k+1)}{4s^2}\ri K| H_{(2s-2k)}\rangle+\frac{k(k+1)}{4s^2(2s-2k-1)} B' \bar{B}'| H_{(2s-2k-2)}\rangle\non\\
    &&+\frac{(2s-k+1)(2s-k+2)}{4s^2(2s-2k+3)} A' \bar{A}'| H_{(2s-2k+2)}\rangle \non\\
    &&+\frac{s+1}{s}| W_{(2s-2k)}\rangle+| V_{(2s-2k)}\rangle-\frac{1}{s}| J_{(2s-2k)}\rangle\bigg\}~, \\
    | E^{(4)}_{(2s-2k)}\rangle&=& \frac{(k-1)!(2s-k)!}{(2s-2k+1)}\non\\
    &&\times \bigg\{-\frac{k(2s-k+1)}{4s^2}\ri \bar{K}| H_{(2s-2k)}\rangle-\frac{k(k+1)}{4s^2(2s-2k-1)} \bar{B}' B'| H_{(2s-2k-2)}\rangle\non\\
    &&-\frac{(2s-k+1)(2s-k+2)}{4s^2(2s-2k+3)} \bar{A}' A'| H_{(2s-2k+2)}\rangle \non\\
    &&+\frac{s+1}{s}| V_{(2s-2k)}\rangle+| W_{(2s-2k)}\rangle-\frac{1}{s}| G_{(2s-2k)}\rangle\bigg\}~.
\eea
\esubeq
The massless gauge transformations are
\bsubeq\label{Gauge Variation Massless}
\bea
    \d| H_{(2s-2k)}\rangle &=& | l_{(2s-2k)}\rangle +| m_{(2s-2k)}\rangle~,\\
    \d| V_{(2s-2k)}\rangle &=& \frac{(2s-k+1)(2s-k+2)}{4s(s+1)(2s-2k+3)} \bar{A}' A'| m_{(2s-2k+2)}\rangle\non\\
    &&+\frac{k(k+1)}{4s(s+1)(2s-2k-1)} \bar{B}' B'| m_{(2s-2k-2)}\rangle\non\\
    &&+\ri \frac{k(2s-k+1)}{4s(s+1)}\bar{K}| m_{(2s-2k)}\rangle~,\\
    \d| W_{(2s-2k)}\rangle &=& -\frac{(2s-k+1)(2s-k+2)}{4s(s+1)(2s-2k+3)} A' \bar{A}'| l_{(2s-2k+2)}\rangle\non\\
    &&-\frac{k(k+1)}{4s(s+1)(2s-2k-1)} B' \bar{B}'| l_{(2s-2k-2)}\rangle\non\\
    &&-\ri \frac{k(2s-k+1)}{4s(s+1)}K| l_{(2s-2k)}\rangle~,\\
    \d| G_{(2s-2k)}\rangle &=& \frac{(2s-k+1)(2s-k+2)}{4s(s+1)(2s-2k+3)} A' \bar{A}'| l_{(2s-2k+2)}\rangle\non\\
    &&+\frac{k(k+1)}{4s(s+1)(2s-2k-1)} B' \bar{B}'| l_{(2s-2k-2)}\rangle\non\\
    &&+\ri \frac{k(2s-k+1)}{4s(s+1)}K| l_{(2s-2k)}\rangle+\frac{(2s-k+1)(2s-k+2)}{2s(2s-2k+3)}P_5| l_{(2s-2k+2)}\rangle\non\\
    &&+\frac{k(k+1)}{2s(2s-2k-1)}P_6| l_{(2s-2k-2)}\rangle~,\\
    \d| J_{(2s-2k)}\rangle &=& -\frac{(2s-k+1)(2s-k+2)}{4s(s+1)(2s-2k+3)} \bar{A}' A'| m_{(2s-2k+2)}\rangle\non\\
    &&-\frac{k(k+1)}{4s(s+1)(2s-2k-1)} \bar{B}' B'| m_{(2s-2k-2)}\rangle\non\\
    &&-\ri \frac{k(2s-k+1)}{4s(s+1)}\bar{K}| m_{(2s-2k)}\rangle-\frac{(2s-k+1)(2s-k+2)}{2s(2s-2k+3)}P_5| m_{(2s-2k+2)}\rangle\non\\
    &&-\frac{k(k+1)}{2s(2s-2k-1)}P_6| m_{(2s-2k-2)}\rangle~,
\eea
\esubeq
with constraints:
\bsubeq 
\bea
     \bar{B}'| G_{(2s-2)}\rangle =0~,\non\\
     \bar{A}'| G_{(2s-2k)}\rangle=\ri~\frac{2s-2k+1}{2s-2k-1} \bar{B}'| G_{(2s-2k-2)}\rangle~,\qquad 1\leq k \leq s-1~,
\eea
and
\bea
     B'| J_{(2s-2)}\rangle =0~,\non\\
     A'| J_{(2s-2k)}\rangle=-\ri~\frac{2s-2k+1}{2s-2k-1} B'| J_{(2s-2k-2)}\rangle~,\qquad 1\leq k \leq s-1~,
\eea
\esubeq
and
\bsubeq \label{gauge constraints massless}
\bea
     \bar{B}'| l_{(2s)}\rangle =0~,\non\\
     \bar{A}'| l_{(2s-2k)}\rangle=\ri~\frac{2s-2k+1}{2s-2k-1} \bar{B}'| l_{(2s-2k-2)}\rangle~,\qquad 0\leq k \leq s-1~,
\eea
and
\bea
     B'| m_{(2s)}\rangle =0~,\non\\
     A'| m_{(2s-2k)}\rangle=-\ri~\frac{2s-2k+1}{2s-2k-1} B'| m_{(2s-2k-2)}\rangle~,\qquad 0\leq k \leq s-1~.
\eea
\esubeq
Since we now have no central charge present, we can take the following reality conditions:
\bea
    | H_{(2s-2k)}\rangle=|\bar{H}_{(2s-2k)}\rangle~,\qquad | m_{(2s-2k)}\rangle=|\bar{l}_{(2s-2k)}\rangle~,\non\\
    | W_{(2s-2k)}\rangle=|\bar{V}_{(2s-2k)}\rangle~,\qquad | J_{(2s-2k)}\rangle=|\bar{G}_{(2s-2k)}\rangle ~.
\eea
From here, the gauge transformations are
\bsubeq\label{Gauge Variation Massless Reality}
\bea
    \d| H_{(2s-2k)}\rangle &=& | l_{(2s-2k)}\rangle +|\bar{l}_{(2s-2k)}\rangle~,\\
    \d| V_{(2s-2k)}\rangle &=& \frac{(2s-k+1)(2s-k+2)}{4s(s+1)(2s-2k+3)} \bar{A}' A'|\bar{l}_{(2s-2k+2)}\rangle\non\\
    &&+\frac{k(k+1)}{4s(s+1)(2s-2k-1)} \bar{B}' B'|\bar{l}_{(2s-2k-2)}\rangle\non\\
    &&+\ri \frac{k(2s-k+1)}{4s(s+1)}\bar{K}|\bar{l}_{(2s-2k)}\rangle~,\\
    \d| G_{(2s-2k)}\rangle &=& \frac{(2s-k+1)(2s-k+2)}{4s(s+1)(2s-2k+3)} A' \bar{A}'| l_{(2s-2k+2)}\rangle\non\\
    &&+\frac{k(k+1)}{4s(s+1)(2s-2k-1)} B' \bar{B}'| l_{(2s-2k-2)}\rangle\non\\
    &&+\ri \frac{k(2s-k+1)}{4s(s+1)}K| l_{(2s-2k)}\rangle+\frac{(2s-k+1)(2s-k+2)}{2s(2s-2k+3)}P_5| l_{(2s-2k+2)}\rangle\non\\
    &&+\frac{k(k+1)}{2s(2s-2k-1)}P_6| l_{(2s-2k-2)}\rangle~.
\eea
\esubeq
We can rewrite this using \eqref{gauge constraints massless}
\bsubeq
\bea
    \d| V_{(2s-2k)}\rangle &=& \frac{2s-k+1}{2s(2s-2k+3)} \bar{A}' A'|\bar{l}_{(2s-2k+2)}\rangle\non\\
    &&+\frac{k}{2s(2s-2k-1)} \bar{B}' B'|\bar{l}_{(2s-2k-2)}\rangle~,\\
    \d| G_{(2s-2k)}\rangle &=& \frac{2s-k+1}{2s(2s-2k+3)} A' \bar{A}'| l_{(2s-2k+2)}\rangle+\frac{k}{2s(2s-2k-1)} B' \bar{B}'| l_{(2s-2k-2)}\rangle\non\\
    &&+\frac{(2s-k+1)(2s-k+2)}{2s(2s-2k+3)}P_5| l_{(2s-2k+2)}\rangle\non\\
    &&+\frac{k(k+1)}{2s(2s-2k-1)}P_6| l_{(2s-2k-2)}\rangle~.
\eea
\esubeq
We now introduce the following reparameterisations:
\bsubeq
\bea
    | X_{(2s-2)}\rangle&=&\ri K| H_{(2s-2)}\rangle+| V_{(2s-2)}\rangle+| \bar{V}_{(2s-2)}\rangle~,\\
    | Y_{(2s-2)}\rangle&=&| \bar{G}_{(2s-2)}\rangle-\frac{1}{2s(2s-3)} B' \bar{B}'| H_{(2s-4)}\rangle~,\\
    | \bar{Y}_{(2s-2)}\rangle&=&| G_{(2s-2)}\rangle+\frac{1}{2s(2s-3)} \bar{B}' B'| H_{(2s-4)}\rangle~,\\
    | Y_{(2s-4)}\rangle&=&| \bar{G}_{(2s-4)}\rangle-\frac{1}{2s} A' \bar{A}'| H_{(2s-2)}\rangle~,\\
    | \bar{Y}_{(2s-4)}\rangle&=&| G_{(2s-4)}\rangle+\frac{1}{2s} \bar{A}' A'| H_{(2s-2)}\rangle~,
\eea
\esubeq
so that the action becomes
\bea
    \bm{S}&=&\bigg(-\frac{1}{2}\bigg)^s\bigg(\frac{1}{s!}\bigg)^2\int\rd^{3|4}z\bigg\{-\frac{(2s)!}{2s-1}\langle X_{(2s-2)}| X_{(2s-2)}\rangle+\langle H_{(2s)}|  E^{(0)}_{(2s)}\rangle+s^2\langle  \bar{Y}_{(2s-2)}|  E^{(1)}_{(2s-2)}\rangle\non\\
    &&~~~~~~~~~~~~~~+s^2\langle  Y_{(2s-2)}|  E^{(2)}_{(2s-2)}\rangle+s^2\langle  V_{(2s-2)}|  E^{(3)}_{(2s-2)}\rangle+s^2\langle  \bar{V}_{(2s-2)}|  E^{(4)}_{(2s-2)}\rangle\bigg\}\non\\
    &&~~~~~~~~~~~~~~~+\bm{S}_{k\geq2},
\eea
where the $\bm{S}_{k\geq2}$ term contains only superfields of rank less than $2s-4$. We note that the field $|X_{(2s-2)}\rangle$ does not couple to any other field and so can easily be integrated out. The remaining action then is decoupled into the $\bm{S}_{k\geq2}$ sector and the top sector
\bea
    \bm{S}_1&=&\bigg(-\frac{1}{2}\bigg)^s\bigg(\frac{1}{s!}\bigg)^2\int\rd^{3|4}z\bigg\{\langle H_{(2s)}|  E^{(0)}_{(2s)}\rangle+s^2\langle  \bar{Y}_{(2s-2)}|  E^{(1)}_{(2s-2)}\rangle+s^2\langle  Y_{(2s-2)}|  E^{(2)}_{(2s-2)}\rangle\non\\
    &&~~~~~~~~~~~~~~+s^2\langle  V_{(2s-2)}|  E^{(3)}_{(2s-2)}\rangle+s^2\langle  \bar{V}_{(2s-2)}|  E^{(4)}_{(2s-2)}\rangle\bigg\}~,
\eea
where
\bsubeq 
\bea
    | E^{(0)}_{(2s-2)}\rangle &=& \frac{(2s)!}{(2s-1)}\bigg\{\frac{1}{4(2s-3)}( B' \bar{B}'| V_{(2s-4)}\rangle- \bar{B}' B'| \bar{V}_{(2s-4)}\rangle)\bigg\}~,\\
    | E^{(1)}_{(2s-2)}\rangle &=& -\frac{1}{s}(2s-2)!| V_{(2s-2)}\rangle~,\\
    | E^{(2)}_{(2s-2)}\rangle &=& -\frac{1}{s}(2s-2)!| W_{(2s-2)}\rangle~,\\
    | E^{(3)}_{(2s-2)}\rangle &=& (2s-2)!\bigg\{\frac{1}{2s} A' \bar{A}'| H_{(2s)}\rangle \non\\
    &&+\frac{2s+1}{2s}| \bar{V}_{(2s-2)}\rangle+\frac{2s-1}{2s}| V_{(2s-2)}\rangle-\frac{1}{s}| Y_{(2s-2)}\rangle\bigg\}~, \\
    | E^{(4)}_{(2s-2)}\rangle&=& (2s-2)!\bigg\{-\frac{1}{2s} \bar{A}' A'| H_{(2s)}\rangle \non\\
    &&+\frac{2s+1}{2s}| V_{(2s-2)}\rangle+\frac{2s-1}{2s}| \bar{V}_{(2s-2)}\rangle-\frac{1}{s}\bar{Y}_{(2s-2)}\rangle\bigg\}~. 
\eea
\esubeq
We can then see that by switching off $| l_{(2s-4)}\rangle$ in this sector, the top field transformations are
\bsubeq
\bea
    \d| H_{(2s)}\rangle &=& | l_{(2s)}\rangle +|\bar{l}_{(2s)}\rangle~,\\
    \d| V_{(2s-2)}\rangle &=& \frac{1}{2s+1} \bar{A}' A'|\bar{l}_{(2s)}\rangle~,\\
    \d| Y_{(2s-2)}\rangle &=& \frac{1}{2s+1} A' \bar{A}'| l_{(2s)}\rangle+P_5| l_{(2s)}\rangle~.
\eea
\esubeq
We see that this sector coincides (up to a scale factor) with the axillary model for gauge-invariant massless superfields of half-integer superspin in $\mathbb{M}^{3|4}$.


\section{Conclusion} \label{Conclusion}

In this paper we constructed, for the first time, 
the gauge-invariant off-shell formulations for massive 3D $\cN=2$ higher-spin supermultiplets with at most two derivatives at the component level. This was achieved by applying the Kaluza-Klein reduction to the off-shell models for 4D $\cN=1$ massless higher half-integer supermultiplets  \cite{KPS}. A similar analysis for the 4D $\cN=1$ massless integer superspin theories \cite{KS, Hutomo:2017nce} will be described elsewhere.

The specific feature of our 3D massive higher-spin supermultiplets is $\cN=2$ supersymmetry with a real central charge.
This is in contrast to the topologically massive 3D $\cN=2$ higher-spin supermultiplets constructed earlier in \cite{KO}.
Our massive $\cN=2$ higher-spin supermultiplets can be reduced to the 3D $\cN=1$ Minkowski superspace by integrating out two Grassmann variables,
and then imposing consistent reality conditions on the superfields.
 As a result, only two supercharges remain unbroken.    
The resulting models will differ from  
the topologically massive 3D $\cN=1$ higher-spin supermultiplets constructed ten years ago \cite{KT}.

To illustrate the above point, let us carry out a brief analysis of the superspace reduction applied to the models in sections \ref{Vector Supermultiplet} and \ref{New Minimal Reduction}. To carry out a SUSY reduction on our models, we must first recast the theory in terms of real Grassmann coordinates by taking
\bsubeq
\bea
    \theta_\a:=\frac{1}{\sqrt{2}}(\theta^{\un{1}}_\a+\ri\theta^{\un{2}}_\a)~,\qquad \bar{\theta}_\a:=\frac{1}{\sqrt{2}}(\theta^{\un{1}}_\a-\ri\theta^{\un{2}}_\a)~,
\eea
and
\bea
    \cD_\a:=\frac{1}{\sqrt{2}}(\cD^{\un{1}}_\a-\ri \cD^{\un{2}}_\a)~,\qquad \DB_\a:=-\frac{1}{\sqrt{2}}(\cD^{\un{1}}_\a+\ri \cD^{\un{2}}_\a)~,
\eea
where $\theta^{\un{\rm{I}}}_\a$ are the real coordinates and 
\bea
    \cD^{\un{1}}_\a:=D^{\un{1}}_\a- \theta^{{\un{2}}}_\a Q~,\qquad \cD^{\un{2}}_\a:=D^{\un{2}}_\a+\theta^{{\un{1}}}_\a Q~,
\eea
\esubeq
where $Q$ is the central charge which is $m$ for unbarred fields and $-m$ for barred fields.
If we then integrate out $\theta_\a^{\un{2}}$ from \eqref{Massive Vector Action}, we end up with the action
\bea\label{Action Dim Reduced Field Strength}
    S&=&-\frac{\ri}{2}\int\rd^{3|2}z~\bigg[\bar{\cW}^\a\cW_\a-m^2\bar{\cH}^\a\cH_\a-\bar{\cX}D^\a D_\a\cX-\frac{1}{4}\bar{\cV}\square D^\a D_\a\cV\non\\
    &&~~~~~~~~~~~~~~~~~~~~+\ri\bigg(m\bar{\cH}^\g D_\g\cX+\frac{1}{2}m\bar{\cV}\pa_{\ab}D^{\a}\cH^\b-\bar{\cV}\square\cX+\rm{h}.\rc.\bigg)\bigg]~,
\eea
invariant under gauge transformations:
\bea
    \d\cV=\l_1~,\qquad \d\cH_\a=\ri D_\a\l_2~,\qquad\d\cX=\frac{\ri}{4} D^\a D_\a\l_1+m\l_2~,
\eea
where $\cW_\a$ is the gauge invariant field strength
\bea
    \cW_\a:=\frac{\ri}{2} D^\b D_\a\cH_\b~,
\eea
and the fields $\{\cV,\cH_\a,\cX\}$ and gauge parameters $\{\l_1,\l_2\}$ are:
\bea
    \cV&:=&\ri  V|_{\theta_2=0}~,\qquad \cH_\a:=\ri\cD^{\un{2}}_\a  V|_{\theta_2=0}~,\qquad \cX:=-\frac{1}{4}\cD^{\un{2}}_\a \cD^{\un{2}\a} V|_{\theta_2=0}~,\non\\
    \l_1&:=&\ri(\L+\r)|_{\theta^{\un{2}}_\a=0}~,\qquad \l_2:=-\ri(\L-\r)|_{\theta^{\un{2}}_\a=0}~.
\eea
We see then that we can gauge away $\cV$ and impose reality conditions $\cX=\bar{\cX},$ $\cH_\a=\bar{\cH}_\a$, $\l_1=\bar{\l}_1$ and $\l_2=\bar{\l}_2$ to get the action
\bea\label{Action Dim Reduced Field Strength Real}
    S&=&-\frac{\ri}{2}\int\rd^{3|2}z~\bigg[\cW^\a\cW_\a-m^2\bar{\cH}^\a\cH_\a-\cX D^\a D_\a\cX+2\ri m\cH^\g D_\g\cX\bigg]~,
\eea
invariant under gauge transformations
\bea
    \d\cV=\l_1~,\qquad \d\cH_\a=\ri D_\a\l_2~,\qquad\d\cX=\frac{\ri}{4} D^\a D_\a\l_1+m\l_2~,
\eea
which is simply the Stueckelberg reformulation of the massive non-gauge vector supermultiplet in $\mathbb{M}^{3|2}$.
If we carry out the same process for the theory in section \ref{New Minimal Reduction} we end up with the action
\bea
    S^{(\rm{II})}&=&\frac{\ri}{4}\int \rd^{3|2}z~\bigg\{\cL^{\ab\d}\bigg(\square\cL_{\a\b\d}-\frac{\ri}{2}\pa_{\d}^\r D^{\g}D_\g\cL_{\r\a\b}+\pa_\ab\pa^{\r\g}\cL_{\d\r\g}-2 m^2\cL_{\d\a\b}\non\\
    &&~~~~~~~~~~~-\frac{2\ri}{3}\pa_{\a\b}D_\d D^{\g}\cJ_{\g}-4\ri mD_\d\cM_\ab\bigg)-4\ri\cM^\ab\pa_\a^\g \cM_{\b\g}\non\\
    &&~~~~~~~~~~~+\frac{2}{3}\cJ^{\a}\bigg(\frac{1}{3}D_\g D^\g D_\a D^\r\cJ_{\r}+4 m^2\cJ_{\a}-8\ri m D^\b\cM_\ab\bigg)\bigg\}~,
\eea
which is invariant under the gauge transformations
\bea
    \d\cJ_\a= -\ri D^{\b}\l_\ab~,\non\\
    \d \cL_{\ab\g}=\ri D_{(\a}\l_{\b\g)}~,\non\\
    \d\cM_\ab= -m\l_{\ab}~.
\eea
Thus we see how a new theory for massive linearised supergravity can be derived in $\mathbb{M}^{3|2}$. The 3D $\cN=2\rightarrow\cN=1$ reduction of the massive higher spin models studied in this paper will be discussed elsewhere.

The approach developed in this paper makes it possible to construct, for the first time, gauge-invariant off-shell actions for massive higher-spin  supermultiplets in four dimensions by carrying out the Kaluza-Klein reduction of the massless 5D $\cN=1$ 
supersymmetric higher-spin models \cite{Buchbinder:2025yef}
which provide a natural generalisation of the off-shell formulations for massless 4D $\cN=2$  higher-spin supermultiplets pioneered in
\cite{Buchbinder:2021ite}.
\\

\noindent
{\bf Acknowledgements:}\\
The work of EIB and SMK is supported in part by the Australian Research Council, project No. DP230101629. The work of AJF is supported by the Australian Government Research Training Program.

\begin{appendices}


\section{Topologically massive higher superspin models}\label{AppendixA}

In this appendix we review the structure of the linearised actions for $\cal N$-extended (super)conformal higher-spin gravity 
and their massive deformations proposed in \cite{BHHK}. These models are formulated in the 3D $\cN$-extended Minkowski superspace ${\mathbb M}^{3|2\cN}$ parametrised by  
real coordinates $z^A= (x^a, \theta^{\alpha}_I)$, where the $R$-symmetry  index of $\theta^{\alpha}_I$
takes $\cN$ values, $I=  {{1}} , {{2}} , \dots , {{\cal N}} $. 
The corresponding spinor covariant derivatives $D^I_\a$ obey the anti-commutation relation
\bea
\{ D^{I}_{\alpha}, D^{J}_{\beta}\} =2 \ri\, \d^{IJ} \partial_{\alpha \beta}~.
\eea

The linearised actions for $\cal N$-extended superconformal higher-spin gravity are formulated in terms of an unconstrained real prepotential $H_{\a(n)}= H_{\a_1 \dots \a_n}= H_{(\a_1 \dots \a_n)}$ 
which is defined modulo the gauge transformation
\bea
\d_\z H_{\a(n )} &=& \ri^n D^I_{(\a_1 } \z^I_{\a_2 \dots \a_n)} ~, \qquad n>0~.
\label{Nngauge}
\eea
Requiring the gauge superfield $H_{\a(n)}$ and the gauge parameter $\z^I_{\a(n-1)}$ to be primary  
fixes the dimension of $H_{\a(n)}$ to be $d_{H_{\a(n)}} = 2-\cN - n/2$.
Associated with $H_{\a(n)}$ is the $\cN$-extended superconformal gauge-invariant action 
\cite{BHHK}:
\bea
 S^{( \cN|n )} [ H_{\alpha(n)} ] = \frac{{\rm i}^n}{ 2}    \int {\rm d}^{3}x  {\rm d}^{2 \cal N} \theta  \, H^{\alpha(n)} 
{W}_{\alpha(n)}(H )~, \qquad n>0~.
\label{NnCSG-action}
\eea
The field strength $W_{\a(n)}$ in \eqref{NnCSG-action} is a local descendant of $H_{\a(n)}$, which is called the linearised 
rank-$n$ super-Cotton tensor.
It is a real completely symmetric 
rank-$n$ spinor, which is required to obey several conditions:
\begin{enumerate}
\item $W_{\a(n)}$ is invariant under the gauge transformation \eqref{Nngauge}, 
${W}_{\a(n) }\big(\d_\z H \big) =0$.
\item $W_{\a(n)}$ is transverse,
\bea
D^{\b I} W_{\b \a_1 \dots \a_{n-1}} =0~.\label{Nn-transverse}
\eea
\item $W_{\a(n)}$ is a primary superconformal multiplet.
The condition \eqref{Nn-transverse} uniquely fixes the dimension 
of ${W}_{\a(n) }$ to be $d_{W_{\a(n)}} = 1 +n/2$.
\end{enumerate}
The above conditions determine $W_{\a(n)}$, modulo an overall numerical factor, to be
\bea
W_{\a(n)} \big(H\big) =  \D^{n+\cN-1}  \P^{\perp}_{[n]}H_{\a(n)}~, \qquad \D = -\frac{\ri}{2\cN} D^{\a I } D^I_\a ~.
\label{super-Cotton}
\eea
Here  the operator $\D$ has the following properties:
\begin{subequations}\label{1.3} 
\bea
    D^{\a I} \J_\a &=& 0 \quad \implies \quad \D \J_\a = \pa_\a{}^\b \J_\b~,\\
    D^{\a I} \J_\a &=& 0 \quad \implies \quad D^{\a I} \D \J_\a = 0~, \\
    \big[\Delta \,, D^{\b I}D^{I}_{\a} \big] &=& 0~.
\eea
\end{subequations}
The operator $ \P^{\perp}_{[n]}$ in \eqref{super-Cotton} is  a transverse projector, 
\bea
 \P_{[n]}^{\perp}  \P_{[n]}^{\perp} =  \P_{[n]}^{\perp}~.
 \label{idempotent}
 \eea
It is defined to 
act on the space of real symmetric  
rank-$n$ spinors $\J_{\a(n)}$
by the rule
\bea
    \P^{\perp}_{[n]} \J_{\a(n)} := \P_{\a_1}{}^{\b_1} \dots \P_{\a_n} {}^{\b_n} 
    \J_{\b_1 \dots \b_n}\equiv \J^{\perp}_{\a_1 \dots \a_n}
    =\J^{\perp}_{(\a_1 \dots \a_n)}~,
    \label{2.18}
\eea 
The operator $\P_\a{}^\b$ has the following universal properties
\bsubeq\label{5.2}
\bea
    D^{\a I} \P_{\a}{}^{\b} &=& 0~,\\
    \P_{\a}{}^{\b} D^{I}_{\b}&=& 0~,\\
    \P_{\a}{}^{\b}\P_{\b}{}^{\g} &=& \P_{\a}{}^{\g}~,\\
    \big[ \P_{\a}{}^{\b}  , \P_{\g}{}^{\d} \big]&=&0~. 
\eea
\esubeq
The superfield $\J^{\perp}_{\a_1 \dots \a_n}$ defined by \eqref{2.18} is transverse\footnote{In general, 
a real symmetric rank-$n$ spinor superfield $T_{\a(n)} $ is called transverse (or divergenceless) if 
$D^{I \b} T_{\b \a_1 \dots \a_{n-1}} =0$.
Then it holds that $\P^{\perp}_{[n]} T_{\a(n)}  = T_{\a(n)} $.}
\bea
    D^{I \b}\J^{\perp}_{\b \a_1 \dots \a_{n-1}} =0 ~.
    \label{Tpro}
\eea
The explicit construction of $ \P_\a{}^\b$ was given in \cite{BHHK} 
for the supersymmetry types   $1\leq \cN \leq 6$. 

In the non-supersymmetric  case ($\cN=0$),  $W_{\alpha(n)}$ was constructed in \cite{PopeTownsend} for even $n$ and \cite{K16} for odd $n$ (see \cite{HHL,HLLMP} for an alternative derivation)).  In AdS$_3$, such tensors were constructed in \cite{Kuzenko:2021hyd}.
In the $\cN=1$ and $\cN=2$ cases, $W_{\alpha(n)}$ was constructed in \cite{K16, KT}  and \cite{KO}, respectively.
Explicit expressions for $W_{\alpha(n)}$ for $\cN>2$ were given in \cite{BHHK}.
The $\cN=1$ AdS supersymmetric extensions of the higher-spin Cotton tensors were derived in \cite{Kuzenko:2021hyd}.

For $n>0$  a massive on-shell $\cN$-extended superfield is defined by 
the conditions
\begin{subequations} \label{EoM}
\bea
D^{\b I} T_{\b \a_1 \dots \a_{n-1}} &=& 0 
~ , 
\label{EoM.a} \\
\D  T_{\a_1 \dots \a_n} &=& m \s T_{\a_1 \dots \a_n}~, 
\qquad \s =\pm 1~,
\label{EoM.b}
\eea
\end{subequations}
of which the former implies 
$\pa^{\b\g} T_{\b\g\a_1\dots \a_{n-2}} =0$ for $n>1$.
This definition generalises those given earlier in the $\cN=1$ and $\cN=2$ 
cases \cite{KNT-M15,KT,KO}.

A gauge-invariant model 
in which the equations of motion are equivalent to \eqref{EoM}, is a deformation 
of the superconformal action \eqref{NnCSG-action} given by \cite{BHHK}
\bea
{S}^{(\cN|n)}_{\rm massive} [  H_{\a(n)}] \propto \frac{ \ri^n}{2} 
   \int \rd^{3|2\cN}z \, H^{\a(n)} \Big( \D - m \s \Big)
{W}_{\a(n)}\big(H\big) ~, 
\qquad n>0~.
\label{action-mass}
\eea
This model is a generalisation of the following massive gauge-invariant higher-spin theories:
(i) the non-supersymmetric models in Minkowski space \cite{BHT,BKRTY,BKLFP}
and anti-de Sitter space AdS${}_3$  \cite{KP18}; 
and (ii) the supersymmetric models in AdS${}_3$
with (1,0) \cite{KP18} and (2,0) \cite{HK18} anti-de Sitter supersymmetry.


\section{Properties of operators in $\mathbb{M}^{3|4}_c$} \label{Properties Section}
In this appendix we outline some of the properties of the operators described in section \ref{3D Oscillator Formalism} which will be useful in calculations throughout. We begin with the identities:
\bea\label{Central Charge Identities}
    \cA^2&=&\bar{\cA}^2=\cA'^2=\bar{\cA}'^2=\cB^2=\bar{\cB}^2=\cB'^2=\bar{\cB}'^2=0~,\non\\
    \cB\cA&=&-\frac{1}{2}N\cD^2~,\qquad \cB'\cA'=-\frac{1}{2} N'\cD^2\qquad \bar{\cB}\bar{\cA}=\frac{1}{2}N\cDB^2~,\qquad \bar{\cB}'\bar{\cA}'=\frac{1}{2} N'\cDB^2\non\\
    \cB\cA'&=&-\frac{1}{2}U\cD^2~,\qquad \cB'\cA=-\frac{1}{2}T\cD^2~,\qquad \bar{\cB}\bar{\cA}'=\frac{1}{2}U\cDB^2~,\qquad \bar{\cB}'\bar{\cA}=\frac{1}{2}T\cD^2~.
\eea
We then note the following useful relations from the operator algebra
\bea \label{Operator Algebra 3D}
    \{\cA,\bar{\cB}\}&=&-2P_1+2\ri mN-\cK~,\qquad \{\cA',\bar{\cB}'\}=-2P_2+2\ri m N'+\cK~,\non\\
    \{\cB',\bar{\cA}'\}&=&-2P_2-2\ri m( N'+2)-\cK~,\qquad \{\cA,\bar{\cA}'\}=-2P_3+2\ri mV~,\non\\
    \{\cB,\bar{\cB}'\}&=&-2P_4-2\ri m W~,\qquad \{\cA',\bar{\cA}'\}=-2P_5~,\qquad \{\cB',\bar{\cB}'\}=-2P_6~,\non\\
    \{\cA,\bar{\cB}'\}&=&\{\cB',\bar{\cA}\}=-2P_7+2\ri mT~,\qquad \{\cB,\bar{\cA}'\}=\{\cA',\bar{\cB}\}=-2P_8-2\ri mU~,\non\\
    ~[V, W]&=&-(N+ N'+2)~,\qquad [\cB,V]=-\cA'~,\qquad [\cB',V]=\cA~,\non\\
    ~[\bar{\cB},V]&=&-\bar{\cA}'~,\qquad [\bar{\cB}',V]=\bar{\cA}~,\non\\
    ~[\cA', W]&=&-\cB~,\qquad [\cA, W]=\cB'~,\qquad \non\\
    ~[\bar{\cA}', W]&=&-\bar{\cB}~,\qquad [\bar{\cA}, W]=\bar{\cB}'~,\qquad \non\\
    ~[N,V]&=&-V~,\qquad [N, W]= W~,\qquad [ N',V]=-V~, \qquad [ N', W]= W~,\non\\
    ~[P_1,V]&=&-P_3~,\qquad [P_1, W]=P_4~,\qquad [P_2,V]=P_3~,\qquad [P_2, W]=-P_4~,\non\\
    ~[P_3, W]&=&P_2-P_1~,\qquad [P_4,V]=P_1-P_2~,\non\\
    ~[P_5, W]&=&-2P_8~,\qquad [P_8, W]=-P_{10}~,\non\\
    ~[P_1,U]&=&P_8~,\qquad [P_2,U]=-P_8~,\qquad [P_3,U]=P_5~,\qquad [P_4,U]=-P_{10}~,\non\\
    ~[\cA,U]&=&\cA'~,\qquad [\cB',U]=-\cB~,\qquad [P_3,U]=P_5~,\non\\
    ~[P_3, W^k]&=&k W^{k-1}(P_2-P_1)-k(k-1) W^{k-2}P_4~,\non\\
    ~[P_5, W^k]&=&-2k W^{k-1}P_8+k(k-1) W^{k-2}P_{10}~,\non\\
    ~[P_1,U^k]&=&kU^{k-1}P_8~,\qquad [P_2,U^k]=-kU^{k-1}P_8~,\non\\
    ~[P_3,U^k]&=&kU^{k-1}P_5~,\qquad [P_4,U^k]=-kU^{k-1}P_{10}\non\\
    ~[V, W^k]&=&-k W^{k-1}(N+ N'+k+1)~,\qquad [V^{k}, W]=-k(N+ N'+k+1)V^{k-1}~,\qquad \non\\
    ~[\cA, W^k]&=&k W^{k-1}\cB'~,\qquad[\cA', W^k]=-k W^{k-1}\cB~,\non\\
    ~[\bar{\cA}, W^k]&=&k W^{k-1}\bar{\cB}'~,\qquad[\bar{\cA}', W^k]=-k W^{k-1}\bar{\cB}~,\non\\
    ~[\cA,U^k]&=&kU^{k-1}\cA'~,\qquad [\cB',U^k]=-kU^{k-1}\cB~,\non\\
    ~[\bar{\cA},U^k]&=&kU^{k-1}\bar{\cA}'~,\qquad [\bar{\cB}',U^k]=-kU^{k-1}\bar{\cB}~.\qquad
\eea


\section{Gauge invariance of massive new linearised SUGRA action}\label{appendixC}

We shall confirm in this appendix the gauge invariance of the action \eqref{New Linearised SUGRA Massive Action} with respect to the gauge transformations \eqref{New Linearised Massive SUGRA Transformations}.

It is worth noting the following identities for the calculations below:
\bea
    \cD_\a \cD_\b=\frac{1}{2}\ve_\ab\cD^2~,\qquad \cDB_\a\cDB_\b=-\frac{1}{2}\ve_\ab\cDB^2~,\non\\
    \cD_\a\cD_\b\cD_\g=\cDB_\a\cDB_\b\cDB_\g=0~,\qquad \cD^\a\cDB^2\cD_\a=\cDB_\a\cD^2\cDB^\a~,\non\\
    ~[\cD^2,\cDB_\a]=2\{\cD_\b,\cDB_\a\}\cD^\b~,\qquad [\cDB^2,\cD_\a]=-2\{\cD_\a,\cDB_\b\}\cDB^\b~.
\eea

Since $E_\ab$ is contracted with a symmetric field, only its symmetric part must be gauge invariant. We consider its variation:
\bea
    \d E_{(\ab)}&=&-\frac{1}{8}\cD^\g\cDB^2 \cD_\g \d H_\ab+\frac{1}{2}\pa_\ab(\pa^{\r\g}\d H_{\r\g}-2\ri m\d H)\non\\
    &&+\frac{1}{8}[\cD_{(\a},\cDB_{\b)}]([\cD^\g,\cDB^\r]\d H_{\g\r}+\ri[\cD^\g,\cDB_\g]\d H)\non\\
    &&+\frac{1}{2}[\cD_{(\a},\cDB_{\b)}]\d F\non\\
    &=&-\frac{1}{8}\cD^\g\cDB^2 \cD_\g (\cDB_{(\a}L_{\b)}-\cD_{(\a} M_{\b)})+\frac{1}{2}\pa_\ab\pa^{\r\g}(\cDB_{(\g}L_{\r)}-\cD_{(\g} M_{\r)})\non\\
    &&+ \frac{1}{2}m\pa_\ab(\cDB^{\g}L_{\g}-\cD_{\g} M^{\g})+\frac{1}{8}[\cD_{(\a},\cDB_{\b)}][\cD^{\g},\cDB^{\r}](\cDB_{(\g}L_{\r)}-\cD_{(\g} M_{\r)})\non\\
    &&-\frac{1}{16}[\cD_{(\a},\cDB_{\b)}][\cD^\g,\cDB_\g](\cDB^{\r}L_{\r}-\cD_{\r} M^{\r})+\frac{1}{8}[\cD_{(\a},\cDB_{\b)}](\cD^\g\cDB^2 L_\g+\cDB_\g \cD^2 M^\g)\non\\
    &=&\frac{1}{8} (\cDB^\g\cD^2 \cDB_\g\cDB_{(\a}L_{\b)}+\cD^\g\cDB^2 \cD_\g\cD_{(\a} M_{\b)})+\frac{1}{2}\pa_\ab\pa^{\r\g}(\cDB_{(\g}L_{\r)}-\cD_{(\g} M_{\r)})\non\\
    &&+ \frac{1}{2}m\pa_\ab(\cDB^{\g}L_{\g}-\cD_{\g} M^{\g})+\frac{1}{8}[\cD_{(\a},\cDB_{\b)}][\cD^{\g},\cDB^{\r}](\cDB_{(\g}L_{\r)}-\cD_{(\g} M_{\r)})\non\\
    &&-\frac{1}{16}[\cD_{(\a},\cDB_{\b)}][\cD^\g,\cDB_\g](\cDB^{\r}L_{\r}-\cD_{\r} M^{\r})+\frac{1}{8}[\cD_{(\a},\cDB_{\b)}](\cD^\g\cDB^2 L_\g+\cDB_\g \cD^2 M^\g)\non\\
    &=&\frac{1}{8} (\cDB_{(\a}\cD_{\b)}\cD_\g\cDB^2L^{\g}+\cD_{(\a}\cDB_{\b)}\cDB_\g\cD^2 M^{\g})-\frac{1}{8}\{\cD_{(\a},\cDB_{\b)}\}\{\cD^{\g},\cDB^{\r}\}(\cDB_{(\g}L_{\r)}-\cD_{(\g} M_{\r)})\non\\
    &&+ \frac{\ri}{4}m\{\cD_{(\a},\cDB_{\b)}\}(\cDB^{\g}L_{\g}-\cD_{\g} M^{\g})+\frac{1}{8}[\cD_{(\a},\cDB_{\b)}][\cD^{\g},\cDB^{\r}](\cDB_{(\g}L_{\r)}-\cD_{(\g} M_{\r)})\non\\
    &&-\frac{1}{16}[\cD_{(\a},\cDB_{\b)}][\cD^\g,\cDB_\g](\cDB^{\r}L_{\r}-\cD_{\r} M^{\r})+\frac{1}{8}[\cD_{(\a},\cDB_{\b)}](\cD^\g\cDB^2 L_\g+\cDB_\g \cD^2 M^\g)\non\\
    &=&\frac{1}{16} \cDB_{(\a}\cD_{\b)}[4\cDB_\g\cD^2 M^\g+2\cD^\g\cDB^2L_{\g}-2\{\cD^\g,\cDB^\r\}(\cDB_{(\g}L_{\r)}-\cD_{(\g} M_{\r})+4\ri m(\cDB^\g L_\g-\cD_\g M^\g)\non\\
    &&+2[\cD^\g,\cDB^\r](\cDB_{(\g}L_{\r)}-\cD_{(\g} M_{\r})-[\cD^\g,\cDB_\g](\cDB^\r L_\r-\cD_\r M^\r)]\non\\
    &&-\frac{1}{16}\cD_{(\a}\cDB_{\b)}[2\cDB_\g\cD^2 M^\g+4\cD^\g\cDB^2L_{\g}+2\{\cD^\g,\cDB^\r\}(\cDB_{(\g}L_{\r)}-\cD_{(\g} M_{\r})-4\ri m(\cDB^\r L_\r-\cD_\r M^\r)\non\\
    &&+2[\cD^\g,\cDB^\r](\cDB_{(\g}L_{\r)}-\cD_{(\g} M_{\r})-[\cD^\g,\cDB_\g](\cDB^\r L_\r-\cD_\r M^\r)]\non\\
    &=&\frac{1}{8} \cDB_{(\a}\cD_{\b)}[2\cDB_\g\cD^2 M^\g+\cD^\g\cDB^2L_{\g}-2\cDB^\r\cD^\g(\cDB_{(\g}L_{\r)}-\cD_{(\g} M_{\r})+\cDB_\g\cD^\g(\cDB^\r L_\r-\cD_\r M^\r)]\non\\
    &&-\frac{1}{8}\cD_{(\a}\cDB_{\b)}[\cDB_\g\cD^2 M^\g+2\cD^\g\cDB^2L_{\g}+2\cD^\g\cDB^\r(\cDB_{(\g}L_{\r)}-\cD_{(\g} M_{\r})-\cD^\g\cDB_\g(\cDB^\r L_\r-\cD_\r M^\r)]\non\\
    &=&\frac{1}{8} \cDB_{(\a}\cD_{\b)}[\cD^\g\cDB^2L_{\g}-2\cDB^\r\cD^\g\cDB_{(\g}L_{\r)}+\cDB_\g\cD^\g\cDB^\r L_\r]\non\\
    &&-\frac{1}{8}\cD_{(\a}\cDB_{\b)}[\cDB_\g\cD^2 M^\g-2\cD^\g\cDB^\r\cD_{(\g} M_{\r}+\cD^\g\cDB_\g\cD_\r M^\r]\non\\
    &=&\frac{1}{8} \cDB_{(\a}\cD_{\b)}[[\cD^\g,\cDB^2]L_{\g}-2\cDB^\r\{\cD^\g,\cDB_{(\g}\}L_{\r)}+\cDB_\g\{\cD^\g,\cDB^\r\} L_\r]\non\\
    &&-\frac{1}{8}\cD_{(\a}\cDB_{\b)}[[\cDB_\g,\cD^2] M^\g-2\cD^\g\{\cDB^\r,\cD_{(\g}\} M_{\r}+\cD^\g\{\cDB_\g,\cD_\r\} M^\r]\non\\
    &=&\frac{1}{8} \cDB_{(\a}\cD_{\b)}[-2\{\cD_\g,\cDB_\r\}\cDB^\r L^{\g}-\{\cD^\g,\cDB_{\g}\}\cDB^\r L_{\r}+\{\cD_\g,\cDB_{\r}\}\cDB^\r L^{\g}+\{\cD_\g,\cDB_\r\}\cDB^\g L^\r]\non\\
    &&-\frac{1}{8}\cD_{(\a}\cDB_{\b)}[-2\{\cD_\r,\cDB_\g\}\cD^\r M^\g+\{\cD^\r,\cDB_\r\}\cD^\g M_{\g}+\{\cD_\g,\cDB_\r\}\cD^\g M^{\r}+\{\cDB_\g,\cD_\r\}\cD^\g M^\r]\non\\
    &=&\frac{1}{8} \cDB_{(\a}\cD_{\b)}[-2\{\cD_{[\g},\cDB_{\r]}\}\cDB^\r L^{\g}-\{\cD^\g,\cDB_{\g}\}\cDB^\r L_{\r}]\non\\
    &&-\frac{1}{8}\cD_{(\a}\cDB_{\b)}[-2\{\cD_{[\r},\cDB_{\g]}\}\cD^\r M^\g+\{\cD^\r,\cDB_\r\}\cD^\g M_{\g}]\non\\
    &=&0~.
\eea
We then consider the variation
\bea
    \d E_0&=& \frac{1}{4} \cD^\g\cDB^2 \cD_\g \d H-\ri m( \pa^{\r\g}\d H_{\r\g}-2\ri m\d H)\non\\
    &&+\frac{1}{8}\ri[\cD^\a,\cDB_\a]([\cD^\g,\cDB^\r]\d H_{\g\r}+\ri[\cD^\g,\cDB_\g]\d H)\non\\
    &&+\frac{\ri}{2}[\cD^\a,\cDB_\a]\d F\non\\
    &=& \frac{\ri}{8} \cD^\g\cDB^2 \cD_\g (\cDB^{\a}L_{\a}-\cD_{\a} M^{\a})-\ri m\pa^{\r\g}(\cDB_{(\g}L_{\r)}-\cD_{(\g} M_{\r)})\non\\
    &&-\ri m^2 (\cDB^{\a}L_{\a}-\cD_{\a} M^{\a})+\frac{1}{8}\ri[\cD^\a,\cDB_\a][\cD^\g,\cDB^\r](\cDB_{(\g}L_{\r)}-\cD_{(\g} M_{\r)})\non\\
    &&-\frac{\ri}{16}[\cD^\a,\cDB_\a][\cD^\g,\cDB_\g](\cDB^{\r}L_{\r}-\cD_{\r} M^{\r})\non\\
    &&+\frac{\ri}{8}[\cD^\a,\cDB_\a](\cD^\b\cDB^2 L_\b+\cDB_\b \cD^2 M^\b)\non\\
    &=& \frac{\ri}{8} \cD^\g\cDB^2 \cD_\g (\cDB^{\a}L_{\a}-\cD_{\a} M^{\a})-\ri m\pa^{\r\g}(\cDB_{(\g}L_{\r)}-\cD_{(\g} M_{\r)})\non\\
    &&-\ri m^2 (\cDB^{\a}L_{\a}-\cD_{\a} M^{\a})+\frac{1}{8}\ri[\cD^\a,\cDB_\a][\cD^\g,\cDB^\r](\cDB_{(\g}L_{\r)}-\cD_{(\g} M_{\r)})\non\\
    &&-\frac{\ri}{16}(2\cD^{\a}\cDB_\a-\{\cD^\a,\cDB_\a\})(\{\cD^\g,\cDB_\g\}-2\cDB_\g\cD^\g)(\cDB^{\r}L_{\r}-\cD_{\r} M^{\r})\non\\
    &&+\frac{\ri}{8}[\cD^\a,\cDB_\a](\cD^\b\cDB^2 L_\b+\cDB_\b \cD^2 M^\b)\non\\
    &=& -\ri m\pa^{\r\g}(\cDB_{(\g}L_{\r)}-\cD_{(\g} M_{\r)})\non\\
    &&+\frac{1}{8}\ri[\cD^\a,\cDB_\a][\cD^\g,\cDB^\r](\cDB_{(\g}L_{\r)}-\cD_{(\g} M_{\r)})\non\\
    &&+\frac{\ri}{8}[\cD^\a,\cDB_\a](\cD^\b\cDB^2 L_\b+\cDB_\b \cD^2 M^\b)\non\\
    &=& \frac{\ri}{4}(2\cDB_\a\cD^\a\cDB^\r\cD^\g-\{\cD^\g,\cDB^\r\}\cDB_\a\cD^\a-\{\cD^\a,\cDB_\a\}\cDB^\g\cD^\r)(\cDB_{(\g}L_{\r)}-\cD_{(\g} M_{\r)})\non\\
    &&+\frac{\ri}{8}[\cD^\a,\cDB_\a](\cD^\b\cDB^2 L_\b+\cDB_\b \cD^2 M^\b)\non\\
    &=& -\frac{\ri}{8}(\cDB^\g\cD^2\cDB^\r+\cD^\g\cDB^2\cD^\r)(\cDB_{(\g}L_{\r)}-\cD_{(\g} M_{\r)})\non\\
    &&+\frac{\ri}{8}[\cD^\a,\cDB_\a](\cD^\b\cDB^2 L_\b+\cDB_\b \cD^2 M^\b)\non\\
    &=& \frac{3\ri}{32}(\cDB^\g\cD^2\cDB^2 L_\g+\cD^\g\cDB^2\cD^2 M_\g)-\frac{\ri}{8}(\cD^\g\cDB^2\cD^\r\cDB_{(\g}L_{\r)}+\cDB^\g\cD^2\cDB^\r\cD_{(\g} M_{\r)})\non\\
    &&+\frac{\ri}{8}[\cD^\a,\cDB_\a](\cD^\b\cDB^2 L_\b+\cDB_\b \cD^2 M^\b)\non\\
    &=& -\frac{\ri}{32}(\cDB^\g\cD^2\cDB^2 L_\g+\cD^\g\cDB^2\cD^2 M_\g)+\frac{\ri}{16}(\cD^\g\cDB^2\{\cD_\r,\cDB_{\g}\}L^{\r}+\cDB^\g\cD^2\{\cD_{\g},\cDB_\r\} M^{\r})\non\\
    &&+\frac{\ri}{16}\{\cD^\a,\cDB_\a\}(\cD^\b\cDB^2 L_\b-\cDB_\b \cD^2 M^\b)\non\\
    &=& -\frac{\ri}{16}\{\cD_{\g},\cDB_\r\}(\cD^\g\cDB^2L^\r+\cDB^\r\cD^2 M^\g)+\frac{\ri}{16}\{\cD_\g,\cDB_{\r}\}(\cD^\r\cDB^2L^{\g}+\cDB^\g\cD^2 M^{\r})\non\\
    &&+\frac{\ri}{16}\{\cD^\a,\cDB_\a\}(\cD^\b\cDB^2 L_\b-\cDB_\b \cD^2 M^\b)
    =0~.
\eea

To see how the term $\bar{F} E_1$ is gauge invariant, we must use the linearity condition on $F$ to decompose it in terms of unconstrained prepotentials. Doing so gives us
\bea
    \bar{F}=\frac{\ri}{4}(\cD^{\a}\cDB^2 \bar{f}_\a+\cDB_{\a}\cD^2 \bar{g}^\a)~.
\eea
As a consequence of this, the expressions $\cDB^2\cD_\a E_1$ and $\cD^2\cDB_\a E_1$ must both be gauge invariant.
Let us start by working out
\bea
    \d E_1&=&\frac{1}{2}[\cD_\g,\cDB_\r]\d H^{\g\r}+\frac{\ri}{2}[\cD^\g,\cDB_\g]\d H+3\d \cDB^2\cD_\a F\non\\
    &=&\frac{1}{2}[\cD^\g,\cDB^\r](\cDB_{(\g}L_{\r)}-\cD_{(\g} M_{\r)})-\frac{1}{4}[\cD^\g,\cDB_\g](\cDB^{\r}L_{\r}-\cD_{\r} M^{\r})\non\\
    &&+\frac{3}{4}(\cD^\g\cDB^2 L_\g+\cDB_\g \cD^2 M^\g)\non\\
    &=&(\cD^\g\cDB^\r\cDB_{(\g}L_{\r)}+\cDB^\r\cD^\g\cD_{(\g} M_{\r)})-\frac{1}{2}\{\cD^\g,\cDB^\r\}(\cDB_{(\g}L_{\r)}+\cD_{(\g} M_{\r)})\non\\
    &&-\frac{1}{2}(\cD^\g\cDB_\g\cDB^{\r}L_{\r}+\cDB_\g\cD^\g\cD_{\r} M^{\r})+\frac{1}{4}\{\cD^\g,\cDB_\g\}(\cDB^{\r}L_{\r}-\cD_{\r} M^{\r})\non\\
    &&+\frac{3}{4}(\cD^\g\cDB^2 L_\g+\cDB_\g \cD^2 M^\g)\non\\
    &=&-\frac{1}{2}\{\cD^\g,\cDB^\r\}(\cDB_{(\g}L_{\r)}+\cD_{(\g} M_{\r)})\non\\
    &&\frac{1}{4}(\cD^\g\cDB^2L_{\g}+\cDB_\g\cD^2 M^{\g})+\frac{1}{4}\{\cD^\g,\cDB_\g\}(\cDB^{\r}L_{\r}-\cD_{\r} M^{\r})\non\\
    &=&\frac{1}{4}(\cDB^2\cD^\g L_{\g}+\cD^2\cDB_\g M^{\g})~,
\eea
at which point, we can clearly see that both $\cDB^2\cD_\a E_1$ and $\cD^2\cDB_\a E_1$ vanish. Hence, the action is gauge invariant.


\section{Gauge invariance of massive higher half-integer superspin action} \label{Gauge Invariance Section}
We now wish to verify that the action \eqref{Action Recap} is indeed invariant under \eqref{Gauge Variation Recap}. The action variation is
\bea\label{Action Varied}
    \d \bm{S}&=&\sum_{k=0}^s\bigg(-\frac{1}{2}\bigg)^s\bigg(\frac{1}{s!}\bigg)^2\int\rd^{3|4}z\bigg\{\langle  H_{(2s-2k)}|\d|  E^{(0)}_{(2s-2k)}\rangle+s^2\langle  G_{(2s-2k)}|\d|  E^{(1)}_{(2s-2k)}\rangle\non\\
    &&~~~~~~~~~~~~~~~~~~~+s^2\langle  J_{(2s-2k)}|\d|  E^{(2)}_{(2s-2k)}\rangle+s^2\langle  V_{(2s-2k)}|\d|  E^{(3)}_{(2s-2k)}\rangle\non\\
    &&~~~~~~~~~~~~~~~~~~~+s^2\langle  W_{(2s-2k)}|\d|  E^{(4)}_{(2s-2k)}\rangle+\text{h}.\rc.\bigg\}~.
\eea
We thus must take a look at the variations of \eqref{EoM Dim Reduced HSUSY Recap}. Taking the variation and inserting \eqref{Gauge Variation Recap} gives us
\bea
    \d| E^{(0)}_{(2s-2k)}\rangle &=& \frac{k!(2s-k+1)!}{(2s-2k+1)}\bigg\{\frac{1}{8}\cC\d| H_{(2s-2k)}\rangle-\frac{\ri}{4}(\cK\d| V_{(2s-2k)}\rangle-\bar{\cK}\d| W_{(2s-2k)}\rangle)\non\\
    &&+\frac{1}{4(2s-2k-1)}(\cB'\bar{\cB}'\d| V_{(2s-2k-2)}\rangle-\bar{\cB}'\cB'\d| W_{(2s-2k-2)}\rangle)\non\\
    &&+\frac{1}{4(2s-2k+3)}(\cA'\bar{\cA}'\d| V_{(2s-2k+2)}\rangle-\bar{\cA}'\cA'\d| W_{(2s-2k+2)}\rangle)\bigg\}\non\\
    &=& \frac{k!(2s-k+1)!}{(2s-2k+1)}\bigg\{\frac{1}{8}\cC(| l_{(2s-2k)}\rangle +| m_{(2s-2k)}\rangle)\non\\
    &&-\frac{\ri}{4}\bigg[\cK\bigg(\frac{(2s-k+1)(2s-k+2)}{4s(s+1)(2s-2k+3)}\bar{\cA}'\cA'| m_{(2s-2k+2)}\rangle\non\\
    &&+\frac{k(k+1)}{4s(s+1)(2s-2k-1)}\bar{\cB}'\cB'| m_{(2s-2k-2)}\rangle\non\\
    &&+\ri \frac{k(2s-k+1)}{4s(s+1)}\bar{\cK}| m_{(2s-2k)}\rangle\bigg)\non\\
    &&-\bar{\cK}\bigg(-\frac{(2s-k+1)(2s-k+2)}{4s(s+1)(2s-2k+3)}\cA'\bar{\cA}'| l_{(2s-2k+2)}\rangle\non\\
    &&-\frac{k(k+1)}{4s(s+1)(2s-2k-1)}\cB'\bar{\cB}'| l_{(2s-2k-2)}\rangle-\ri \frac{k(2s-k+1)}{4s(s+1)}\cK| l_{(2s-2k)}\rangle\bigg)\bigg]\non\\
    &&+\frac{1}{4(2s-2k-1)}\bigg[\cB'\bar{\cB}'\bigg(\frac{(2s-k)(2s-k+1)}{4s(s+1)(2s-2k+1)}\bar{\cA}'\cA'| m_{(2s-2k)}\rangle\non\\
    &&+\frac{(k+1)(k+2)}{4s(s+1)(2s-2k-3)}\bar{\cB}'\cB'| m_{(2s-2k-4)}\rangle\non\\
    &&+\ri \frac{(k+1)(2s-k)}{4s(s+1)}\bar{\cK}| m_{(2s-2k-2)}\rangle\bigg)\non\\
    &&-\bar{\cB}'\cB'\bigg(-\frac{(2s-k)(2s-k+1)}{4s(s+1)(2s-2k+1)}\cA'\bar{\cA}'| l_{(2s-2k)}\rangle\non\\
    &&-\frac{(k+1)(k+2)}{4s(s+1)(2s-2k-3)}\cB'\bar{\cB}'| l_{(2s-2k-4)}\rangle-\ri \frac{(k+1)(2s-k)}{4s(s+1)}\cK| l_{(2s-2k-2)}\rangle\bigg)\big]\non\\
    &&+\frac{1}{4(2s-2k+3)}\bigg[\cA'\bar{\cA}'\bigg(\frac{(2s-k+2)(2s-k+3)}{4s(s+1)(2s-2k+5)}\bar{\cA}'\cA'| m_{(2s-2k+4)}\rangle\non\\
    &&+\frac{k(k-1)}{4s(s+1)(2s-2k+1)}\bar{\cB}'\cB'| m_{(2s-2k)}\rangle\non\\
    &&+\ri \frac{(k-1)(2s-k+2)}{4s(s+1)}\bar{\cK}| m_{(2s-2k+2)}\rangle\bigg)\non\\
    &&-\bar{\cA}'\cA'\bigg(-\frac{(2s-k+2)(2s-k+3)}{4s(s+1)(2s-2k+5)}\cA'\bar{\cA}'| l_{(2s-2k+4)}\rangle\non\\
    &&-\frac{k(k-1)}{4s(s+1)(2s-2k+1)}\cB'\bar{\cB}'| l_{(2s-2k)}\rangle-\ri \frac{(k-1)(2s-k+2)}{4s(s+1)}\cK| l_{(2s-2k+2)}\rangle\bigg)\bigg]\bigg\}~,\non\\
\eea
after which point we can make use of properties from \eqref{Operator Action 3D} and \eqref{Central Charge Identities} to simplify this and eventually find
\bea
    \d| E^{(0)}_{(2s-2k)}\rangle &=& 0~.
\eea
We then do the same for $| E^{(3)}_{(2s-2k)}\rangle$ to get
\bea
    \d| E^{(3)}_{(2s-2k)}\rangle &=& \frac{(k-1)!(2s-k)!}{(2s-2k+1)}\non\\
    &&\times \bigg\{\frac{k(2s-k+1)}{4s^2}\ri \cK\d| H_{(2s-2k)}\rangle+\frac{k(k+1)}{4s^2(2s-2k-1)}\cB'\bar{\cB}'\d| H_{(2s-2k-2)}\rangle\non\\
    &&+\frac{(2s-k+1)(2s-k+2)}{4s^2(2s-2k+3)}\cA'\bar{\cA}'\d| H_{(2s-2k+2)}\rangle \non\\
    &&+\frac{s+1}{s}\d| W_{(2s-2k)}\rangle+\d| V_{(2s-2k)}\rangle-\frac{1}{s}\d| J_{(2s-2k)}\rangle\bigg\}\non\\
    &=& \frac{(k-1)!(2s-k)!}{(2s-2k+1)}\non\\
    &&\times \bigg\{\frac{k(2s-k+1)}{4s^2}\ri \cK(| l_{(2s-2k)}\rangle +| m_{(2s-2k)}\rangle)\non\\
    &&+\frac{k(k+1)}{4s^2(2s-2k-1)}\cB'\bar{\cB}'(| l_{(2s-2k-2)}\rangle +| m_{(2s-2k-2)}\rangle)\non\\
    &&+\frac{(2s-k+1)(2s-k+2)}{4s^2(2s-2k+3)}\cA'\bar{\cA}'(| l_{(2s-2k+2)}\rangle +| m_{(2s-2k+2)}\rangle) \non\\
    &&-\frac{(2s-k+1)(2s-k+2)}{4s^2(2s-2k+3)}\cA'\bar{\cA}'| l_{(2s-2k+2)}\rangle-\frac{k(k+1)}{4s^2(2s-2k-1)}\cB'\bar{\cB}'| l_{(2s-2k-2)}\rangle\non\\
    &&-\ri \frac{k(2s-k+1)}{4s^2}\cK| l_{(2s-2k)}\rangle+\frac{(2s-k+1)(2s-k+2)}{4s(s+1)(2s-2k+3)}\bar{\cA}'\cA'| m_{(2s-2k+2)}\rangle\non\\
    &&+\frac{k(k+1)}{4s(s+1)(2s-2k-1)}\bar{\cB}'\cB'| m_{(2s-2k-2)}\rangle+\ri \frac{k(2s-k+1)}{4s(s+1)}\bar{\cK}| m_{(2s-2k)}\rangle\non\\
    &&+\frac{(2s-k+1)(2s-k+2)}{4s^2(s+1)(2s-2k+3)}\bar{\cA}'\cA'| m_{(2s-2k+2)}\rangle\non\\
    &&+\frac{k(k+1)}{4s^2(s+1)(2s-2k-1)}\bar{\cB}'\cB'| m_{(2s-2k-2)}\rangle\non\\
    &&+\ri \frac{k(2s-k+1)}{4s^2(s+1)}\bar{\cK}| m_{(2s-2k)}\rangle+\frac{(2s-k+1)(2s-k+2)}{2s^2(2s-2k+3)}P_5| m_{(2s-2k+2)}\rangle\non\\
    &&+\frac{k(k+1)}{2s^2(2s-2k-1)}P_6| m_{(2s-2k-2)}\rangle+m \frac{k(2s-k+1)}{s^2}| m_{(2s-2k)}\rangle\bigg\}~,
\eea
where we once again use properties from \eqref{Operator Action 3D} and \eqref{Central Charge Identities} to get
\bea
    \d| E^{(3)}_{(2s-2k)}\rangle &=& 0~,
\eea
and through a similar process for $| E^{(4)}_{(2s-2k)}\rangle$ we find
\bea
    \d| E^{(4)}_{(2s-2k)}\rangle &=& \frac{(k-1)!(2s-k)!}{(2s-2k+1)}\non\\
    &&\times \bigg\{-\frac{k(2s-k+1)}{4s^2}\ri \bar{\cK}\d| H_{(2s-2k)}\rangle-\frac{k(k+1)}{4s^2(2s-2k-1)}\bar{\cB}'\cB'\d| H_{(2s-2k-2)}\rangle\non\\
    &&-\frac{(2s-k+1)(2s-k+2)}{4s^2(2s-2k+3)}\bar{\cA}'\cA'\d| H_{(2s-2k+2)}\rangle \non\\
    &&+\frac{s+1}{s}\d| V_{(2s-2k)}\rangle+\d| W_{(2s-2k)}\rangle-\frac{1}{s}\d| G_{(2s-2k)}\rangle\bigg\}
    =0~.
\eea
This leaves us with
\bea
    \d \bm{S}&=&\sum_{k=0}^s\bigg(-\frac{1}{2}\bigg)^s\bigg(\frac{1}{s!}\bigg)^2\int\rd^{3|4}z\bigg\{s^2\langle  G_{(2s-2k)}|\d|  E^{(1)}_{(2s-2k)}\rangle\non\\
    &&~~~~~~~~~~~~~~~~~~~+s^2\langle  J_{(2s-2k)}|\d|  E^{(2)}_{(2s-2k)}\rangle+\text{h}.\rc.\bigg\}~,
\eea
for which it is useful to write the bras in terms of prepotentials:
\bea
    \d \bm{S}&=&\sum_{k=0}^s\bigg(-\frac{1}{2}\bigg)^s\bigg(\frac{1}{s!}\bigg)^2\int\rd^{3|4}z\bigg\{s^2(-\langle g_{(2s-2k-1)}|\cA'+\ri\langle g_{(2s-2k+1)}|\cB')\d|  E^{(1)}_{(2s-2k)}\rangle\non\\
    &&~~~~~~~~~~~~~~~~~~~-s^2 (\langle j_{(2s-2k-1)}|\bar{\cA}'+\ri\langle j_{(2s-2k+1)}|\bar{\cB}')\d|  E^{(2)}_{(2s-2k)}\rangle+\text{h}.\rc.\bigg\}~,
\eea
which we can rearrange to get
\bea
    \d \bm{S}&=&\sum_{k=0}^s\bigg(-\frac{1}{2}\bigg)^s\bigg(\frac{1}{s!}\bigg)^2\int\rd^{3|4}z\bigg\{-s^2\langle g_{(2s-2k-1)}|(\cA'\d|  E^{(1)}_{(2s-2k)}\rangle-\ri\cB'\d|  E^{(1)}_{(2s-2k-2)}\rangle)\non\\
    &&~~~~~~~~~~~~~~~~~~~-s^2 \langle j_{(2s-2k-1)}|(\bar{\cA}'\d|  E^{(2)}_{(2s-2k)}\rangle+\ri\bar{\cB}'\d|  E^{(2)}_{(2s-2k-2)}\rangle)+\text{h}.\rc.\bigg\}~.\non\\
\eea
We then shall compute
\bea
    &&(\cA'\d|  E^{(1)}_{(2s-2k)}\rangle-\ri\cB'\d|  E^{(1)}_{(2s-2k-2)}\rangle)\non\\
    &&= -\frac{(k-1)!(2s-k-1)!}{s}\bigg(\frac{2s-k}{2s-2k+1}\cA'\d| W_{(2s-2k)}\rangle-\ri\frac{k}{2s-2k-1}\cB'\d| W_{(2s-2k-2)}\rangle\bigg)\non\\
    &&= -\frac{(k-1)!(2s-k-1)!}{s}\bigg[\frac{2s-k}{2s-2k+1}\cA'\bigg(-\frac{(2s-k+1)(2s-k+2)}{4s(s+1)(2s-2k+3)}\cA'\bar{\cA}'| l_{(2s-2k+2)}\rangle\non\\
    &&-\frac{k(k+1)}{4s(s+1)(2s-2k-1)}\cB'\bar{\cB}'| l_{(2s-2k-2)}\rangle\non\\
    &&-\ri \frac{k(2s-k+1)}{4s(s+1)}\cK| l_{(2s-2k)}\rangle\bigg)-\ri \frac{k}{2s-2k-1}\cB'\bigg(-\frac{(2s-k)(2s-k+1)}{4s(s+1)(2s-2k+1)}\cA'\bar{\cA}'| l_{(2s-2k)}\rangle\non\\
    &&-\frac{(k+1)(k+2)}{4s(s+1)(2s-2k-3)}\cB'\bar{\cB}'| l_{(2s-2k-4)}\rangle\non\\
    &&-\ri \frac{(k+1)(2s-k)}{4s(s+1)}\cK| l_{(2s-2k-2)}\rangle\bigg)\bigg]~,
\eea
where we once more make use of properties from \eqref{Operator Action 3D} and \eqref{Central Charge Identities} to arrive at
\bea
    (\cA'\d|  E^{(1)}_{(2s-2k)}\rangle-\ri\cB'\d|  E^{(1)}_{(2s-2k-2)}\rangle) = 0~.
\eea
A similar calculation gives us
\bea
    (\bar{\cA}'\d|  E^{(2)}_{(2s-2k)}\rangle+\ri\bar{\cB}'\d|  E^{(2)}_{(2s-2k-2)}\rangle)=0~,
\eea
hence, the action \eqref{Action Recap} is invariant under \eqref{Gauge Variation Recap}.

\end{appendices}
 
\end{document}